\documentclass[onecolumn]{els-mrw} 

\usepackage{amsmath,amssymb,amsfonts,amsthm,makeidx,graphicx}
\usepackage{txfonts}
\usepackage{helvet}
\usepackage{bbold}
\usepackage{subfigure}

\begin{document}

\chapter{Hamiltonian Chaos}
\label{hamilton}

\author[1]{Steven Tomsovic}%

\address[1]{\orgname{Washington State University}, \orgdiv{Department of Physics and Astronomy}, \orgaddress{Pullman, WA 99164-2814 USA}}

\articletag{Chapter Article tagline: Chaos}

\maketitle

\begin{glossary}[Keywords]
Chaos, unstable and stable manifolds, heteroclinic and homoclinic tangles, symbolic dynamics, periodic trajectories, Maslov indices, perturbations, classical actions, complex trajectories
\end{glossary}
 
\begin{abstract}[Abstract]
Through semiclassical methods the subject of quantum chaos motivates and depends on Hamiltonian chaos research.  Presented here is a selection of some of the most important Hamiltonian chaos topics that in this way get directly related to any of a variety of quantum chaos research problems.  The overall emphasis is on intuitive explanations and illustrations of various ideas with the references containing more mathematically rigorous expositions.  The chapter begins with a description of various useful theoretical and computational tools of chaos research, e.g.~surfaces of section, paradigms of chaos, stability analysis, and symbolic dynamics...  This is followed by discussions regarding the geometry of chaos, how chaotic systems respond to perturbations, and the complexification of Hamiltonian dynamics.  Most of the contents concern topics that are long established and well known by now.  The exceptions involve complications that arise from the complexification of phase space variables and time.  There are relatively recent results summarized regarding the appropriate branches of determinant phases and how to distinguish complex trajectories that are associated with classically allowed processes versus those associated with non-allowed ones.  
\end{abstract}

\section{Introduction}
\label{Intro}

~\indent {\it ``The arranger of disorder with your strange and simple rules...''}, Gypsy, Solitude Standing, Suzanne Vega (1987)\\ 

The research field with the moniker {\it quantum chaos}~\cite{Brody81, GutzwillerBook, Chaosbook1, Stockmannbook, Dalessio16, Haake18, Richter22, Altland23} has come to include an extremely broad range of topics, involving atomic, molecular, nuclear, spin, and condensed matter systems, as well as quantum gravity and others.  Of its three theoretical pillars~\cite{Richter22}, semiclassical theory, random matrix theory, and the theory of disordered systems, the first provides the direct connection to classical Hamiltonian chaos.  Thus, the principal motivation for this chapter is the study of quantum systems possessing classical analogs with purely chaotic dynamics.  Semiclassical theory gives a construction of asymptotic quantum properties (and wave mechanical system properties too), often denoted in shorthand by $\hbar \rightarrow 0$ (wave vector $k\to \infty$), based on exclusively classical mechanical quantities, and they give the deepest insight into the Correspondence Principle~\cite{Bohr13a} connecting quantum and classical mechanics.  If a quantum system's classical analog possesses a chaotic dynamics, then it becomes necessary to have a deep understanding of Hamiltonian chaos in order to understand the asymptotic behaviors of such a system.

The subject of Hamiltonian chaos took a great leap forward in the 1890's with Poincar\'e's three volume tome on the three-body problem in celestial mechanics~\cite{Poincare92, Poincare93, Poincare99}; also very important is the publication of Lyapunov's thesis in the same decade~\cite{Lyapunov92}.  Poincar\'e described extremely important geometrical properties of chaos for the first time.  For example, he noted that there exist trajectories that converge towards each other in the infinite future whose complete set of phase points collectively comprise a stable manifold.  Likewise, backward in time converging trajectories' phase points comprise the unstable manifold.  There are trajectories living on the infinity of intersections of an unstable and stable manifold known as either heteroclinic or homoclinic trajectories, depending on whether the unstable and stable manifolds belong to the same trajectory.  The wildly complicated patterns arising from each of these manifolds (either unstable or stable) foliate the chaotic region of phase space and their joint structure is known as a heteroclinic (homoclinic) tangle.  He had some quite remarkable insights before the advent of powerful computers about just how complicated these patterns had to be.  Lyaponuv introduced a stability analysis of solutions to ordinary differential equations, which includes the exponential instability whose exponents carry his name and which are intimately related to Poincar{\'e}'s unstable and stable manifolds.  These manifolds, exponents, trajectories, and tangles have since been foundational in a great deal of research.   More generally, the topic of chaos has since grown into an immense subject and it is vital here to have a rather restricted focus.  Thus, the subjects included in this chapter are selected from those which can be directly connected to quantum properties through semiclassical analysis.  Many subjects are largely excluded, e.g.~Lyapunov exponents, which measure the multiple exponential rates at which trajectories are separating, dynamics in random media, stochastic processes, or subjects somewhat related to Hamiltonian chaos, e.g. strange attractors, fractals, etc...  

The foundations of semiclassical methods arise from the approximations of stationary phase and the method of steepest descents typically applied to evolving wave functions,  Green functions, and/or path integrals~\cite{Gutzwiller70, Gutzwiller71, Balian71, Maslov81}.  The stationary phase and saddle points emerging from these analyses have properties completely determined by classical trajectories that satisfy two-point boundary value constraints, e.g.~the periodic trajectories appearing in the Gutzwiller trace formula~\cite{Gutzwiller71}.  The essential classical trajectory properties involved are: i) Hamilton's principal and characteristic functions~\cite{Goldstein80}, known colloquially as classical actions; ii) various determinants of stability matrices~\cite{Gutzwiller71} and sub-blocks thereof~\cite{Heller91, Tomsovic18b}; and iii) geometric indices, which are loosely called Maslov indices here~\cite{Keller58}.  Some indications are given ahead concerning particular properties of Hamiltonian chaos that are directly involved in some quantum chaos topic.  

The dynamical possibilities of Hamiltonian systems lie somewhere between the two extremes of integrability and full chaos~\cite{Poincare92, Birkhoff27, Lichtenberg92}.  The former occurs where there are as many constants of the motion in involution as degrees of freedom, and nearly all the trajectories in such systems exhibit stable (or marginally stable) evolution, which naturally is reflected in the properties of their corresponding stability matrices.  Under perturbation, the constants of motion no longer hold perfectly, and systems transition toward chaotic dynamics.  The transition to chaos is extremely interesting leading to Kolmogorov-Arnol'd-Moser (KAM) theory~\cite{Kolmogorov54, Arnold63, Moser62} and the subject of periodic trajectory bifurcations/catastrophe theory~\cite{Thom89, Arnold92}, but for the most part those subjects are also not included here.  Instead, the latter possibility, i.e.~full chaos, is assumed.  It comes with a possible hierarchy of increasingly strong ``chaotic'' features~\cite{Ozoriobook, Ullmo16}, namely ergodicity~\cite{Boltzmann71} whereby a typical trajectory explores arbitrarily close to all the available phase points, mixing~\cite{Koopman32} in which one view focusses on neighboring trajectories doing their exploration of phase space independently beyond some time, Kolmogorov (K-) system~\cite{Kolmogorov58} in which neighboring trajectories diverge exponentially rapidly,  and other stronger features as seen in Bernoulli systems~\cite{Sinai59}, and Anosov systems~\cite{Anosov67}.  As the Bohigas-Giannoni-Schmit conjecture connecting the appearance of random matrix theory fluctuation properties in the spectra of quantized chaotic dynamical systems specifically invoked K-system properties~\cite{Bohigas84}, which also implies ergodicity and mixing, here the term chaos is understood to be shorthand for dynamical systems possessing at least ergodicity, mixing, and K-system properties (exponential divergence/positive Lyapunov exponents).  Thus, nearly all the trajectories explore the entire available phase space, do so in an independent way, and are exponentially unstable with $D-1$ positive Lyapunov exponents, where $D$ is the number of degrees of freedom and, unless otherwise stated, it is assumed that the total energy of the system is the only constant of the motion (there is no real loss of generality in this assumption).

A paradoxical feature of Hamiltonian dynamics is that although the individual trajectories of integrable systems are stable and marginally stable, integrable systems are structurally unstable~\cite{Lichtenberg92}.  On the other hand, individual chaotic trajectories are exponentially unstable, and yet chaotic systems possess a kind of strong structural stability~\cite{Andronov37, Devaney22}.  For example, in integrable systems small perturbations introduce a small denominator problem and generically alter continuous parameter families of periodic trajectories on a given energy surface (they form tori with rational frequency ratios) into resonances possessing isolated stable and unstable periodic trajectories.  On the other hand, in a chaotic dynamical system, the periodic trajectories are isolated to begin with and under small perturbations remain so and can be continuously followed through deformation generating slightly modified classical actions.  Bifurcations are relatively exceptional.  Furthermore, the foliation of phase space by unstable and stable manifolds and the partitioning of phase space into like evolving trajectories for finite times are barely altered.  It turns out that for chaotic systems that this structural stability guarantees that a first order semiclassical perturbation theory capturing quantum perturbation effects can be based on a first order correction to classical actions, which is discussed below.

A beautiful property of quantum mechanics is that there are phenomena which do not occur in the classical dynamics.  A quintessential example is that of quantum tunneling~\cite{Messiah61, Merzbacher02}, which is inherently defined by its contrast to behavior found in the quantum system's classical analog.  Implicit in this contrasting perspective is that classical mechanics is built of exclusively real positions, momenta, and time.  However, properly analytically continuing these quantities to their complex counterparts enables the description of quantum ``classically non-allowed'' processes via complex trajectories~\cite{Miller72, Creagh98}.  In addition, under any circumstance where developing the quantum asymptotic behavior requires applying the method of steepest descents (saddle point approximation), as opposed to the stationary phase approximation, the solutions generally require complexified trajectories~\cite{Klauder78, Baranger01}.  This includes the important cases of wave packet propagation~\cite{Huber87, Huber88, Pal16} and using mean field trajectories for many-body bosonic systems~\cite{Tomsovic18, Tomsovic18b}.  Therefore, some general discussion of complexified classical dynamics is included even if not all of it is exclusively related to chaos.  There are a number of complications introduced by analytic continuation that necessarily impact the chaotic case as well.

Not surprisingly, many of the subjects that appear ahead, explicitly or implicitly, also appear in a significant number of the other contributions to this collection.  Subjects including, but not limited to, ergodicity, periodic trajectories, uniformity principle, action correlations (encounters), unstable and stable manifolds, exponential sensitivity to initial conditions, Peres' fidelity, tunneling, complex classical mechanics, and systems such as the stadium billiard and kicked rotor figure prominently in many other chapters in this volume, e.g.~\cite{Nonnenmacher26, Urbina26, Mueller26, Harayama26, Lakshminarayan26, Hentschel26, Novaes26, Garcia26, Shudo26, Dietz26, Benenti26}.

Here we emphasize illustration and intuitive descriptions as opposed to mathematical rigor for which the reader can consult many of the references included.  It is structured as follows: Sec.~\ref{sec:back} begins with some discussion of theoretical tools useful for the study of chaotic dynamical systems followed by remarks on essential properties of chaotic trajectories.  The next section covers structural and geometric features of chaos.  This is followed by Sec.~\ref{sec:strucstab} which considers how chaotic systems respond to small perturbations.  Section~\ref{sec:complex} covers a few of the complications that come with analytically continuing the dynamics to complex positions and momenta.  Finally, there are some concluding remarks.

\section{Background}
\label{sec:back}

The subject begins with the seemingly innocent looking Hamilton's equations acting in a phase space of $(\vec q, \vec p)$,
\begin{equation}
\label{eq:Hameq}
\frac{{\rm d} q_j }{{\rm dt}} = \frac{\partial H(\vec q, \vec p; t)}{\partial p_j} \ , \qquad
\frac{{\rm d} p_j }{{\rm dt}} = -\frac{\partial H(\vec q, \vec p; t)}{\partial q_j} 
\end{equation}
where the pairs ($q_j, p_j$) are components of canonically conjugate position and momentum variables, respectively, and $j=1, 2, 3,... D$.  The solutions are trajectories, $(\vec q(t), \vec p(t))$, determined uniquely by their initial conditions, $(\vec q(0), \vec p(0))$.  If the partial derivatives of the Hamiltonian function are nonlinear, then chaos may result.  For autonomous systems, Poisson brackets ($\{...\}$) may be used to determine whether there are conserved quantities under the dynamics (constants of the motion),
\begin{equation}
\label{eq:com}
\frac{{\rm d} f(\vec q, \vec p)}{{\rm d}t} = \sum_{j=1}^D \frac{\partial f(\vec q, \vec p)}{\partial q_j} \frac{{\rm d} q_j }{{\rm dt}} + \frac{\partial f(\vec q, \vec p)}{\partial p_j} \frac{{\rm d} p_j }{{\rm dt}} = \sum_{j=1}^D \frac{\partial f(\vec q, \vec p)}{\partial q_j} \frac{\partial H(\vec q, \vec p)}{\partial p_j} - \frac{\partial f(\vec q, \vec p)}{\partial p_j} \frac{\partial H(\vec q, \vec p)}{\partial q_j} = \{f(\vec q, \vec p),H(\vec q, \vec p)\} \ ,
\end{equation}
which vanishes at all the available phase points, $(\vec q, \vec p)$, for globally conserved quantities.  It is assumed here for the most part that fully chaotic systems have no conserved quantities independent of the Hamiltonian itself.  Note also that the Poisson bracket of a canonical pair of variable components equals unity.

From these equations for fully chaotic systems an extremely rich and complicated dynamics emerges, far more interesting in many ways than the heavily constrained (by $D$ or more constants of the motion) regular dynamics of integrable systems.  Indeed, Poincar\'e's discovery and description in the late 1890's~\cite{Poincare99} of the geometry of structures such as unstable and stable manifolds, their foliations of the available phase space, their intersections involving heteroclinic (homoclinic) trajectories, etc... makes for fascinating reading.

\subsection{Useful tools}
\label{sec:tools}

With a few exceptions, such as the baker's map discussed ahead, Hamilton's equations are not analytically solvable for all the trajectories of a chaotic system.  Many studies have largely proceeded through numerical calculations, which leads to an interesting conundrum about whether it is even possible to follow numerically a ``true'' trajectory for long times~\cite{Sauer97}.  Nevertheless, a significant number of useful tools have been developed in the process, which are introduced next.

\subsubsection{Poincar\'e surface of section and mappings}
\label{sec:poincare}

For systems with two degrees of freedom, the phase space is four dimensional, and if the Hamiltonian does not explicitly depend on time (an autonomous system), the trajectories remain within a three dimensional, presumably complicated, constant energy surface.  This can be rather difficult to visualize.  Instead, a Poincar\'e surface of section~\cite{Poincare92} can be constructed in which one records the location of a trajectory only when it crosses a particular surface within the volume of the constant energy surface (one dimension lower) and from a particular orientation, e.g.~for a hyperplane only record the phase point if crossing the surface from, say, the left, but not the right.  The resulting two-dimensional image can give a rather complete description of the dynamics.  The idea of a surface of section extends to any number of degrees of freedom, but it becomes difficult to visualize for systems with more than two degrees of freedom; see however~\cite{Richter14, Firmbach18}.

Conceptually in this way the dynamics of a continuous time system can be converted into a discrete mapping, i.e.~the mapping that takes any intersection point and maps it to its next intersection point.  This generates a symplectic map and in some sense generalizes Hamiltonian dynamics.  Every continuous time Hamiltonian system can be converted to a variety of symplectic maps, however, not every symplectic map has a correspondence to a continuous time Hamiltonian system.

As an example of a connection between a continuous time Hamiltonian system and a symplectic mapping, consider the kicked rotor~\cite{Chirikov79}.  It serves as the quintessential example of how the introduction of a perturbation that creates resonances in the dynamics leads to chaos.  It consists of a mechanical particle constrained to move on a ring, kicked at discrete unit time intervals, 
\begin{equation}
H(q,p) = \frac{p^2}{2} - \frac{K}{4\pi^2} \sum_{n=-\infty}^\infty \cos (2\pi q)\delta(t-n) \ .
\end{equation}
By recording $(q,p)$ only at integer times, kicking first and with free propagation afterward, it leads to the standard map
\begin{eqnarray}
p_{t+1} &=& p_t- {K\over 2 \pi} \sin 2 \pi q_t \qquad \mbox{mod}\ 1 \nonumber \\
q_{t+1} &=& q_t + p_{t+1} \qquad\qquad\quad \mbox{mod}\ 1 \ ,
\end{eqnarray}
\begin{figure}[t]
\centering
\includegraphics[width=.33\textwidth]{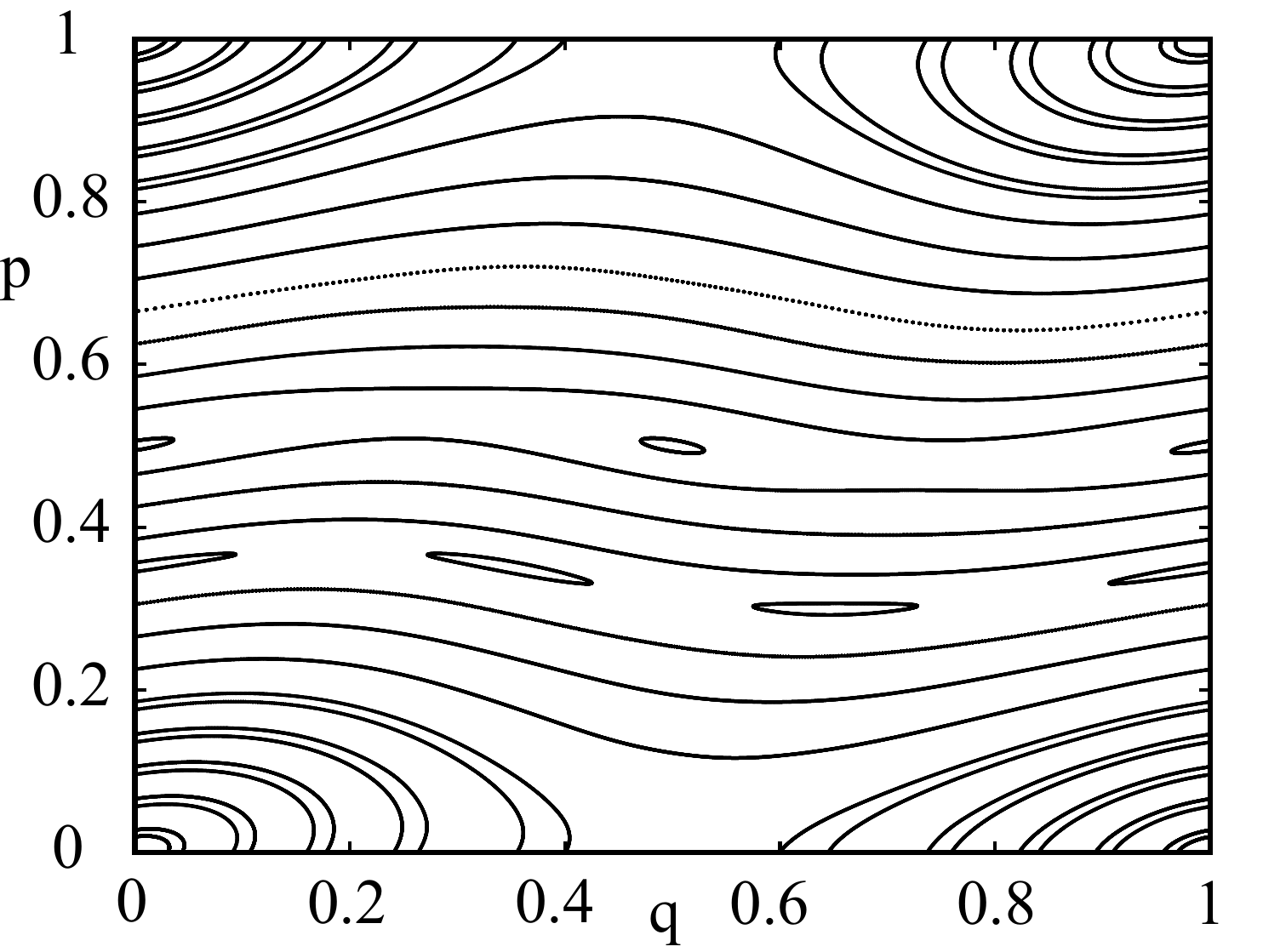}\includegraphics[width=.33\textwidth]{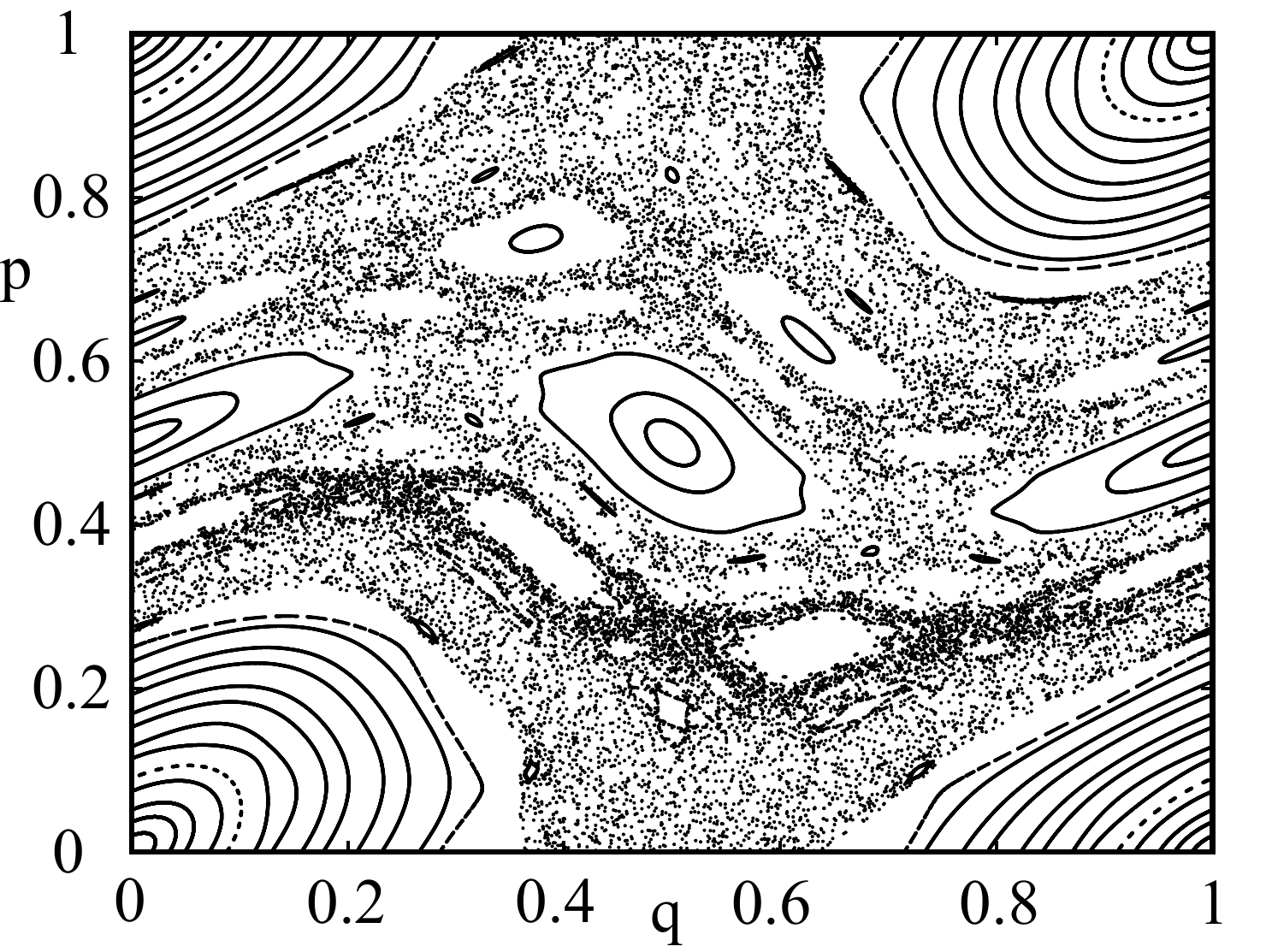}\includegraphics[width=.33\textwidth]{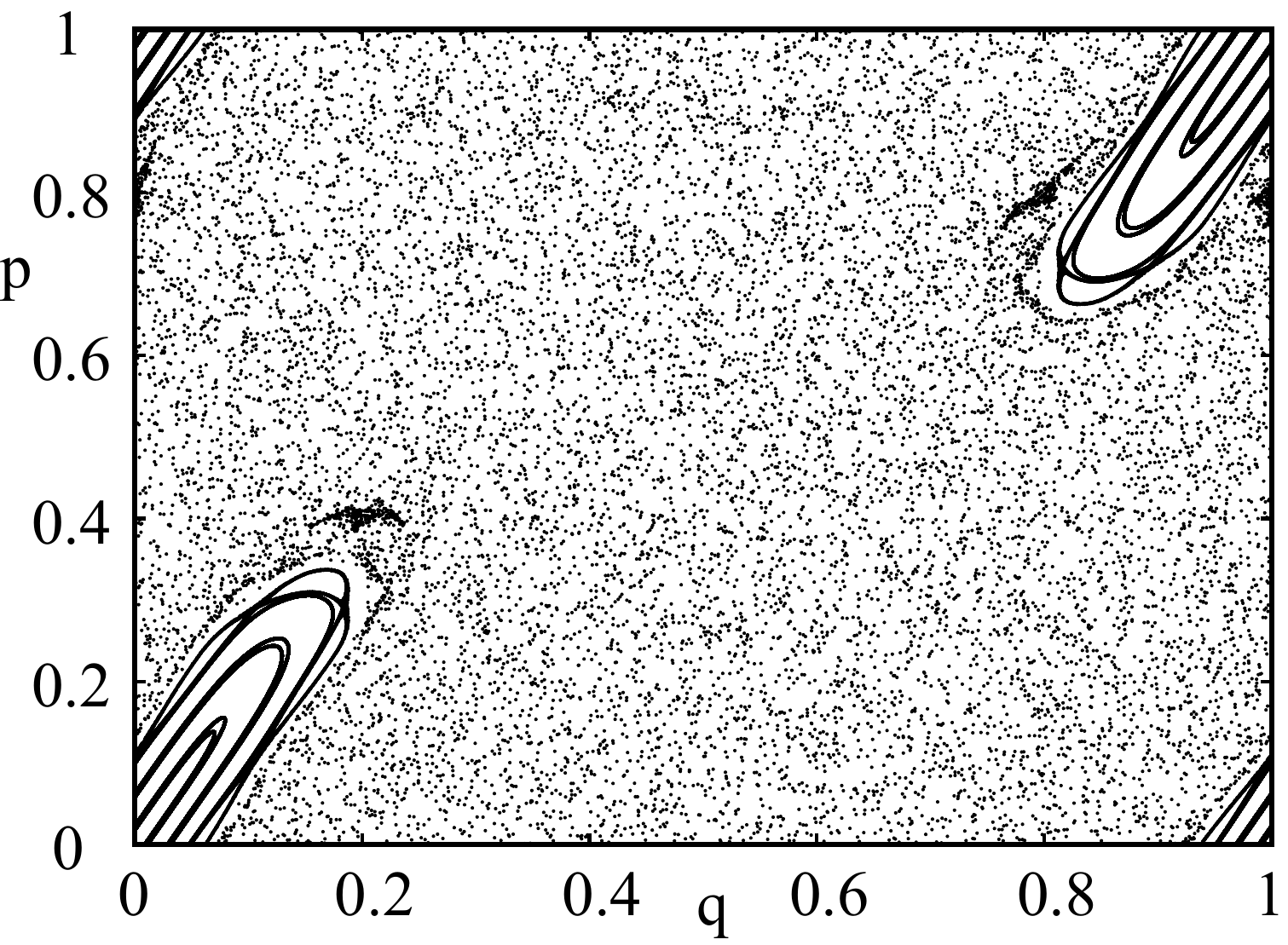}
\caption{Three phase portraits of the standard map. The values of $K$ are $0.4$, $1.1$, and $4.0$, left to right, respectively.  Those regions filled with regular trajectories are easily seen as individual trajectories that lie on lines, whereas chaotic trajectories fill two-dimensional regions.  The left panel example is in a near-integrable dynamical regime, the middle panel more of a mixed phase space regime, and the right panel is furthest toward full chaos of these examples.  Typical of smooth Hamiltonian systems, there are both regions of stable trajectories (regular) and unstable trajectories (chaotic).  Sticky phase space regions (higher density of points) can be seen in the middle and right panels due to transport barriers, either effects of cantori or small fluxes through turnstiles.  This is an artifact of finite time propagation, eventually the chaotic regions would fill in uniformly.}
\label{fig:soskr}
\end{figure}
if considered on the unit phase space torus.  If the mod(1) condition is dropped from the momentum equation, then the phase space has the geometry of an infinite cylinder.  This dynamical system makes a transition from integrability for $K=0$ to dominantly chaotic dynamics for values of $K\gtrsim 2\pi$.  In Fig.~\ref{fig:soskr} the transition is illustrated from the near-integrable regime to approaching the predominantly chaotic regime, left to right.  A number of distinct features are visible.  In the near-integrable regime (left panel), the vast majority of the trajectories are confined to move on reduced dimensional structures called tori~\cite{Arnold67}, which are somewhat deformed if compared to the integrable $K=0$ case (which would appear as horizontal lines in the surface of section).  These are known as KAM tori in the literature~\cite{Kolmogorov54, Arnold63, Moser62}, and they have the property of being Lagrangian (sub)manifolds ($D$-dimensional sets of phase points)~\cite{Arnold78}, which has important consequences in classical, semiclassical, and thus quantum mechanics~\cite{Maslov81}.  In addition, a resonance is immediately created with its stable periodic trajectory at $(q,p)=(0.0,0.0)$ and unstable one at $(q,p)=(0.5,0.0)$.  Increasing $K$ just beyond the destruction of the last KAM torus ($\gtrsim 0.971635$)~\cite{Greene79}, generates the middle panel surface of section.  A significant chaotic region has emerged, but not all of the KAM islands of regular motion are destroyed.  The interfaces between regular islands and chaotic seas are extremely complicated, fractal in nature~\cite{Greene81}, and there exist transport barriers in the form of cantori~\cite{MacKay84a, MacKay84b} and turnstiles~\cite{Channon80, Bensimon84, MacKay87}; see Ref.~\cite{Meiss15} and references therein.  A large significant sticky region is visible as a locally higher density of points in the middle panel of Fig.~\ref{fig:soskr}.  Such structures have been linked to a variety of localizing phenomena involving eigenstates of the quantized version of the system~\cite{Geisel86, Brown86, Radons88, Bohigas93}.  Increasing $K$ further reduces the remaining KAM islands as seen in the right panel, though perhaps never attaining fully chaotic behavior in the strict sense that remaining stable and marginally stable trajectories are of measure zero.

There are further interesting dynamical features of the kicked rotor that are not immediately visible in these surfaces of section, especially for the phase space cylinder version.  For some $K$ values, there exist KAM islands associated with so-called accelerator modes because the momentum increases ballistically and leads to anomalous diffusion~\cite{Karney77, Chirikov79}.  There are also analogous quantum modes that lead to coherent acceleration of wave packets~\cite{Oberthaler99, Fishman03}.  In regimes with standard classical diffusion, a momentum representation of the quantized system leads to the Lloyd model of Anderson localization~\cite{Fishman82, Lloyd69, Moore95}, and a quantum suppression of  diffusion~\cite{Shepelyansky83}.

The absence of mathematically rigorous full chaos as occurs with the kicked rotor also seems to be typical of the dynamics of smooth Hamiltonians.  One smooth potential thought to be fully chaotic for quite awhile, the $x^2y^2$-potential, nevertheless turned out to have very small islands of regularity~\cite{Dahlqvist90}.  The author is not aware of a bounded smooth system for which a proof of zero measure stable and marginally stable trajectories exists.  For smooth chaotic systems, it seems rather typical that there remains non-zero measure islands of stable motion, though they may be tiny and extremely hard to find without some sophisticated search method.  Even so such islands may influence the behaviors of the chaotic trajectories in some significant ways~\cite{Tomsovic07, Manchein09a} leading to, for example, anomalous time scalings in the fluctuations of finite time stability exponents.  It is worth noting that a system of many coupled kicked rotors, as a toy model of a many-body system, lends itself to analytic predictions for the full spectrum of positive Lyapunov exponents~\cite{Lakshminarayan11},  and the Kolmogorov-Sinai (KS) metric entropy (in this system equal to the sum of positive Lyapunov exponents)~\cite{Kolmogorov58, Kolmogorov59, Sinai59, Pesin77}.  Other models of many-body systems, such as the Bose-Hubbard model, also exhibit chaos~\cite{Kolovsky04}, but the full extent of their chaos tends to be difficult to assess and not  really as ``pur et dur'' (as an old saying goes, perhaps due to M.~C.~Gutzwiller) as some of the simpler models.

\subsubsection{Chaos paradigms}
\label{sec:paradigm}

Examples of chaos abound in the solar system, inner planet orbits and orientations, especially Mercury, the tumbling of Saturn's moon Hyperion, motions of objects in the asteroid and Kuiper belts, to give a few examples~\cite{Wisdom87, Sussman92, Laskar94, Lecar01}.  Their origins are due to overlapping resonances from three or more bodies interacting.  However, the solar system studies can be extremely complicated, and in order to understand the essence of chaos or to investigate experimentally (quantum) chaos in the laboratory a wide variety of much simpler paradigms have been introduced.  The double pendulum provides a relatively simple classroom demonstration of chaos~\cite{Shinbrot92}, for example.  Three early quantum systems with predominantly chaotic classical analogs motivating experimental and theoretical studies include, the anisotropic Kepler problem~\cite{Gutzwiller73, Devaney78}, diamagnetic Rydberg atoms~\cite{Delande86, Holle88, Friedrich89, Iu91}, and the Hydrogen atom in a microwave cavity~\cite{Bayfield74, Galvez88, Koch95}.

\begin{figure}[t]
\centering
\includegraphics[width=0.9\textwidth]{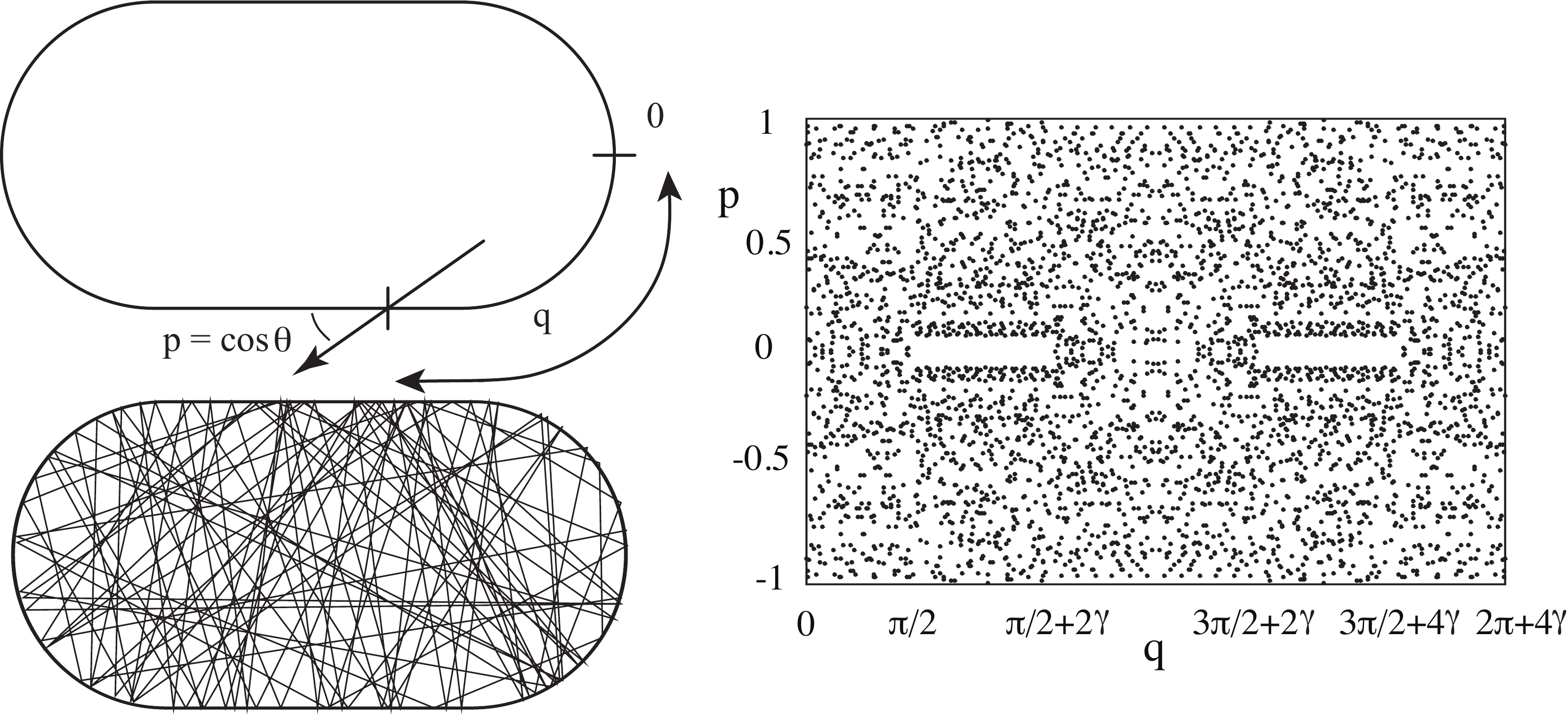}
\caption{A drawing of a chaotic trajectory in the stadium billiard and its surface of section image.  The semicircle radius of curvature is unity and the straight edge length divided by the diameter, $\gamma$, equals unity (i.e.~$L=2R=2$).   The trajectory for initial condition, $(5.0, 0.04)$, is shown for the first hundred reflections in the lower left, and the first thousand phase points in the surface of section is shown to the right.}
\label{fig:sos1000}
\end{figure}

Even simpler systems have been introduced that sometimes lend themselves to rigorous mathematical treatments.  One important class of dynamical systems is billiards, a particle with fixed kinetic energy moving within a domain and specularly reflecting at hard walls.  An important simplicity is their homogeneous nature.  Any trajectory on one energy surface can be scaled to any other energy surface.  Thus, they only require study on a single energy surface, and the rest are connected by a simple scaling.  Whether a billiard is chaotic or not is only dependent on its shape.  A couple of useful examples for which proofs of chaotic properties exist include Sinai's billiard~\cite{Sinai70} (a unit cell of a periodic Lorentz gas~\cite{Lorentz05a, Lorentz05b, Lorentz05c}) and the Bunimovich stadium billiard~\cite{Bunimovich74} pictured in Fig.~\ref{fig:sos1000}.  The latter is used extensively here to illustrate many generic properties of fully chaotic systems.  

Two degree-of-freedom billiards are typically straightforward to convert into dynamical maps.  The coordinate system is illustrated in Fig.~\ref{fig:sos1000} for the stadium.   A conjugate pair of phase space variables that constitute a surface of section are given by position along the boundary and the cosine of the angle as pictured in the upper left.   They are used for the surface of section illustration of a single chaotic trajectory on the right side.
As the number of reflections tends to infinity, this trajectory fills out the surface of section in a uniformly dense way to as small a region as one wishes to investigate.   However, it is also clear that for finite times (finite numbers of reflections), there are a great deal of fluctuations, leading to two regions that are essentially avoided, at least for the first thousand reflections.  It turns out that there is a set of measure zero trajectories known as bouncing ball orbits, which are periodic after two bounces and reflect normally from the straight edges.  The dynamics is very nearly stable in this region of phase space, and although it takes a long time to enter this region, once there, a trajectory remains a long time.  In the end those two aspects balance out to generate an ergodicity.

Finally, it is worth mentioning perhaps the simplest chaotic dynamical systems, symplectic maps that do not derive from continuous time Hamiltonian counterparts.  The Arnol'd cat map~\cite{Arnold67}, and the baker's map~\cite{Hopf37} are two excellent examples.  Ahead, the baker's map in the unit square is used for further discussion as essentially everything is known about its dynamics analytically.  

\subsection{Hamilton's Principal and Characteristic Functions}
\label{sec:hpcf}

To be a bit more definite about the origin of classical quantities in semiclassical theories, consider first the expression for the van Vleck-Gutzwiller propagator~\cite{Vanvleck28, Keller58, Gutzwiller70}
\begin{equation}
\label{eq:vanvleck}
G({\bf q},{\bf q}^\prime;t) \approx \left({1\over 2\pi i\hbar}\right)^{D/2} \sum_j \bigg|{\rm Det}
\bigg( {\partial^2 S_j({\bf q},{\bf q}^\prime;t)\over \partial {\bf q}
\partial {\bf q}^\prime}\bigg)\bigg|^{1/2} \exp 
\left(iS_j({\bf q},{\bf q}^\prime;t)/\hbar
-{i \pi\nu_j \over 2}\right) \ ,
\end{equation} 
which expresses the time dependent Green function in terms of a summation over properties of trajectories that evolve from ${\bf q}^\prime$ to ${\bf q}$ in a time $t$ labelled by the index $j$.  The function $S_j({\bf q},{\bf q}^\prime;t)$ is Hamilton's principal function for the trajectory and $\nu_j$ the geometric index (Maslov index), i.e.~count of caustics the trajectory passes through\footnote{Throughout this contribution, the Greek letter $\nu$ is reserved for geometric indices and $\mu$ for stability exponents.}.  The determinant can be expressed in terms of a sub-block of the stability matrix
\begin{equation}
{\rm Det}\left[ {\partial^2 S_j \left({\bf q},{\bf q}^\prime;t\right) \over  \partial {\bf q}
\partial {\bf q}^\prime} \right] = - \left[{\rm Det}\left({\bf M}_{21} \right)\right]^{-1} \ ,
\end{equation}
where the stability matrix and sub-block notation is defined ahead in Eqs.~\eqref{eq:stabm} and \eqref{eq:stabder}.  Similarly, the Gutzwiller trace formula~\cite{Gutzwiller71} expresses the quantum density of states, i.e.~the trace of the energy Green function, as a mean density plus a sum over properties of primitive periodic trajectories (first repetition) and their repetitions in order to capture the fluctuations about the mean,
\begin{equation}
\label{eq:gtf}
\rho(E) = -\frac{1}{\pi} Im\ {\rm Tr}\left[G({\bf q},{\bf q}^\prime;E)\right] \approx \overline{\rho(E)} + \frac{1}{i\hbar} \sum_{p} \tau_{p}\sum_{r=1}^\infty \frac{1}{\left|{\rm Det}\left(M^r_{p}-\mathbb{1}\right)\right|^{1/2}} \exp\left[ ir\left(\frac{W_{p}}{\hbar} - { \pi\nu_p \over 2}\right)\right] \ ,
\end{equation}
where $W_{p}$ is Hamilton's characteristic function, $\tau_p$ is the period, and ${\bf M}_p$ is the stability matrix for a primitive periodic trajectory on the $E$ energy surface.  The main quantities in the phase terms above, Hamilton's principal and characteristic functions, result from the stationary phase approximations used to obtain those semiclassical expressions. They appear in Hamilton-Jacobi and reduced Hamilton-Jacobi theory~\cite{Goldstein80}, respectively.  Their presence is not surprising given, for example, that the principle of least action states that the action, i.e.~Hamilton's principal function, is stationary for classical trajectories.  It is relatively common to see both of these functions referred to as classical actions.  

Similarly, in time-dependent WKB theory, Hamilton's principal function integrated along a stationary phase trajectory is the main contributor to the phase which appears in that term, i.e.~it shows up multiplied by $i/\hbar$ in the argument of an exponential.  Hamilton's principal function integrated along such a trajectory, labelled by $j$, is expressed as an integral over the Lagrangian, $\cal L$,
\begin{equation}
\label{eq:hpf}
S_j = \int_j  {\cal L}\ {\rm d}t\ .
\end{equation}
In Einstein-Brillouin-Keller (EBK) theory~\cite{Einstein17, Brillouin26, Keller58} or torus quantization Hamilton's characteristic function is integrated along a phase space curve, $\cal C$, restricted to lie on a torus, which is a Lagrangian manifold.  It may be some segment of a trajectory,
\begin{equation}
\label{eq:hcf}
W_{\cal C} = \int_ {\cal C} {\bf p}\cdot {\rm d}{\bf q} \ .
\end{equation}
but typically is not.  This is because a useful property of Lagrangian manifolds is the invariance of $W_{\cal C}$ under the deformation of paths with the same endpoints (and topological features such as winding numbers) so long as they remain on the Lagrangian manifold of the original path.  For example, the quantization conditions most simply apply to paths chosen to follow the fundamental cycles of the torus, not necessarily along a trajectory.  In more than one degree of freedom, trajectories generally do not follow the cycles.

Even for chaotic systems, where EBK theory cannot apply~\cite{Einstein17}, the invariance of $W_{\cal C} $ under path deformation on Lagrangian manifolds is a critically useful property in a variety of ways.  One crucial example is that although trajectories do not lie on tori, they do make up unstable and stable manifolds, which are Lagrangian manifolds.  This property and the action principle of~\cite{MacKay84a} can be exploited to derive a multitude of relations amongst periodic, heteroclinic, and homoclinic trajectories~\cite{MacKay84a, Li18}, between periodic, heteroclinic, and homoclinic trajectories~\cite{MacKay84a, Ozorio89, Li17}, action corrections to cycle expansions~\cite{Cvitanovic88, Cvitanovic89, Li18}, and Sieber-Richter action relations~\cite{Sieber01, Li17b}; see Sec.~\ref{sec:geometry} and comments near the end of Sec.~\ref{sec:sets}.

Although, often somewhat less of a primary concern in classical physics contexts, in the context of a semiclassical perspective of quantum chaos, classical actions are the major contributor to phases, and hence the main determinant of interference phenomena, which depend critically on the relative phases of the many stationary phase contributions.  There is an important correction to the phases determined above, however, arising from classical caustics~\cite{Keller58, Maslov81}.  In the course of performing the above integrations, it is quite typical for a trajectory to pass through turning points, focal points, any of a variety of classical catastrophes~\cite{Arnold92}, each one leading to a $\pi/2$ phase correction.  The count of caustics, $\nu$, which is called here the Maslov index, depends on the presence of zeros in determinants based on stability matrices; caustics may may come with a multiplicity factor greater than unity, which also must be reflected in the count $\nu$.  These matrices are discussed next. 

\subsection{Stability analysis}
\label{sec:stability}

It is commonplace to see positive Lyapunov exponents mentioned in any discussion of K-systems and the characterizations of just how unstable their dynamics may be~\cite{Lichtenberg92,Ott97}.  They are strictly speaking an infinite time limiting property of the chaotic dynamics.  For example, the notion that almost all initial conditions of unstable trajectories in the same chaotic region of phase space lead to exactly the same set of Lyapunov exponents is only true in the infinite time limit.  For finite-time trajectory segments there are immense fluctuations seen in how trajectories deviate from one another~\cite{Wolfson01}, and there is quite a bit of literature in a variety of contexts regarding so-called finite-time Lyapunov exponents~\cite{Ott97, Grassberger88, Sepulveda89, Amitrano92, Amitrano93, Crisanti93, Schomerus02, Tomsovic07, Beims07}.  However, they are not what appears in semiclassical theories.  Instead, there is an alternative description of instability, specified by the stability matrix and its stability exponents, which is precisely what is needed.  For example, in trace formulas, the determinant ${\rm Det}(\mathbb{1}-M_t)$ appears where $M_t$ is a stability matrix~\cite{Selberg56, Gutzwiller71}.    Of course, stability and Lyapunov exponents are extremely closely related, but for finite times they typically differ by $O\left(t^{-1}\right)$ corrections~\cite{Tomsovic07} and possess large fluctuations.    

The unit determinant stability matrix, ${\bf M}_t\left[\vec q(0), \vec p(0)\right]$, represents a linear canonical transformation describing the spreading of neighboring trajectories in the immediate vicinity of some particular reference trajectory with initial condition $[\vec q(0), \vec p(0)]$.  At any fixed propagation time, $t$,
\begin{equation}
\label{eq:stabm}
\left(\begin{matrix}
\delta \vec p(t) \\
\delta \vec q(t)
\end{matrix}
\right)
= {\bf M}_t\left[\vec q(0), \vec p(0)\right]
\left(
\begin{matrix}
\delta \vec p(0) \\
\delta \vec q(0)
\end{matrix}
\right)
\; ,
\end{equation}
where~\cite{Tabor89a}
\begin{equation}
\label{eq:stabder}
{\bf M}_t\left[\vec q(0), \vec p(0)\right] =
\left(
\begin{matrix}
{\bf M}_{11} & {\bf M}_{12}\\ {\bf M}_{21} & {\bf M}_{22}
\end{matrix}
\right)
= \left(
\begin{matrix}
\left. \frac{\partial \vec p(t)}{\partial \vec p(0)} \right|_{\vec q(0)} &
\left. \frac{\partial \vec p(t)}{\partial \vec q(0)} \right|_{\vec p(0)} \\
\left. \frac{\partial \vec q(t)}{\partial \vec p(0)} \right|_{\vec q(0)}  &
\left. \frac{\partial \vec q(t)}{\partial \vec q(0)} \right|_{\vec p(0)}
\end{matrix}
\right)
\ ,
\end{equation}
and $[\delta \vec q(0), \delta \vec p(0)]$ is the phase space deviation from the reference trajectory's initial condition, like-wise $[\delta \vec q(t), \delta \vec p(t)]$ is the deviation from the trajectory endpoint at time $t$, $[\vec q(t), \vec p(t)]$.  The initial condition argument is dropped for brevity of notation, but clearly each
initial condition can lead to a unique time-dependent stability matrix as a function of propagation time.  The differential equation for the stability matrix evolution is
\begin{equation}
\label{eq:mevol}
\frac{{\rm d} {\bf M}_t}{{\rm d} t} = {\bf K}_t {\bf M}_t
\end{equation}
where ${\bf M}_{t=0}$  is the identity matrix, and
\begin{equation}
\label{K}
{\bf K}_t  =
\left(\begin{matrix}
-{\partial ^2 H \over \partial \vec q \partial \vec p} &
-{\partial ^2 H \over \partial \vec q^2} \\
{\partial ^2 H \over \partial \vec p^2} &
{\partial ^2 H \over \partial \vec q \partial \vec p}
\end{matrix}
\right)\ ,
\end{equation}
with the derivatives in ${\bf K}_t$ evaluated at the phase space points along the reference trajectory as a function of time.  ${\bf M}_t$ describes both stability properties and contains information about the Maslov index.  For example, it turns out that in the semiclassical expression for the time-dependent Green function, a trajectory passes through a caustic when ${\rm Det}\left({\bf M}_{21}\right)$ vanishes.   The index in this case advances (or decreases, depending on circumstances) by the number of zero roots at each caustic~\cite{Keller58}.

There are three basic behaviors for each degree of freedom resulting from such a stability analysis, i.e.~the stable, marginally stable, and hyperbolic cases, the latter being associated with chaos.  However, there are some subtleties in extracting information from the stability matrix about these behaviors.  To begin with, generally speaking, global coordinate changes can change the form of the stability matrix.  According to a theorem of Oseledec~\cite{Oseledec68}, there exists canonical coordinate transformations for which ${\bf M}_t$ can be separated into blocks associated with stable (including marginally stable) and unstable dynamics~\cite{Tall24}.  Using action-angle variables for all the constants of motion leads to the $2 \times 2$ stability matrix form for each constant of the motion,
\begin{equation}
\label{shear}
\begin{pmatrix}
\delta {I}_t \\ \delta \theta_t
\end{pmatrix} = \begin{pmatrix} 1 & 0 \\ \frac{{\rm d}\omega(I_0)}{{\rm d}I_0} t & 1 \end{pmatrix} \begin{pmatrix}
\delta I_0 \\ \delta \theta_0
\end{pmatrix}
\end{equation}
In such a representation, the distinction between stable and marginally stable is that the shearing rate, ${\rm d}\omega(I_0)/{\rm d}I_0$, vanishes in the stable case.  Also, note that for a stable degree of freedom, but not expressed in action-angle variables, ${\bf M}_t$ takes on a slightly different form.  The harmonic oscillator gives
\begin{equation}
\label{eq:harmM}
{\bf M}_t=\left(\begin{matrix}
\cos \omega t & - m\omega \sin\omega t \\  \frac{1}{m\omega}\sin\omega t & \cos\omega t 
\end{matrix}\right) \ ,
\end{equation}
in position/momentum coordinates.

After separating out the stable and marginally stable degrees of freedom from the unstable ones, a reduced dimensional stability matrix remains in a subspace that contains all the information about directions of exponential stretching and contraction and behaves locally hyperbolic with non-vanishing positive and negative finite-time stability exponents.  The remaining reduced dimensional ${\bf M}_t$ cannot, in general,  be diagonalized by a similarity transformation.  Indeed, there is a transformation of the kind, ${\bf M}^\prime_t = {\bf S}_t {\bf M}_t {\bf S}^{-1}_0$, that brings it closest to a diagonal form for the global coordinate system for the remaining space whereby ${\bf S}_t$ is a linear canonical transformation of the local final coordinates, $[\delta \vec q(t), \delta \vec p(t)]$ and ${\bf S}_0$ is a linear canonical transformation of the local initial coordinates, $[\delta \vec q(0), \delta \vec p(0)]$.   However, ${\bf M}_t {\bf M}_t^T$ is guaranteed to be diagonalizable with eigenvalues coming in pairs for each remaining degree of freedom, taking on the form
\begin{equation}
\label{eq:stabft}
\begin{pmatrix} \text{e}^{-2\mu_t} & 0 \\ 0 & \text{e}^{2\mu_t} \end{pmatrix}
\end{equation}
where $\mu_t$ is interpreted as the finite-time stability exponent for that particular canonical pair of coordinates.  The eigenvectors give the tangents to the component of the unstable and stable manifolds associated with $\mu_t$ at the end phase point of the trajectory $[\vec q(t), \vec p(t)]$ that reflect the orientations of the exponential expansion and contraction, respectively.  The orientation of the manifolds at the initial point $[\vec q(0), \vec p(0)]$ can be determined with the inverse of the stability matrix, effectively identifying the directions that evolve into the final directions.

\subsection{Periodic trajectories}
\label{sec:phht}

In a fully chaotic system, typical trajectories (measure one) ergodically wander the available phase space, yet there is an important exceptional class, periodic trajectories, which are not ergodic.  As put forth by Cvitanovic~\cite{Cvitanovic91}, these trajectories provide a ``skeleton'' for the complete dynamics.  In a sense, all the dynamical possibilities reside within the set.  In addition, using semiclassical theory for the trace of the Green function, Gutzwiller~\cite{Gutzwiller71} expressed the quantum density of states as a sum over periodic trajectories and earlier the trace formula of Selberg for surfaces of constant negative curvature was derived~\cite{Selberg56}, further increasing their significance.

\begin{figure}[t]
\centering
\includegraphics[width=0.9\textwidth]{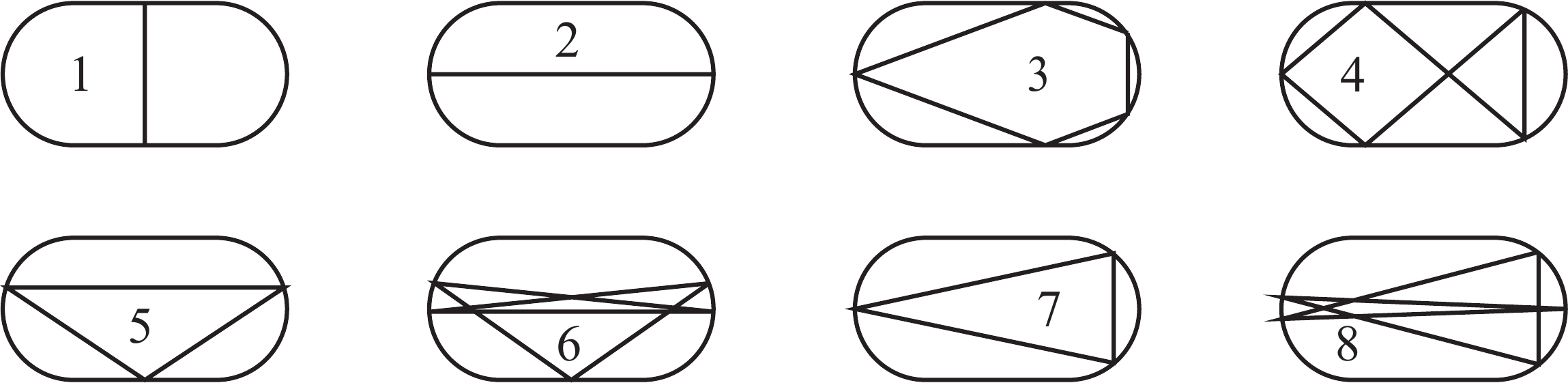}
\caption{Example periodic trajectories of two, three, and five bounces.  No.~1 is the so-called bouncing ball orbit, no.~2 is the shortest and most unstable trajectory, nos.~3 \& 4 are two arbitrary 5-bounce periodic trajectories, and nos.~5-8 are periodic trajectories that follow the excursions of the two primary homoclinic trajectories of trajectory no.~2; see Fig.~\ref{fig:soshet}.}
\label{fig:sospo8}
\end{figure}

Example periodic trajectories of the stadium billiard are shown in Fig.~\ref{fig:sospo8}.
On a fixed energy surface, the magnitude of the momentum (velocity) is fixed and the period of motion is just the geometric length of the trajectory divided by the velocity.   However, when considering the stadium as a mapping of its surface of section, it's convenient to imagine a pseudo-time variable that just counts the number of iterations of the map, i.e.~the number of bounces a trajectory makes with the boundary.  Each periodic trajectory of period $n$ intersects the surface of section $n$ times.   Any of these phase points returns to itself after $n$ bounces, and thus appears as a fixed point under $n$ iterations of the map.  

\begin{figure}[t]
\centering
\includegraphics[width=0.49\textwidth]{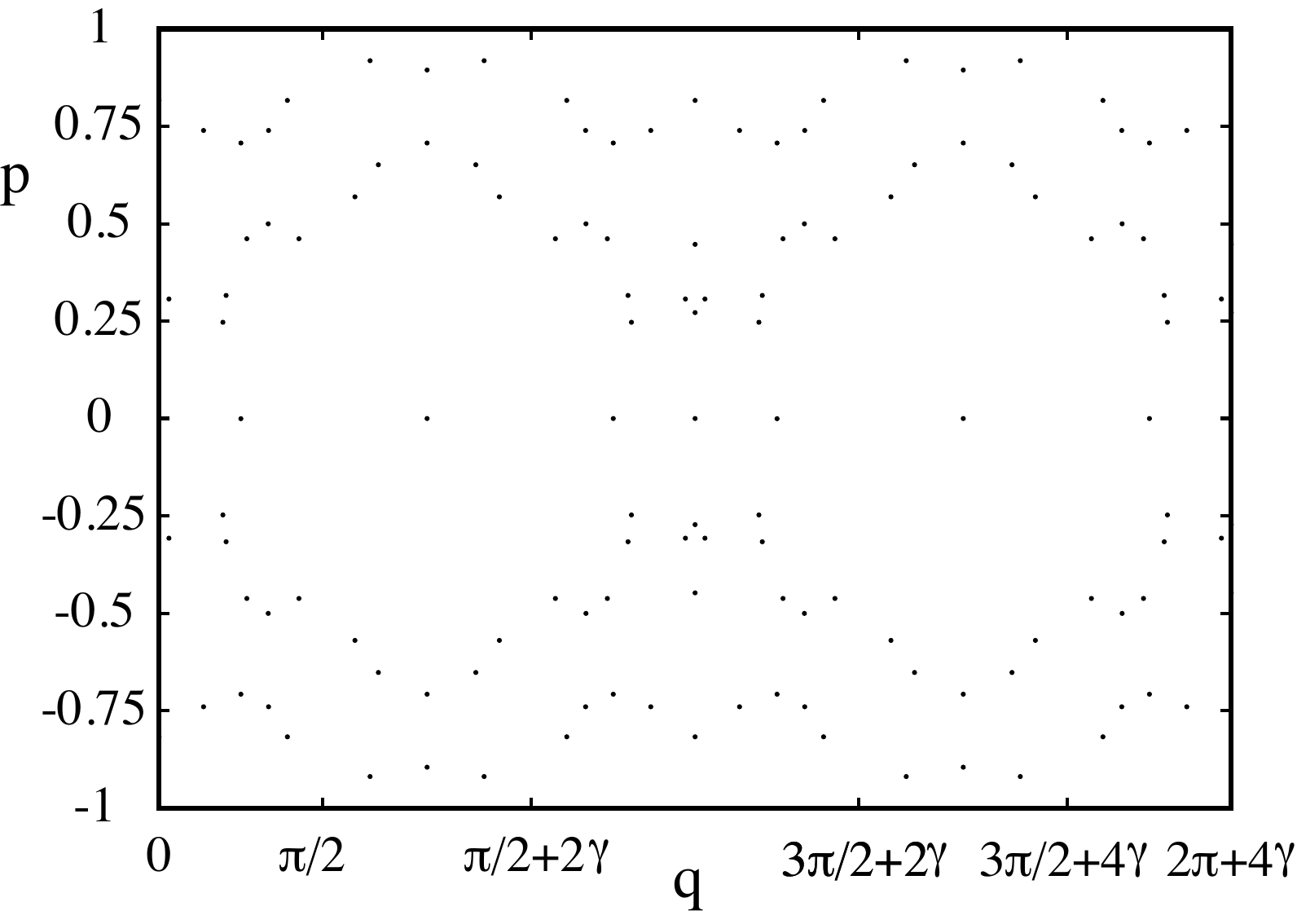}\includegraphics[width=0.49\textwidth]{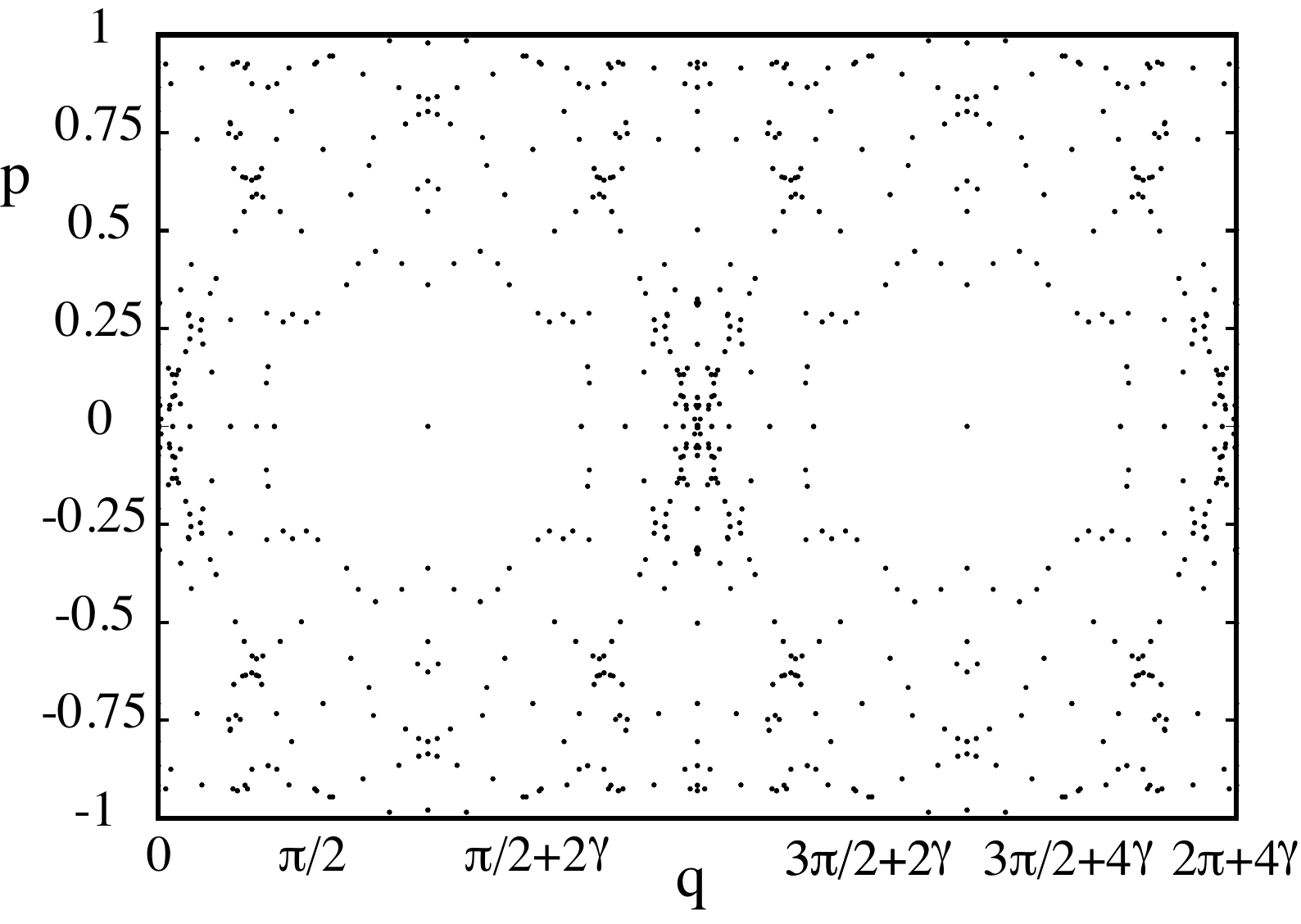}
\includegraphics[width=0.49\textwidth]{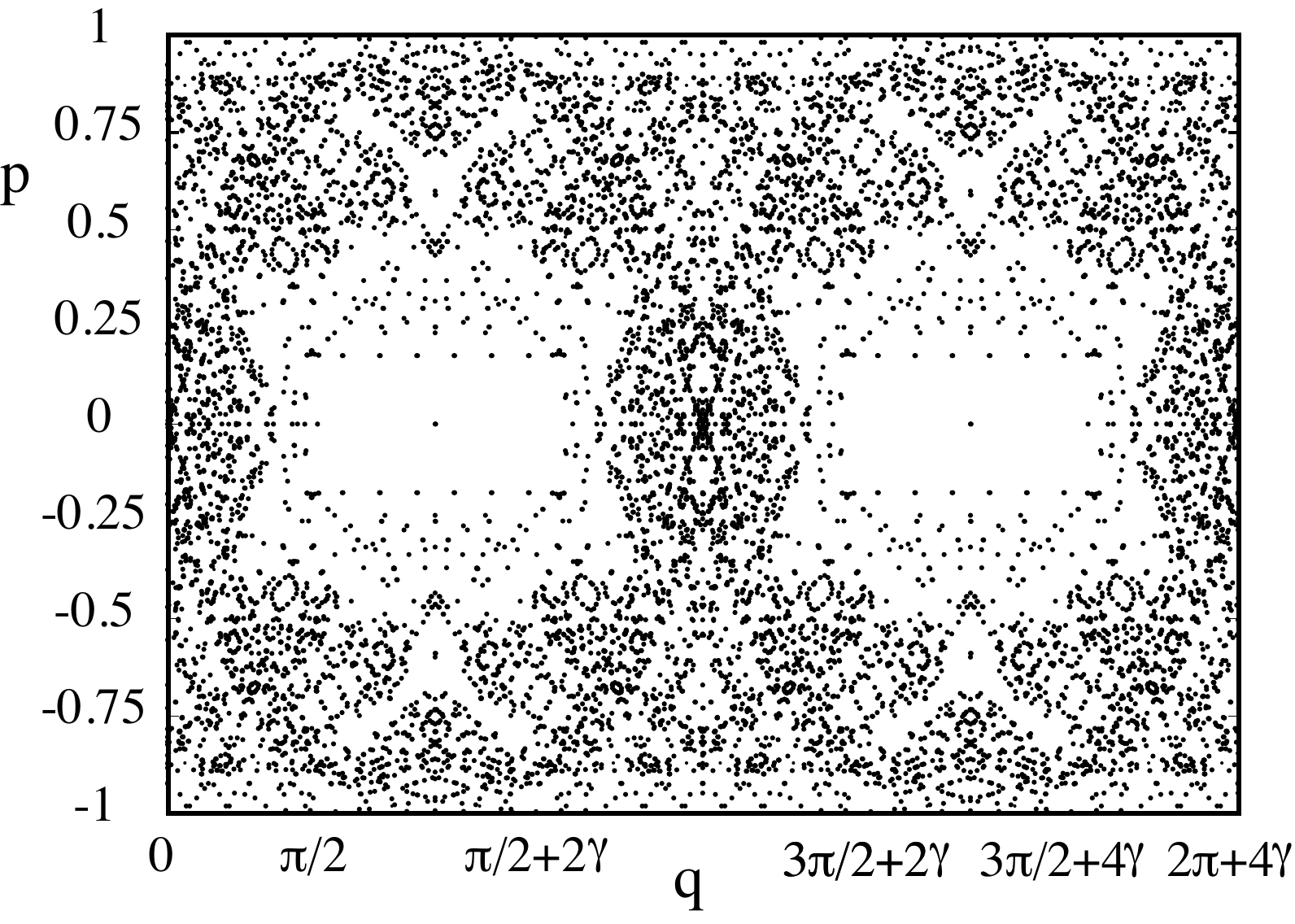}\includegraphics[width=0.49\textwidth]{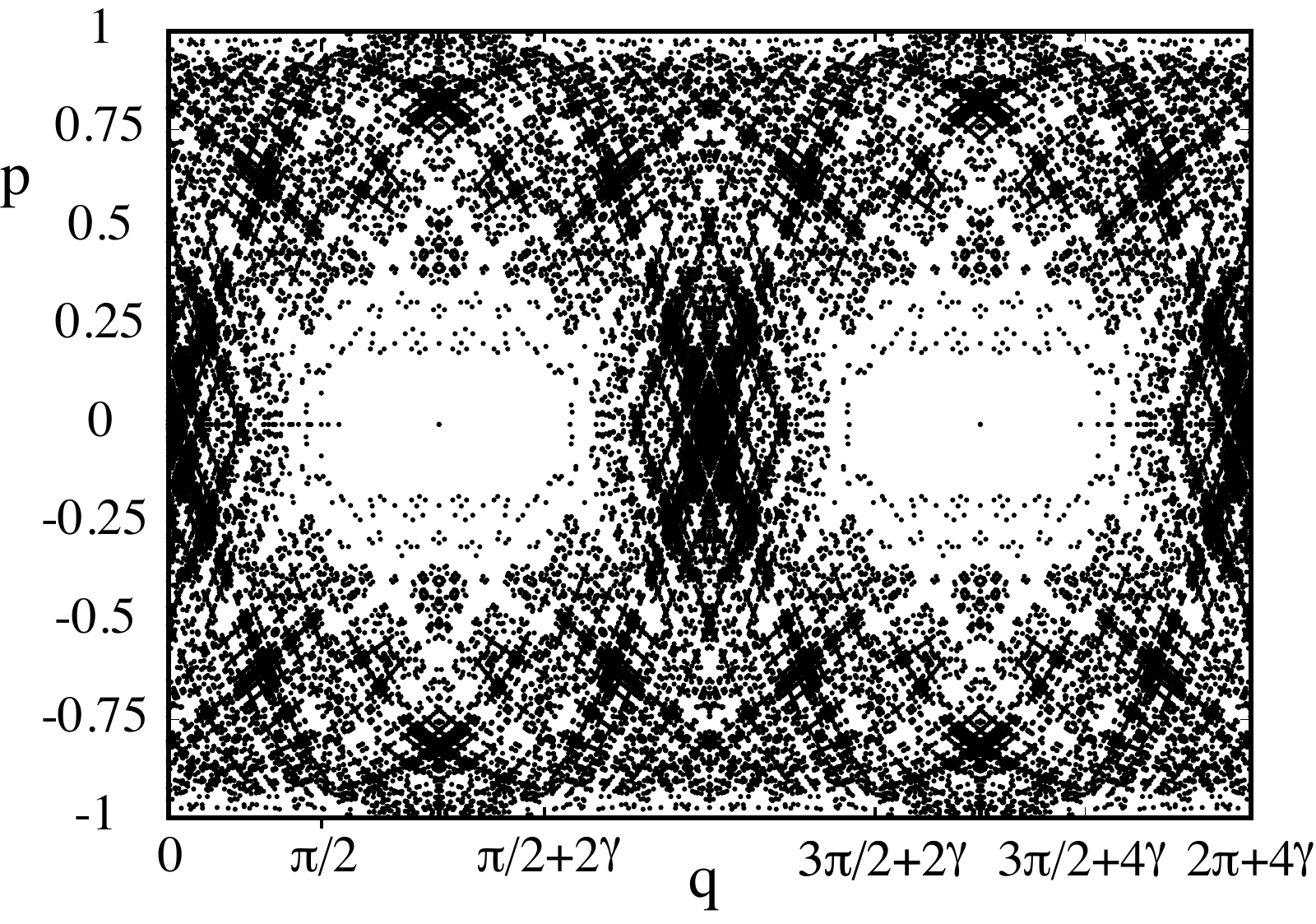}
\caption{Fixed points of the stadium map with $\gamma=1$.  From top left to bottom right, the fixed points of all the periodic trajectories are shown for 4, 6, 8, and 10 iterations of the map, respectively.  The lowest density of points reflects the difficulty of entering the phase space neighborhood of the bouncing ball trajectories (no.~1 of Fig.~\ref{fig:sospo8}).  The highest density is in the neighborhood of the shortest most unstable periodic trajectory (no.~2 of Fig.~\ref{fig:sospo8}) and along its unstable and stable manifolds.}
\label{fig:sospo10}
\end{figure}

Pictured trajectories nos.~1 $\&$ 2 play further special roles.  Trajectory no.~1 and its identical counterparts through horizontal translations are the only marginally stable trajectories that exist.  All others are exponentially unstable.  The phase space region immediately surrounding these so-called `bouncing ball orbits' is rather difficult to enter and exit as mentioned just above in discussing Fig.~\ref{fig:sos1000}, and creates a sticky region of phase space even without the presence of a KAM island or some obvious transport barrier~\cite{Tanner97}.  It slows down the exploration of phase space by ergodic trajectories.  This can be seen in the surface of section for the ergodic trajectory of Fig.~\ref{fig:sos1000}, where two narrow, approximately rectangular regions, are devoid of points even after a thousand iterations (bounces).  It also has an important influence on the location of fixed points where it is seen in Fig.~\ref{fig:sospo10} that there are far fewer fixed points in those regions as well. 
This region is responsible for the relatively well localized bouncing ball quantum eigenstates~\cite{McDonald79, McDonaldthesis, Tanner97}, which is distinct from Heller's eigenstate scarring on unstable periodic trajectories~\cite{Heller84}.

Excluding the marginally stable trajectories, no.~2 is the shortest periodic trajectory and the most unstable, periodic or otherwise, per bounce; see comments near the end of Sec.~\ref{sec:sd}.  Its unstable and stable manifolds foliate the phase space and ahead its homoclinic trajectories that lie at these manifold's intersections are discussed further.   Its greater instability leads to a greater local density of neighboring fixed points, which is visible in Fig.~\ref{fig:sospo10} where all the fixed points are plotted separately for $4,6,8,10$ iterations, respectively.  With increasing $n$, the number of such fixed points increases exponentially, also clearly visible in Fig.~\ref{fig:sospo10}, and corresponds to the rate of dynamical information production, which is measured by the metric KS entropy mentioned very briefly at the end of Sec.~\ref{sec:poincare}.  In this case, it is also equal to the positive Lyapunov exponent or more generally for a broad class of systems, it is equal to the sum of positive Lyapunov exponents~\cite{Pesin77}.  As a final note, the stadium billiard has the following discrete symmetries, reflection about $p=0$ (time reversal symmetry), and reflections separately about the $x=0$ and $y=0$ axes.  The $x$-axis intersects the surface of section at $q=0,\pi+2\gamma$, and the $y$-axis at $q=\gamma+\pi/2, 3\gamma+3\pi/2$.  Periodic trajectories without a discrete symmetry come in $8$ copies.  Each of their fixed points belongs to a set of $8$ symmetry related points.  This generates in Fig.~\ref{fig:sospo10} the eight-fold symmetry that is visible.  These symmetries have a pronounced effect on quantum eigenstate scarring.  Accounting for reduced density of states within a single irreducible representation of the symmetry leads to an enhancement in Heller's scarring criterion~\cite{Heller84}, was likely instrumental in Heller's original observation of scarring, and may in some cases lead to symmetry enhanced scarring (in the original sense) surviving the large many-body system limit~\cite{Hummel23}.\\
 
\subsection{Classical sum rules: Uniformity Principle}
\label{sec:up}
 
 \noindent Periodic trajectories (and other exceptional sets) in a fully chaotic system lend themselves to classical sum rules~\cite{Hannay84, Sieber99}, some motivated by semiclassical expressions for physical observables~\cite{Argaman96, Richter02}.  The most basic and influential one is the Hannay-Ozorio sum rule, which derives from their uniformity principle~\cite{Hannay84, Ozoriobook}.  The sum rule was originally expressed as a summation over all periodic trajectories of fixed period weighted by the absolute square of the amplitude in the Gutzwiller trace formula.  It is a critical input for Berry's derivation of spectral rigidity for quantized chaotic dynamical systems~\cite{Berry85}.  However, the uniformity principle reflects a stronger condition on chaotic dynamics which applies in a more local way in phase space.   This has a rather interesting consequence applied to fixed points of a repeatedly iterated chaotic map.  It is expressed in terms of a natural generalization of the stability matrix for continuous dynamical systems, ${\bf M}_t$, to its counterpart in Poincar{\'e} surface of section variables for maps (or symplectic maps in general), ${\bf M}_n$, where the number of map iterations $n$ plays the role of time.
 
 The perhaps surprising feature is that unlike a single wandering ergodic trajectory densely filling the surface of section uniformly even with increasing $n \to \infty$, the fixed points do not do so.  They must be weighted properly according to the uniformity principle to achieve a uniform density.    Consider the following summation over fixed points (f.p.) contained in some phase space volume denoted $\Delta {\cal V}(\alpha)$,
 \begin{equation}
 \label{eq:uniform1}
{\cal F}_n(\alpha) =  \sum_{f.p.\in \Delta {\cal V}(\alpha)} \frac{1}{\left|{\rm Det}\left({\bf M}_n-\mathbb{1}\right)\right|}
 \end{equation}
where ${\bf M}_n$ is the stability matrix (discussed in Sec.~\ref{sec:stability}) of an $n$-iteration fixed point contained in the surface of section volume $\Delta {\cal V}(\alpha)$.  The parameter $\alpha$ is just a label specifying where the particular volume is located, and $\Delta {\cal V}$ is the value of the volume.  The uniformity principle adjusted for the distinction between a sum over trajectories and a sum of fixed points states that
\begin{equation}
\frac{\Delta {\cal V}}{\cal V} = \lim_{n\to \infty} {\cal F}_n(\alpha)
\end{equation}
where the ratio becomes independent of the location of the volume in the given limit and the $\alpha$ argument can be dropped on the volume $\Delta \cal V$; the ratio just equals whatever volume is involved (no matter where it is located) divided by the total available phase space volume $\cal V$.  This is the idea behind uniformity, and it is an asymptotic relation that, strictly speaking, only holds in the $n\to \infty$ limit.

There are a number of interesting properties related to this sum rule.  Although the Lyapunov exponents are the same for all the ergodic trajectories, the fluctuations in ${\rm Det}\left({\bf M}_n-\mathbb{1}\right)$ from one fixed point to another grow infinitely large in the $n\to \infty$ limit~\cite{Elton10} except for systems, such as the baker's map or Arnol'd cat map, for which there are no finite time stability exponent fluctuations at all.  Hence, there are large fluctuations growing to infinity in the long time limit in the number of fixed points with volumes in different phase space locations.  This feature is clearly visible in Fig.~\ref{fig:sospo10}.  Nevertheless, the fluctuations in ${\cal F}_n(\alpha)$ are expected to decrease exponentially with increasing time.  It turns out that the density of fixed points is higher in the regions of greatest instability in such a way as to balance out the greater value of ${\rm Det}\left({\bf M}_n-\mathbb{1}\right)$.  In addition, there is a rate at which the volumes $ \Delta {\cal V}(\alpha)$ can be shrunk with increasing time and the uniformity still holds~\cite{Pollicott11}, but at the expense of increased fluctuations for fixed time.  This rate can depend on the location in phase space~\cite{Elton10}.  In one chaotic paradigm, known as the lazy baker's map~\cite{Lakshminarayan93}, the convergence is exponential as expected, can be calculated analytically~\cite{Elton10}, and it is dominated by the Pollicott-Ruelle resonances (which are intrinsic an intrinsic property of the dynamical system that measures decay rates and their oscillatory fluctuations)~\cite{Pollicott85, Pollicott86, Ruelle86, Ruelle87}; there is also an additional error term in this particular model not given by a Pollicott-Ruelle resonance which gives a lower order correction.  Applied to the stadium billiard, where analytic results do not exist, it is not possible to see the exponential convergence of the summation over the full surface of section by $10$ bounces.  There are still significant fluctuations, a strong odd-even fluctuation effect, and apparently, the existence of marginally stable periodic trajectories (which must be excluded) and varying numbers of periodic trajectories of different discrete symmetry properties conspire to slow down the convergence.

\subsection{Unstable and stable manifolds}
\label{sec:sets}

\begin{figure}[t]
\centering
\includegraphics[width=0.48\textwidth]{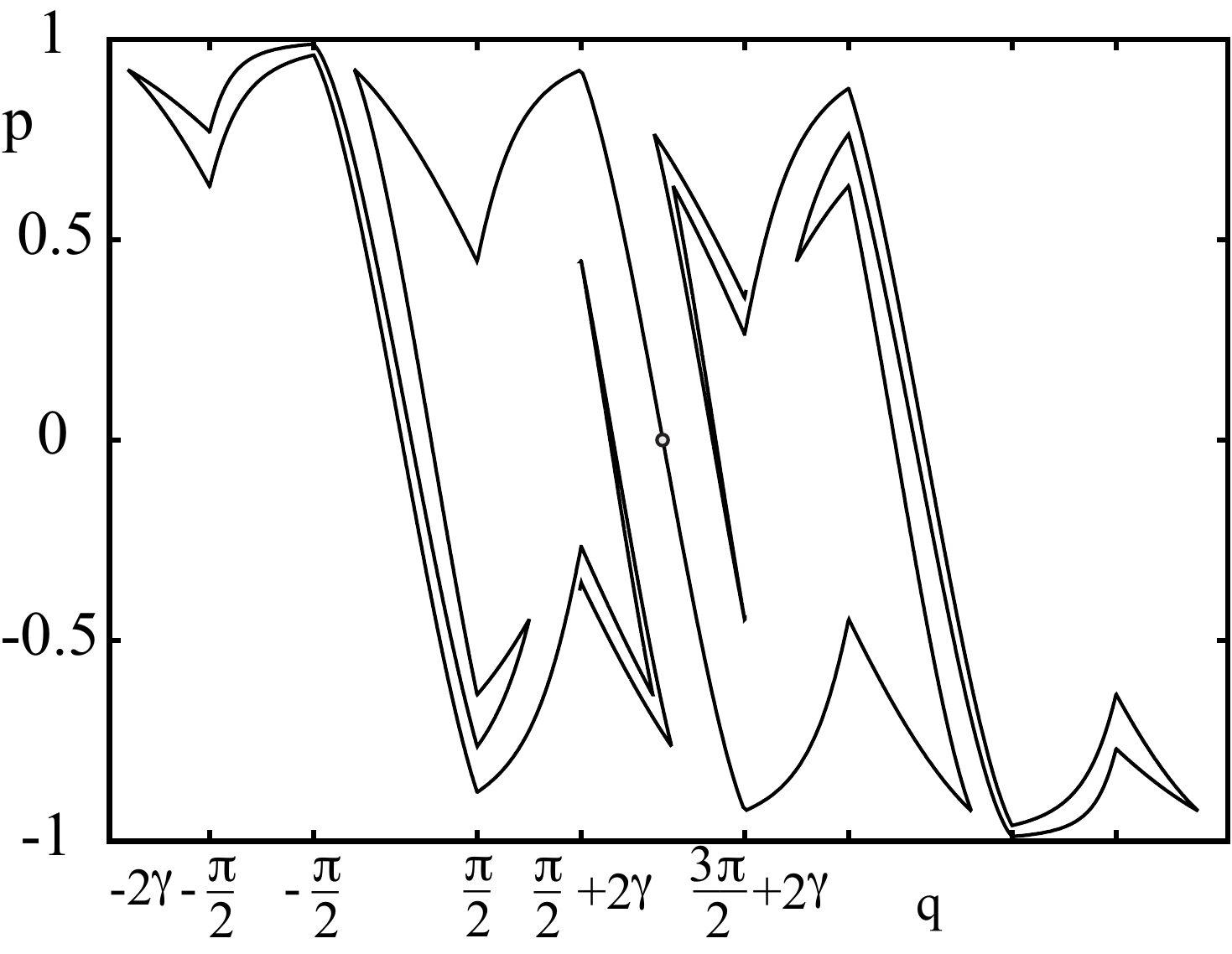} \ \ \includegraphics[width=0.48\textwidth]{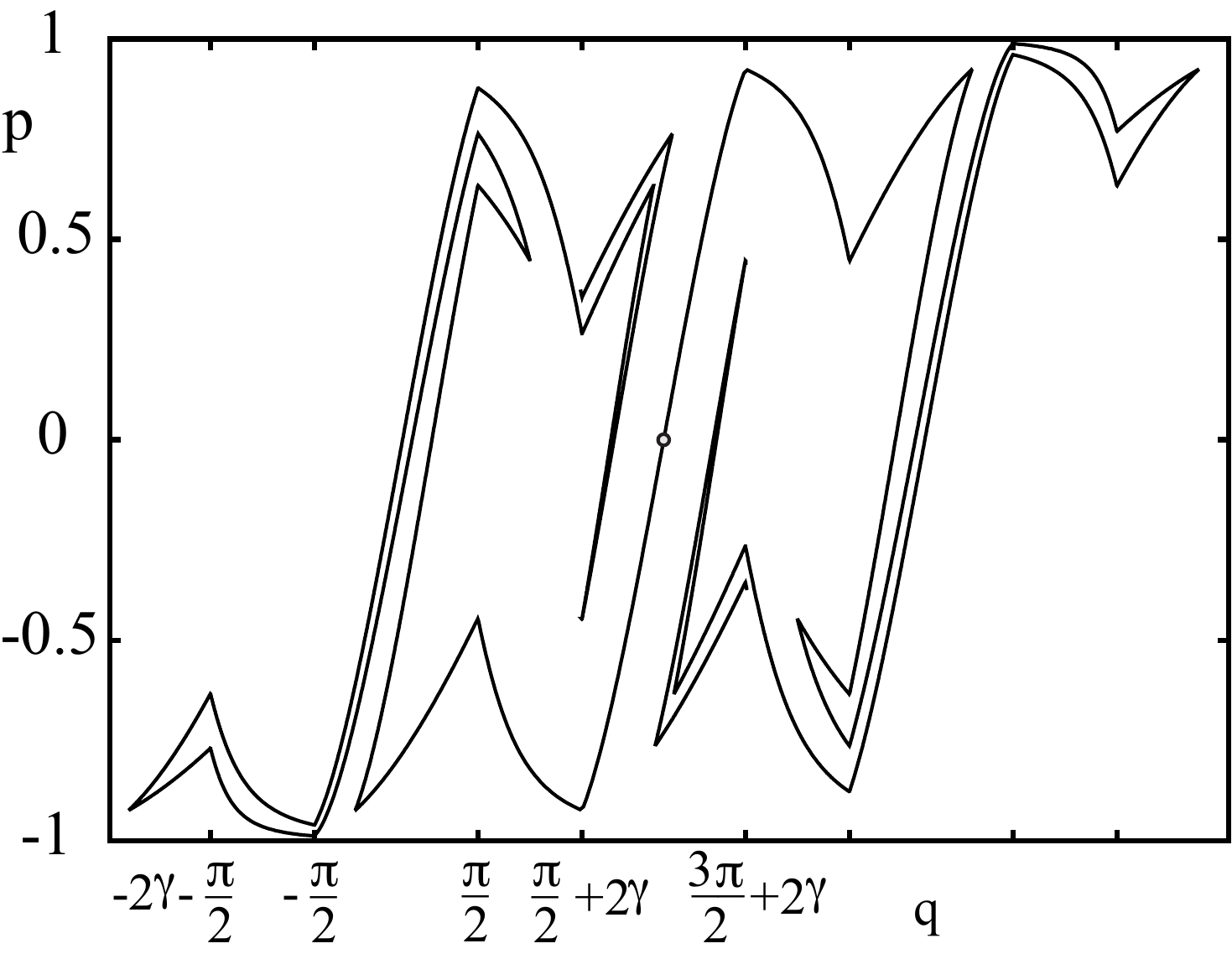}
\caption{Commencement of unstable and stable manifolds for the shortest unstable trajectory of the stadium billiard (no. $2$ of Fig.~\ref{fig:sospo8}).   The small circle near the center denotes the fixed point.  As defined in Fig.~\ref{fig:sos1000} the $q$ coordinate lies in the interval $[0,2\pi+4\gamma)$, however, the manifolds can be ``unfolded'' by adding or subtracting the perimeter appropriately to maintain continuity across the point where $q=0$ meets $q=2\pi+4\gamma$, which makes it visually easier to follow the manifolds away from the fixed point.  }
\label{fig:man}
\end{figure}
The trajectories which converge to each other in the infinite past or infinite future~\cite{Poincare99} creating unstable and stable manifolds generate invariant structures under the dynamics.  They have a quite complicated topology since they are non-self-intersecting and must approach all of the available phase space arbitrarily closely.  Consider the shortest unstable periodic trajectory of a system, such as trajectory no.~2 of Fig.~\ref{fig:sospo8}, and all the trajectories that converge to it in the infinite past.  A starting portion of their surface of section intersections are pictured in the left panel of Fig.~\ref{fig:man}.  The full set constitutes the unstable manifold of that trajectory, and is clearly invariant as it includes the full history of each converging trajectory~\cite{Arnold78, Wiggins88}.  Similarly, the invariant set formed by the trajectories converging in the infinite future constitutes the stable manifold, of which a portion is pictured in the right panel of Fig.~\ref{fig:man}.  These finite portions lengthen roughly with the Lyapunov exponent with each additional mapping.  They rapidly become so convoluted that it is more or less impossible to follow by eye from the starting fixed point.  

\subsubsection{Birkhoff normal coordinates}
\label{sec:bnc}

\begin{figure}[t]
\centering
\includegraphics[width=0.24\textwidth]{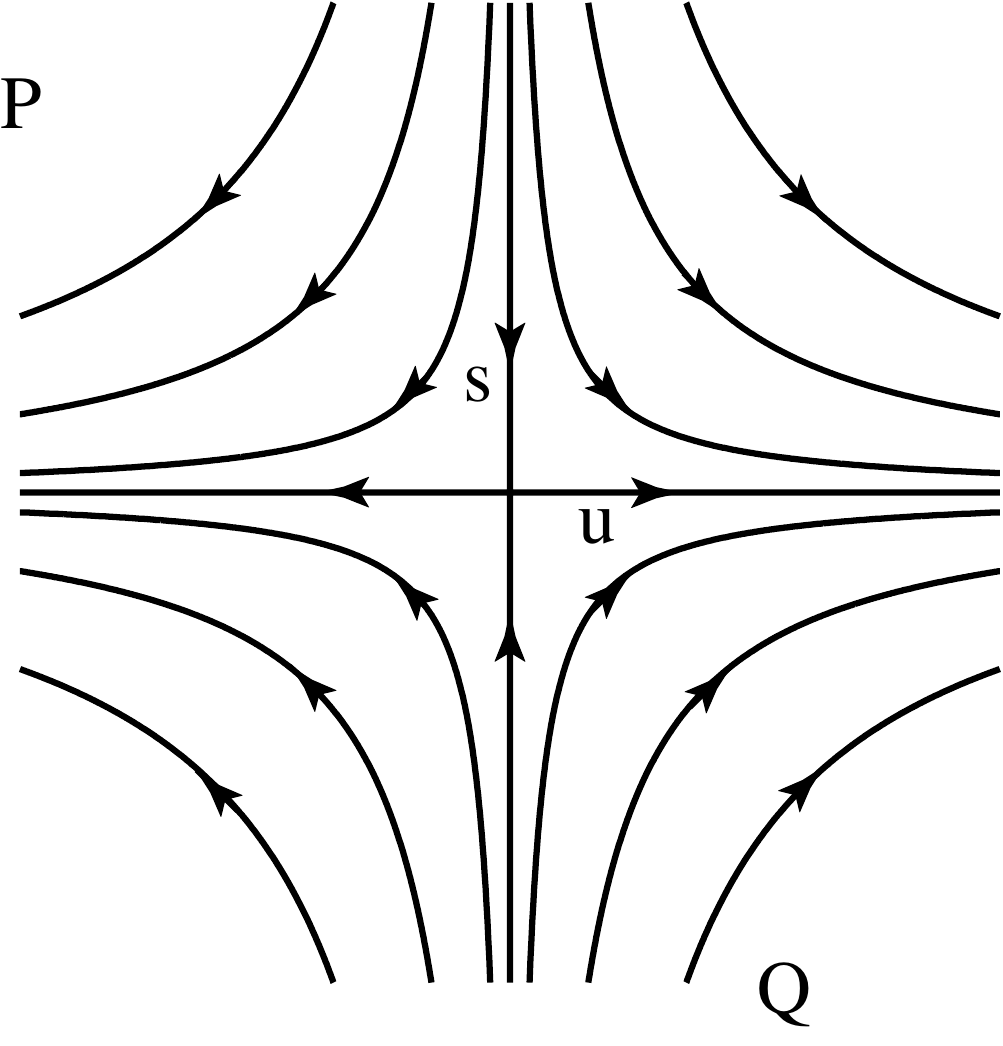}\quad \includegraphics[width=0.38\textwidth]{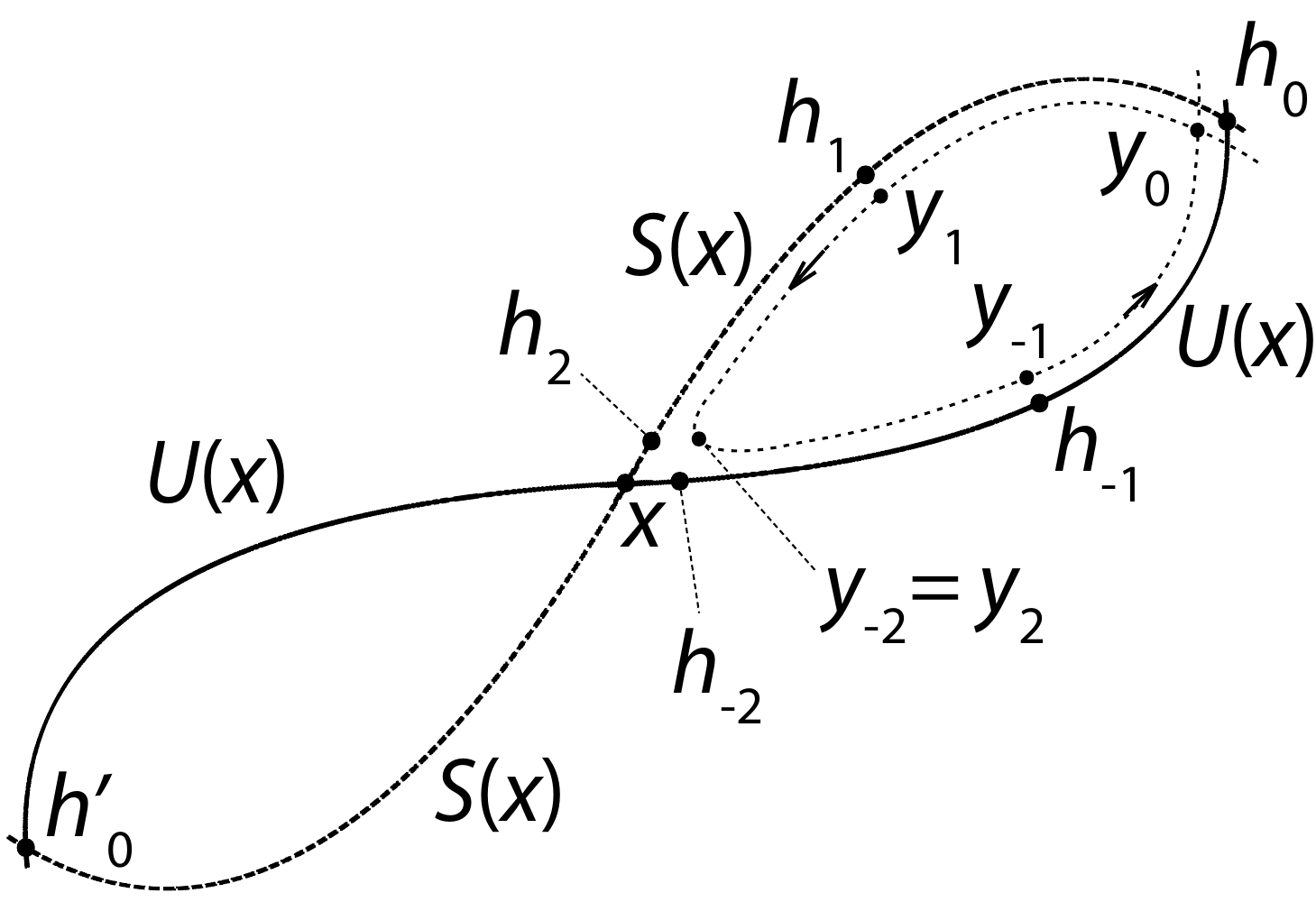} \includegraphics[width=0.35\textwidth]{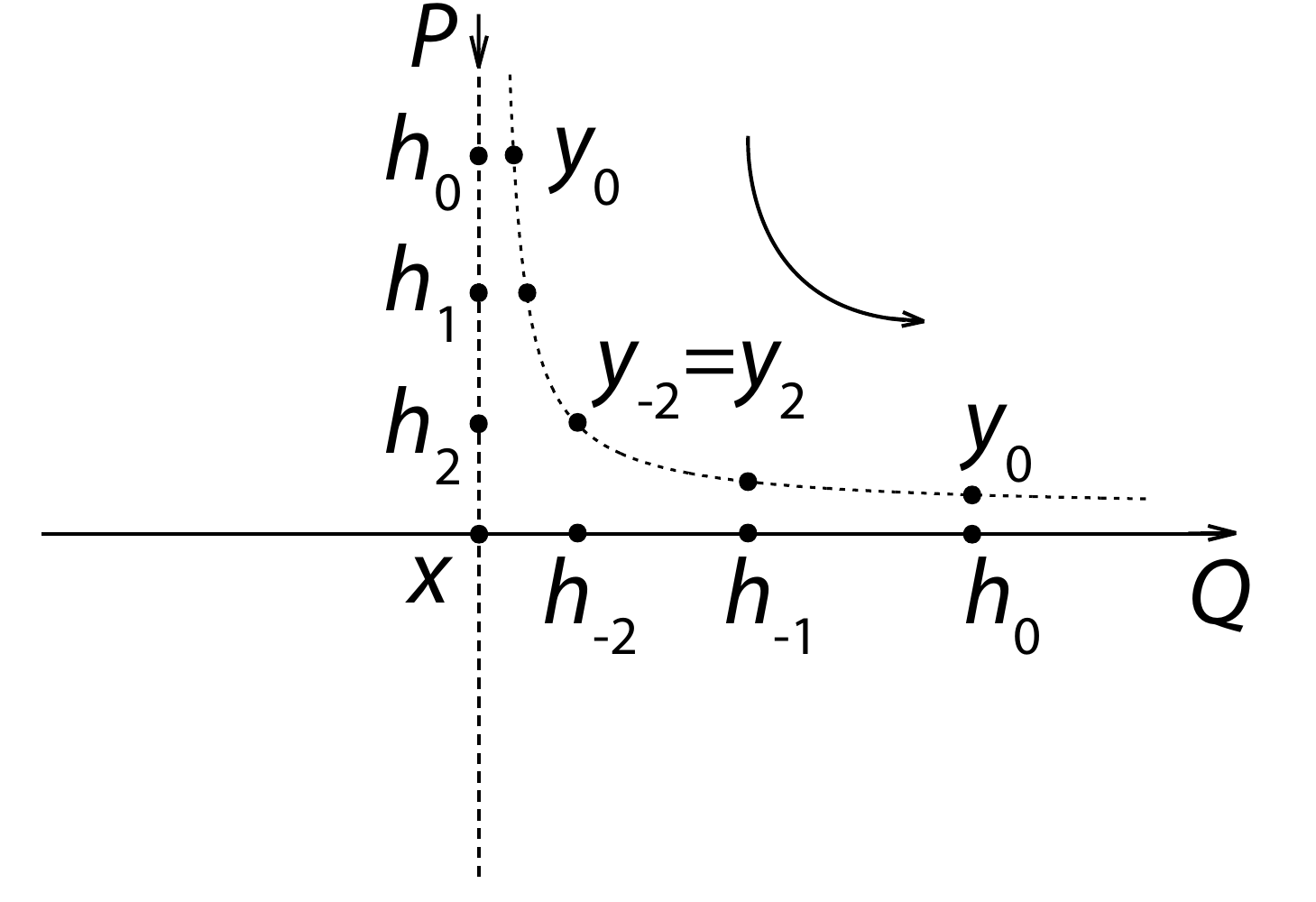}
\caption{Phase portraits in Birkhoff normal coordinates of the stable and unstable manifolds, and local hyperbolic motion.   In the left panel, after a canonical transformation with new coordinates $(Q,P)$ the stable manifold is mapped onto the vertical $Q=0$ axis, and the unstable manifold is mapped onto the horizontal $P=0$ axis.  Schematically following some reference trajectory continuously in time, transverse coordinates of neighboring trajectories follow the arrows along hyperbolas defined by $H(p_0q_0)=H(pq)$ where $H(p_0q_0)$ is some constant determined by initial conditions defining which particular hyperbola the trajectory is on.  A discrete mapping example is shown in the middle and right panels of the comparison of the local dynamics in the original and normal coordinates.  The symbol $x$ refers to a fixed point, $h$ to a homoclinic trajectory, and $y$ to a periodic trajectory.  The middle and right panels are slightly modified from Fig.~6 of~\cite{Li17a}.}
\label{fig:sos}
\end{figure}

Roughly a century ago, Birkhoff introduced a rather useful series of successively improving approximations to a canonical transformation with special properties. In the case of an unstable fixed point of a smooth map, the Hamiltonian is converted locally in new coordinates $(Q,P)$ to $H(PQ)$.  This has the consequence that the neighboring trajectories lie on hyperbolic surfaces~\cite{Birkhoff27, Ozorio88}.  Effectively, the unstable and stable manifolds are in a sense unfolded onto the axes in a $2D$ mapping.  The left panel of Fig.~\ref{fig:sos} illustrates the generic resulting structure.   This illustration is for an ordinary hyperbolic point.  There is an additional possibility, that of hyperbolicity with reflection.  In that case, each intersection with the surface of section toggles back and forth between opposing quadrants ($I \leftrightarrow III,\ \ II \leftrightarrow IV$).

Convergence within a region of phase points near the origin was proven by Moser~\cite{Moser56}, which was later proven to apply also to that region propagated forward and backward in time~\cite{Silva87}.  Hence, there are hyperbolas within which the dynamics is guaranteed to converge to the stable manifold backward in time and onto the unstable manifold forward in time.  These hyperbolas or their counterparts in the original coordinates are known as Moser invariant curves.  Any Liouvillian density of phase points wholly within the convergence zone must respect these constraints.  This implies a simple form for transport as a summation over heteroclinic trajectories.  The $n^{th}$ repeated mapping, $T^n$, of an initially localized Liouvillian density of phase points, $\rho_i$, projected on a final density, $\rho_f$, can be decomposed as a summation over locally linearized mappings in the neighborhood of heteroclinic trajectory segments of length $n$, i.e.
\begin{equation}
(\rho_f, T^n\rho_i) = \sum_{\gamma}(\rho_f, T_\gamma\rho_i)
\label{eq:hetsumg}
\end{equation}
where $T_\gamma$ is a locally linearized mapping about the heteroclinic trajectory segment labelled by $\gamma$ (defined by the stability matrix $M_n$ for the segment)~\cite{Tomsovic93}.  If the final density is equal to the initial density, then the summation would be over homoclinic trajectory segments.  This property features prominently in the transport discussion of Sec.~\ref{sec:complex}.  The calculational stability of the unstable and stable manifolds also underlies a very accurate way of calculating heteroclinic and homoclinic trajectories~\cite{Li17}.  Instead of using Hamilton's equations to propagate some phase point identified as belonging to such a trajectory, rather it is possible to identify successive intersections of manifolds in a much more accurate way since they are exponentially converged everywhere along their existence in the sense that any deviation collapses exponentially onto the respective manifold, i.e.~forward in time, the unstable manifold, reversed propagation, the stable manifold.  The constructions and hence also their intersections converge exponentially rapidly.

\subsubsection{Periodic trajectories imitating homoclinic trajectories}
\label{sec:ptiht}

A consequence of the hyperbolic structure is the existence of an infinite sequence of periodic trajectories that mimic for longer and longer times the excursion of a particular homoclinic trajectory~\cite{Ozorio89}.  The middle panel of Fig.~\ref{fig:sos} shows short segments of an unstable and stable manifold cutoff at the homoclinic phase point $h_0$, which backward and forward in time converges to the fixed point $x$.  Its image in normal coordinates is shown in the right panel.  Note that $h_0$ has two images since it coexists on the unstable and stable manifold.  The dashed line hyperbola comes from the dashed line in the middle panel.  Notice how it has to cross itself.  It has a four iteration periodic trajectory marked and labelled by the $y$-points.  Any trajectory that starts at the crossing point and iterates back to the crossing point after some number of iterations must necessarily be periodic.  This only happens for a countable infinity of hyperbolas.  Since it takes an infinite time, forward or backward, for the homoclinic trajectory to arrive at the fixed point $x$, the closer the hyperbola is to the unstable and stable manifolds, the more iterations it takes to pass by the local neighborhood of $x$.  Somewhere interior to the four iteration periodic trajectory's hyperbola is another which contains a five iteration periodic trajectory, next six, ... ad infinitum.  The sequence of points $y_0(n)$ for increasing numbers of iterations, $n$, converge toward the point $h_0$ as $n\to \infty$.  Recall trajectory nos.~$5-8$ in Fig.~\ref{fig:sospo8}, which provide an illustration of this feature.  The first two, nos. $5\ \&\ 6$, are the three and five bounce periodic trajectories approaching the primary homoclinic trajectory a of Fig.~\ref{fig:soshet}.  The second two, nos. $7\ \&\ 8$, are the three and five bounce periodic trajectories approaching the other primary homoclinic trajectory b.  This sequence of periodic trajectories with points approaching a homoclinic point were used in Ozorio's work on homoclinic quantization~\cite{Ozorio89}.

\subsubsection{Resonances and turnstiles}
\label{sec:rzt}

In addition to the tori which provide the support for EBK quantization, there are other phase space structures  significantly impacting quantum mechanical properties.  For example, as noted above, tori with rational frequency ratios imply the existence of near degeneracies in the spectrum.  Under perturbation, they generate bifurcations that lead to resonance zones associated with resonance-assisted tunneling~\cite{Brodier01, Brodier02}.  Generally speaking, bifurcations also give rise to unstable periodic trajectories whose unstable and stable manifolds may restrict transport through associated turnstiles~\cite{MacKay84a, MacKay84b, Channon80, Bensimon84, MacKay87} in such a way as to partially localize the eigenfunctions of the system~\cite{Geisel86, Brown86, Radons88, Bohigas93}.  Resonances and turnstiles thus leave important imprints in quantized counterparts.

Perhaps the simplest example of a resonance is provided by the pendulum.  In phase space, there exists a separatrix, which is a dividing surface between trajectories that oscillate back and forth, which are within the resonance zone, and those that rotate continuously either clockwise or counterclockwise.  The separatrix is comprised of the two trajectories that take an infinite amount of time to fall away from the stationary vertical position and rotate once clockwise or counterclockwise and then take an infinite time to approach the vertical position again.  In a surface of section for a $2D$ integrable system, a resonance has the same appearance as a string of pendulum separatrices that tie together in a loop.  The separatrix in this case is associated with an unstable periodic trajectory that typically pierces the surface of section in multiple points, and a set of trajectories that approach the periodic trajectory forward and backward in time.  There is again a set of trajectories that remain trapped within the resonance zone forever.

In a $2D$ chaotic system where measure one of the trajectories must visit arbitrarily close to any of the available points in phase space, it is not possible to have a dividing surface like a separatrix.  Nevertheless, there is a close analog to the resonance zone found in integrable systems.  It is formed by the unstable and stable manifolds of an unstable periodic trajectory.  In the middle panel of Fig.~\ref{fig:sos}, a segment each of the unstable and stable manifolds associated with the unstable fixed point $x$, $U(x)$ and $S(x)$, respectively, are drawn up to a homoclinic trajectory intersection point, $h_0$.  The interior region of the loop is the resonance zone associated with $x$, except now in a chaotic dynamical system.  The phase points neighboring $U(x)$ and $S(x)$ follow the deformed hyperbola as indicated by the dotted line in the interior region.  The biggest difference with the integrable case where the interior trajectories remain within the resonance zone for all times, is that in the chaotic case there is a mechanism for trajectories to escape and enter.    The only place where that can happen is in the neighborhood of $h_0$.  Trajectories near where the dotted line crosses $U(x)$ are entering the zone and those near the dotted line intersects $S(x)$ are exiting.

\begin{figure}[t]
\centering
\includegraphics[width=.99\textwidth]{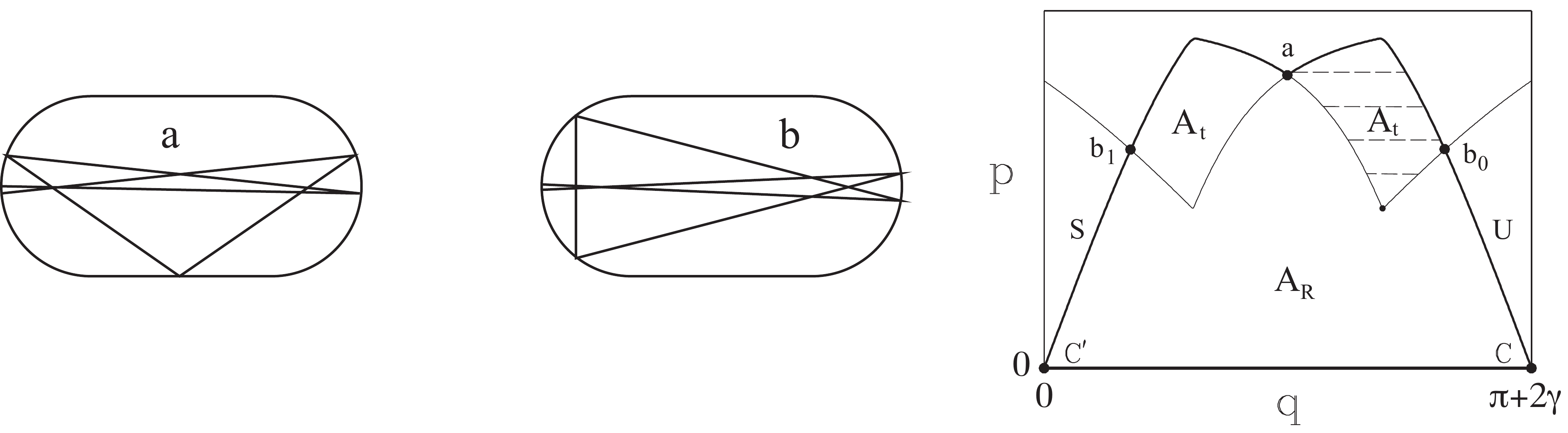}
\caption{Illustration of the resonance structure and turnstile of trajectory no.~$2$ in Fig.~\ref{fig:sospo8}.  Only a quarter portion of the surface of section including the associated unstable and stable manifold segments is drawn beside the primary homoclinic trajectories.  The points labelled $C$ and $C^\prime$ are the intersection points of the periodic trajectory.  The labels U and S denote the darkened segments of its unstable and stable manifolds, respectively, that intersect at point a, which  labels the primary homoclinic trajectory illustrated on the left.  Similarly, b labels the other primary homoclinic trajectory illustrated just to the right of a, and the subscript indicates that the point ${\rm b}_0$ propagates to ${\rm b}_1$.  The darkened segments along with a reflected copy inverted about the $p=0$ line enclose a chaotic resonance zone in phase space (the roles of U and S are swapped left to right in the inverted segments).   The area labelled $A_R$ is half the area of the resonance zone.  The unstable and stable manifold segments connecting points ${\rm b}_0$ and ${\rm b}_1$ form a turnstile.  The turnstile area, $A_t$, marked by the dashed horizontal lines and its counterpart $A_t$ are equal.  The phase points in the dashed area $A_t$ exit the chaotic resonance zone under mapping, and the phase points in the blank area $A_t$ just entered the resonance zone.}
\label{fig:soshet}
\end{figure}
The transport mechanism in and out of a resonance zone is known as a turnstile~\cite{MacKay84a}.  Segments of the unstable and stable manifolds form the structure.  Although, a little more complicated due to reflection symmetry, the turnstile of the resonance zone associated with the shortest unstable periodic trajectory is illustrated for the stadium billiard in Fig.~\ref{fig:soshet}.  There are two primary homoclinic trajectories labelled a and b, i.e.~those with the shortest excursions, drawn to the left of the surface of section.  Extending the $U$ segment from a to ${\rm b}_1$ and  the $S$ segment from a to ${\rm b}_0$ delineates the turnstile, the right half exiting the resonance zone, and the left half just entering.  The only way in and out of the resonance zone is by ending up in the clear area $A_t$ and the dashed $A_t$, respectively, (and their reflected counterparts).  Thus, under circumstances in which the area $A_t <<  A_R$, typical trajectories get trapped within the resonance zone for long times, and that region appears sticky for finite times, similar to the region with a denser set of points in the middle panel of Fig.~\ref{fig:soskr}.  This mechanism and another similar one related to the existence of cantori~\cite{Percival80} can have significant localizing effects on quantum eigenstates or in quantum transport~~\cite{Geisel86, Brown86, Radons88, Bohigas93, Michler12}.

\subsection{Symbolic dynamics}
\label{sec:sd}

For the purposes of studying Hamiltonian chaos the enterprise of symbolic dynamics is the construction of a unique string of symbols for each trajectory.  The sequence being like a discretization of time dovetails best with Poincar\'e surfaces of section and dynamical maps.  Such a coding of the dynamics generates a great deal of information about the system, i.e.~periodic trajectories have repeating symbolic sequences, homoclinic and heteroclinic trajectories have limiting symbols sequences in their infinite past and future with ``excursions'' in between, etc...  The broad general applicability of symbolic coding to genuine chaotic systems, not just simple paradigms, is manifest in Smale's homoclinic theorem~\cite{Smale65} and horseshoe~\cite{Smale67}.  A good discussion can be found in Chapter $8$ of~\cite{Robinson99}.

A number of chaotic systems have explicit constructions by now, e.g.~the baker's map~\cite{Morse38}, the standard map~\cite{Christiansen96}, the stadium billiard~\cite{Biham92, Hansen95}, many others, as well as abstract constructions such as the Smale horseshoe~\cite{Smale67}.  The seeds of the subject seems to have been planted by Hadamard in 1898~\cite{Hadamard1898}, but it is argued by a couple authors~\cite{Coven08} that Hedlund's 1944 article~\cite{Hedlund44} marks the beginning of the modern era, not the earlier papers including Morse~\cite{Morse21, Morse38}.  There is a recent review in~\cite{Hirata23}.

Perhaps the absolute simplest symbolic dynamics appears in the baker's map.  The map is defined by the equations,
\begin{figure}[t]
\centering
\includegraphics[width=0.95\textwidth]{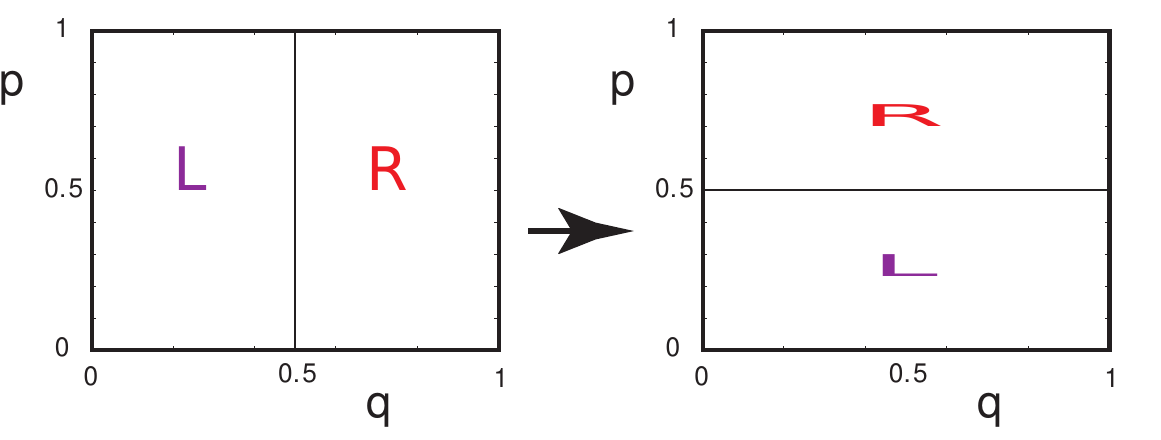}
\caption{Illustration of the baker's map.  The space is compressed a factor two vertically and stretched a factor two horizontally.  The points outside the unit square are cut and placed above the remaining points.}
\label{fig:bm}
\end{figure}
\begin{eqnarray}
\label{eq:bm}
q_{n+1} &=& 2 q_n - \left[ 2 q_n\right] \\
p_{n+1} &=& \frac{p_n}{2} +\frac{ \left[ 2 q_n\right]}{2}\ ,
\end{eqnarray}
which is illustrated in Fig.~\ref{fig:bm}.  It is uniformly hyperbolic and is nongeneric in the sense that there are absolutely no fluctuations at all in the stability exponents of any trajectory.  On the other hand, it has the simplest hyperbolic structure possible.

It just so happens that a good symbolic dynamics can be generated by assigning the letter $\cal L$ to a phase point to the left of $q=1/2$, and the letter $\cal R$ otherwise.  A unique infinite symbol sequence of $\cal L$'s $\&$ $\cal R$'s determined by considering both backward and forward propagation defines a unique trajectory, and all possible trajectories are given in the set of all possible infinite sequences.  It is customary to place a decimal point somewhere in the sequence to denote the present.  A step forward in time amounts to shifting the decimal point to the right by one symbol, and a step backward a step to the left.

There is a straightforward way to connect the symbol sequence to a trajectory.  Consider first the representation of the position variable in base two.  Doubling its value amounts to shifting the decimal place once to the right.  Removing the integer part leaves it in the unit domain.  The mapping has essentially shifted the decimal point to the right one digit and removed any digits to the left of the decimal point.  A $q$ value that is on the left has a first digit equaling zero and greater than or equal to $1/2$ a digit of one.  This continues ad infinitum and associating $\cal L$ to $0$ and $\cal R$ to $1$ generates the present value of $q$  by using exclusively the symbols to the right of the decimal point.   Similarly, for the binary representation of the momentum, the decimal point shifts to the left with the open digit is replaced by the one dropped from the position binary sequence under the mapping.  This suggests writing the momentum binary string of digits in reverse to the left of the decimal point with the same association of $\cal L$'s $\&$ $\cal R$'s to $0$'s and $1$'s.  With $\gamma_q$ representing the infinite string of $0$'s and $1$'s, i.e.~$q=.\gamma_q$, and similarly $p=.\gamma_p$, the entire history of the initial condition $(q,p)$ forward and backward in time is given by moving the decimal point right and left, respectively, as
\begin{equation}
\label{eq:bmbits}
\gamma_p^R ~_{\overset{\bullet} {\Leftrightarrow} }\gamma_q \qquad {\rm e.g.}\  \ ...0001010000111_{\overset{\bullet} {\Longleftrightarrow} }0011100011...
\end{equation}
where $\gamma_p^R$ is just $\gamma_p$ with the bits reversed.

The identification of all periodic trajectories is immediate.  Let $\gamma_n$ be any string with $n$-digits and let the overline denote infinite repetition.  All the $(q,p)$ values, 
\begin{align}
q &= .\overline{\gamma_n} =\frac{I_{\gamma_n}}{2^n-1}\nonumber \\
p &= .\overline{\gamma_n^R} =\frac{I_{\gamma_n^R}}{2^n-1} \ ,
\end{align}
where $I_{\gamma_n}$ is the integer value of the bit string $\gamma_n$,  lead to periodic trajectories, and there are no others.  Similarly, it is possible to identify all homoclinic or heteroclinic trajectories associated with other trajectories.  They just require the ends to be identical.  The trajectory
\begin{equation}
\overline{\gamma_n}\gamma_k~_{\overset{\bullet}{}}\gamma_j\overline{\gamma_n}
\end{equation}
is homoclinic to $\overline{\gamma_n}~_{\overset{\bullet}{}}\overline{\gamma_n}$ with an excursion of length $j+k$ mappings.

\begin{figure}[t]
\centering
\includegraphics[width=0.33\textwidth]{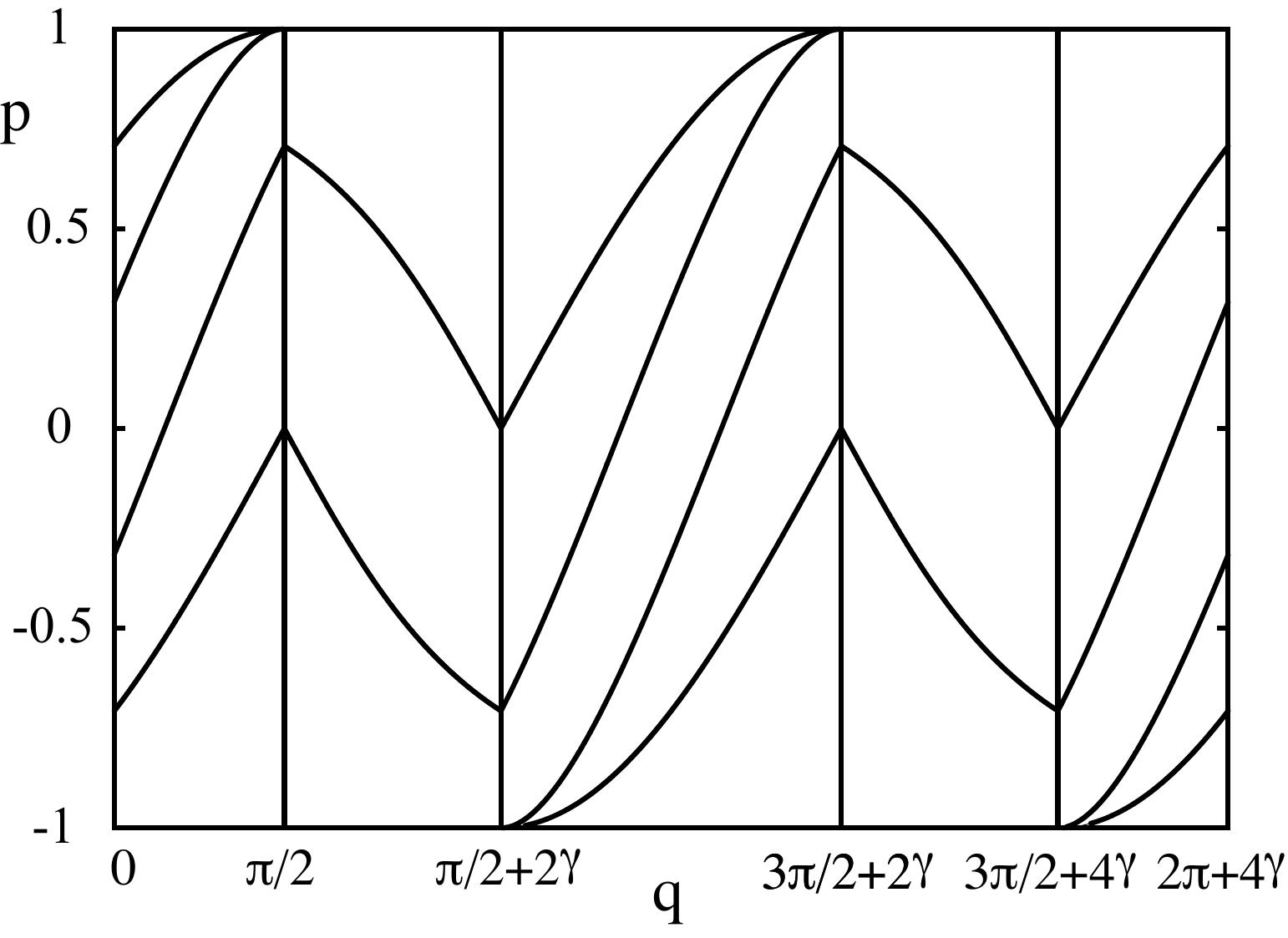}\includegraphics[width=0.33\textwidth]{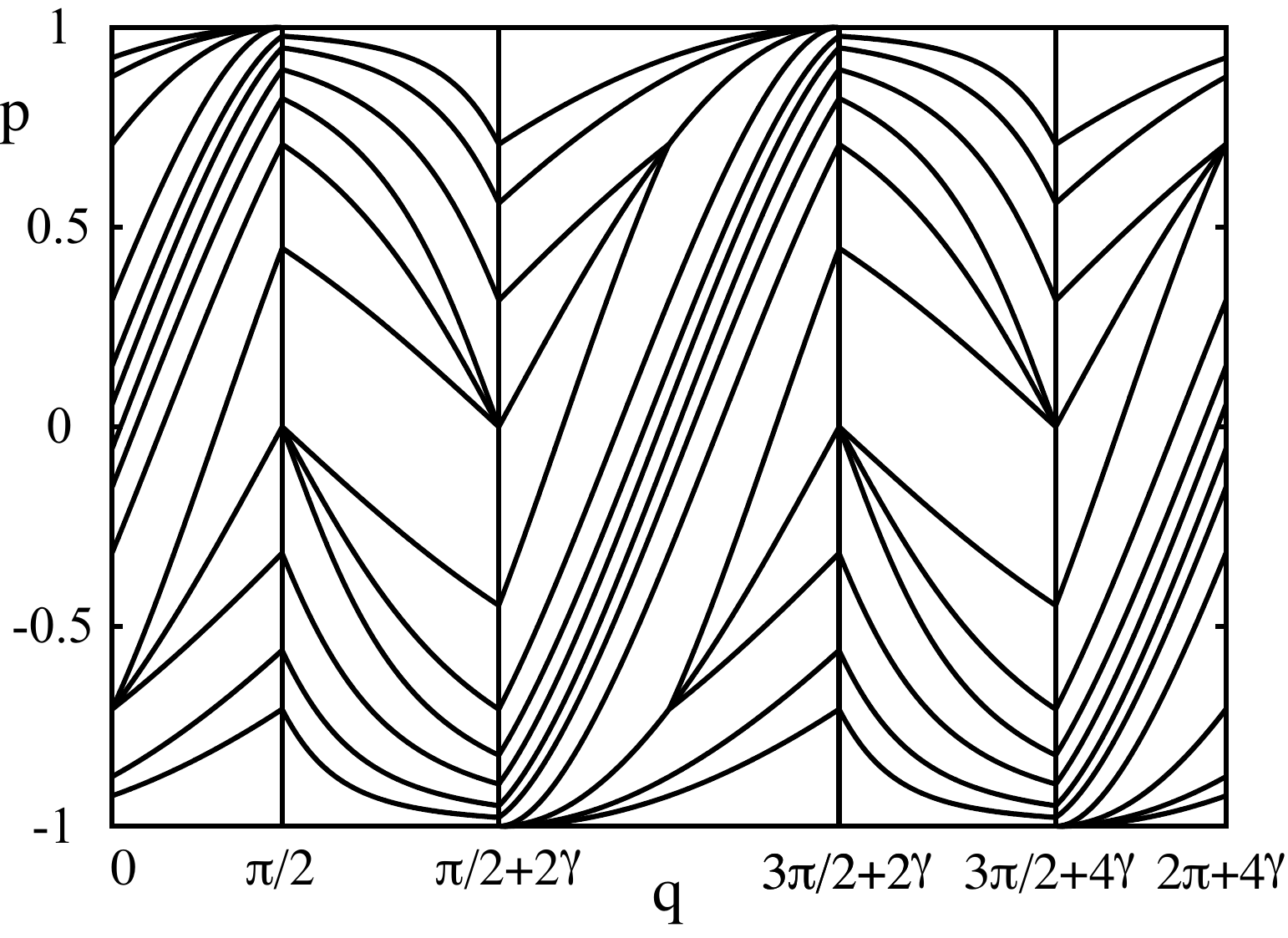}\includegraphics[width=0.33\textwidth]{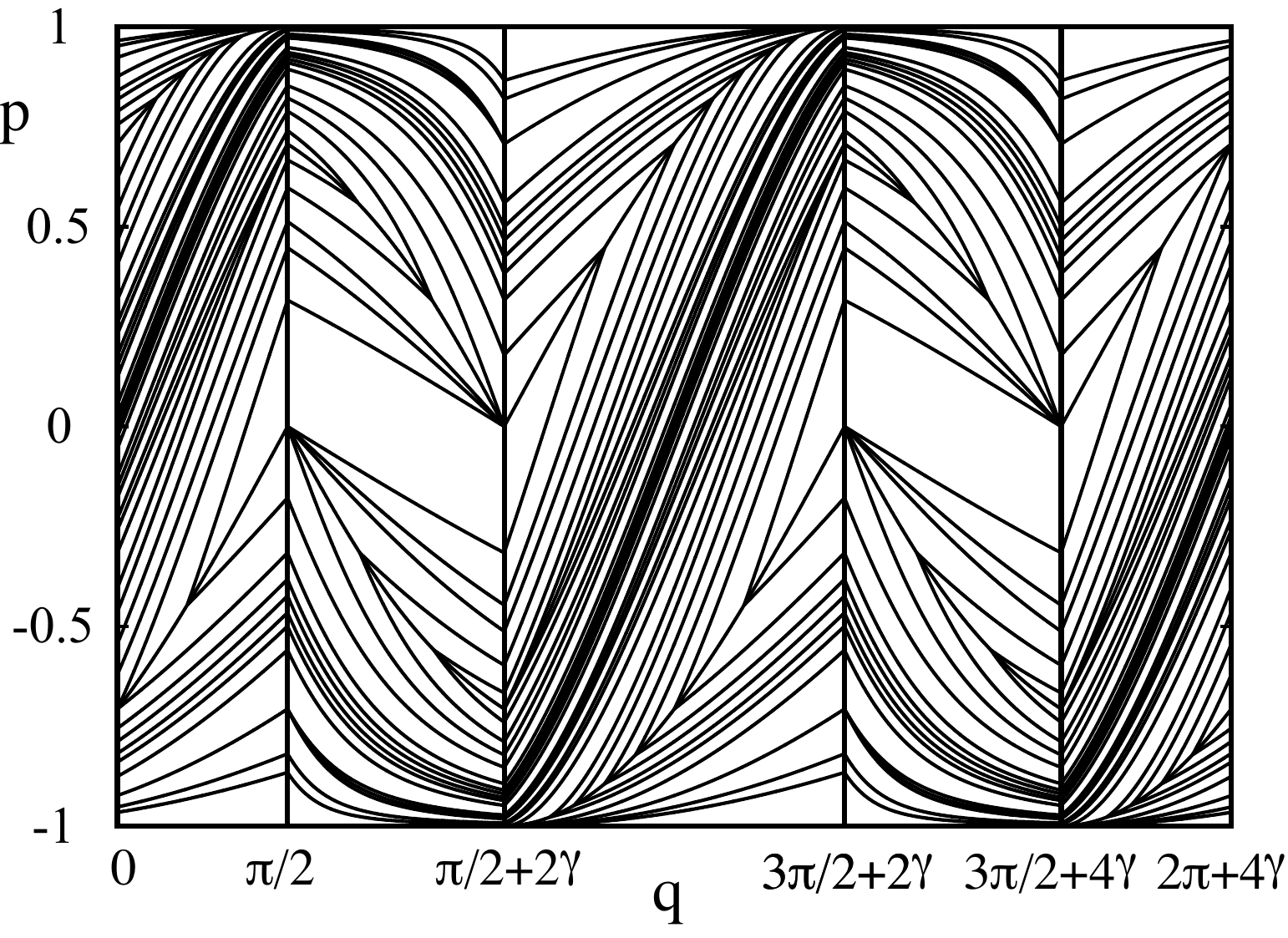}
\caption{Illustration of a partitioning of the phase space into collections of similarly behaving trajectories.  Within each delimited zone, all the trajectories hit in the same sequence of particular straight edges and semicircles.  The left panel results from one iteration of the map, the middle panel from two, and the right panel from three iterations.  The number of partitions is 16, 60, and 192, respectively.}
\label{fig:mrk}
\end{figure}
A much more difficult symbolic dynamics to construct is the map associated with the stadium billiard~\cite{Biham92, Hansen95}.  It is instructive as it gives an idea of some of the difficulties that can arise, such as uncertainty about the optimal symbol set and pruning fronts (grammar rules).  An initial guess could be that four symbols are needed, one for each straight edge and semicircle.  This cannot work though.  One way to grasp the difficulty is to consider what can divide trajectories into separate groups associated with finite symbol sequences.  As two neighboring trajectories evolve, if they happen to hit the boundary on opposite sides of the joint where the straight edges meet the semicircles, they cannot possess the same symbol sequence.  Hence, a first step is separating the trajectories out into strictly contiguous collections that begin on a given side and upon one iteration of the map hit the same side.  This is illustrated in the left panel of Fig.~\ref{fig:mrk}.  Consider the bottom straight edge of the stadium, which is represented by the position interval $\pi/2 \le q \le \pi/2+2\gamma$.  There are three zones.  At the top are the trajectories coming from the right semicircle to the straight edge, in the middle are trajectories coming from the upper straight edge, and at the bottom trajectories coming from the left semicircle.  There is no problem yet with the four symbol idea, although a trajectory cannot start on a straight edge and return to itself on the first bounce (there are only three regions, not four).  This has to be pruned out of the symbol sequence whatever final form it takes; a straight edge symbol cannot repeat.  Next consider the trajectories that end up in the left semicircle, $\pi/2+2\gamma \le q \le 3\pi/2+2\gamma$.  There are five zones instead of four.  It turns out not to be sufficient just to label the four side segments.  The middle three zones come from the two straight edges and the right semicircle, which leaves the top and bottom zones.  In the top zone, the trajectories are advancing in a clockwise direction, whereas at the bottom, they are advancing in a counterclockwise direction.  Although, they have the same sequence of hitting the semicircle, the zones are not contiguous and must be distinguished.  This suggests six symbols as a starting point, two symbols for each semicircle, one for clockwise advancing trajectories, another for counterclockwise.  Another pruning rule comes to mind immediately with further iterations, a trajectory cannot return to the same semicircle in two iterations if it hits a straight edge on the first.  There are others as well.  Note also that there is a clear symmetry in that the structure of the two semicircles and of the two straight edges, being identical.  It would be desirable to generate a coding which reflected all symmetries so that they would be immediately visible in the coding.

In an efficient symbolic coding with no pruning or grammar rules excluding certain symbol sequences, each additional mapping would proliferate the number of partitions by the number of symbols in the coding.  Here there are 16 partitions of the phase space in the left panel, and 60 partitions in the middle panel, which itself results from following two iterations of the map.  The number 60 is far fewer than $6\times16$ (it's actually $3.75\times 16$).  This raises the question of whether there might be a more efficient set of symbols, perhaps with only four symbols and far fewer pruning rules.  The right panel resulting from three iterations of the map has $192$ partitions, which is only $3.2$ times larger than the number in the middle panel, even a smaller ratio.  Armed with just this meager amount of information, it appears that the positive Lyapunov exponent and KS entropy are likely to be less than $\ln 3.2\approx 1.16$ (this is still larger than much better estimates).  Compare this to trajectory no.~2 of Fig.~\ref{fig:sospo8}, for which $2\cosh(2\mu t)=34$ or $\mu\approx 1.76$.  The stadium is much more generic than the baker's map, and the spatial behavior of the finite time stability exponent fluctuations are reflected in the large range of areas of the partitions in Fig.~\ref{fig:mrk}.  Suffice it to say that a typical chaotic system, as opposed to some rather simple paradigm is likely to require quite a bit of work to identify an optimal or relatively efficient coding.

\section{Skeletons of chaos}
\label{sec:geometry}

Just as periodic trajectories are regarded as skeletons of chaos~\cite{Cvitanovic91}, so too and equally well are homoclinic and heteroclinic trajectories.  There are a number of advantages to the use of these trajectories.  To start with, the calculation of unstable and stable manifolds can be done extremely accurately because neighboring trajectories collapse onto the manifolds propagating either forward and backward in time, respectively.  This manifold stability implies that homoclinic and heteroclinic trajectories can be identified in extremely stable ways by calculating unstable and stable manifold crossings instead of using a method that relies on propagating Hamilton's equations directly~\cite{Li17}.  Furthermore, as discussed in Sec.~\ref{sec:sets}, the manifolds foliate the available phase space, bound resonance zones and turnstiles, and the periodic and homoclinic trajectories imitate each other. This is true to such an extent that a numerical method for obtaining all periodic trajectories of a certain time (length code) can be based off of obtaining the homoclinic trajectories at unstable and stable manifold intersections~\cite{Tomsovic93}.  Knowledge of the homoclinic trajectories enables the exact calculation of classical action differences between periodic trajectories and homoclinic trajectories, and even the classical actions of finite segments of any trajectory in terms of phase space areas bounded by segments of invariant curves~\cite{Li20}.  Inevitably, the dominant contribution to the action differences can be related to segments of unstable and stable manifolds, with only exponentially small corrections.   The fact that the manifolds are Lagrangian plays a critical role in the far-reaching relationships between them~\cite{Li17a, Li18}.   These relationships arise due to the classical action differences between pairs of trajectories being related to areas in phase space delineated by unstable and stable manifolds, and Moser invariant curves~\cite{Moser56, Ozorio89}, thus providing a very geometric structure underlying the classical actions that connect quantum and classical dynamics of chaotic systems.  The precise areas follow by deforming paths along the Lagrangian manifolds or using the MacKay-Meiss-Percival action principle~\cite{MacKay84a}.   These areas turn up repeatedly amongst various trajectory pairs, and generate an exact decomposition of all the possibilities.  A great simplification results in the sense that an exponentially proliferating number of action differences can be expressed as certain sums over a less-than-exponential set of areas~\cite{Li19}.  It also turns out that Maslov indices and the relationships between stability coefficients for pairs of trajectories can be decomposed, simplifying the description of those differences (fluctuations) as well~\cite{Li19a}.  All the classical ingredients of trace formulas or semiclassical constructions of quantum propagators and Green functions adhere to the relationships imposed by this geometry; recall Eqs.~\eqref{eq:vanvleck} and \eqref{eq:gtf}, for example.  This alternate perspective, which focusses on phase space areas, gives a direct insight into the structural stability of chaotic dynamics discussed in Sec.~\ref{sec:strucstab} ahead.  Although,  a particular trajectory is exponentially unstable under any kind of perturbation, areas bounded by unstable and stable manifolds are very weakly altered, and they control quantum interferences.  This perspective also lends itself to calculating curvature corrections to cycle expansions~\cite{Chaosbook1} and giving exact action differences in Sieber-Richter pairs~\cite{Li17b}.  These two applications are discussed further below.

Perhaps the simplest phase space area-action difference relation results between a periodic trajectory and one of its primary homoclinic trajectories.  It gives the resonance zone area discussed in Sec.~\ref{sec:rzt}.  In Fig.~\ref{fig:soshet}, the resonance area $A_R$ delimited by the unstable manifold from point $C$ to point a and from there along the stable manifold to $C^\prime$ and back along the $p=0$ line, is given by calculating the infinite limiting action difference between the horizontal bounce periodic trajectory (no.~2 in Fig.~\ref{fig:sospo8}) and its primary homoclinic trajectory (labelled a in Fig.~\ref{fig:soshet}).  For the $\gamma=1$ stadium billiard the action difference on the energy surface $|p|=1$ equals $3.36839$, precisely the resonance area $A_R$.  Had the other primary homoclinic trajectory been chosen (labelled b in Fig.~\ref{fig:soshet}) the difference would have been $2.991143$.  In that case, the switch from the unstable to stable manifold must be taken at point $b_0$ or one of its iterates, which encloses a smaller area, the difference being given by the turnstile area $A_t$.  Hence, this also establishes that the limiting action difference between the pair of primary homoclinic trajectories gives the flux through the turnstile, $A_t=0.377248$ and determines transport properties.  

For more complicated periodic trajectories and/or homoclinic trajectories with longer, more complicated excursions, it can become much more difficult to follow the manifolds and see which areas are relevant.  However, with a symbolic dynamics established for a chaotic system, the entire enterprise can be elucidated with the symbolic sequences.  An interesting consequence of the action difference/area relations is related to the escape time transport studies based on homoclinic tangles of Mitchell et al.~\cite{Mitchell03a, Mitchell03b, Mitchell06}.  The tangle areas they identify governing transport can be expressed in terms of action differences of specific combinations of homoclinic trajectories, which gives an alternate way to calculate the areas directly.  

\subsection{Cycle expansions}
\label{sec:cycles}

Closely related to the Gutzwiller trace formula, Eq.~\eqref{eq:gtf}, there are expressions between the spectrum and periodic trajectories in the form of dynamical zeta functions and spectral determinants that can have greatly accelerated convergence properties~\cite{Cvitanovic88, Cvitanovic89, Chaosbook1}.  In these expressions, it is possible to make expansions, called cycle expansions, that group certain related periodic trajectories in such a way as to create significant cancelling of contributions.  In order to understand what groups of trajectories can be advantageously combined, consider the fact that it is possible to make small deviations from one or more shorter periodic trajectories in order to create a longer one that imitates or shadows a sequence of the shorter ones.  In Fig.~\ref{fig:sospo8}, there are two visible, simple examples, i.e.~there exists small shifts from nos.~$2$ and $5$ that lead to no.~$6$, and likewise, nos. $2$ and $7$ to no.~$8$.  The combined classical actions and stability properties of the longer trajectories closely match the combination from the shorter trajectories that the longer one shadows.  In this way, the shorter, more ``primitive'' periodic trajectories generate the dominant contributions, as there are no other shadowing trajectories with which to cancel.  However, the longer trajectories that do shadow the primitive ones, have grouped contributions that do nearly cancel.  The contribution remainders are known as curvature corrections, and though small, they can be important and essential for accurate approximations of the spectrum.  This is the mechanism behind the greatly accelerated convergence.  The terminology dubs the longer trajectories cycles, and their primitive shadowed components pseudo-cycles.  The system's symbolic dynamics provides an essential tool for generating all the possibilities, and it is related to the cycle expansion itself.  Given that the longer periodic trajectories are shadowing the shorter ones, their symbol sequences follow from joining together (concatenating) the shorter sequences of the primitive trajectories.  It is straightforward to verify that this works for the two examples mentioned just above.

\begin{figure}[t]
\centering
\includegraphics[width=0.99\textwidth]{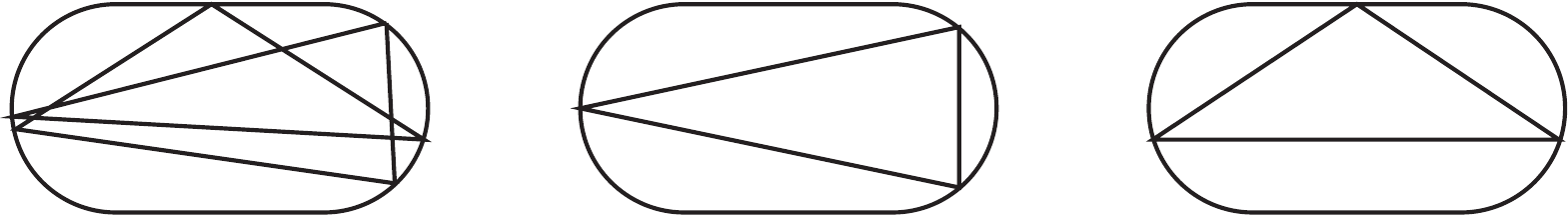}
\caption{Example from the $\gamma=1$ stadium billiard of a longer periodic trajectory sequentially shadowing two shorter primitive periodic trajectories.  The two triangular $3$-bounce periodic trajectories themselves shadow the two primitive homoclinic trajectories of the horizontal bounce periodic trajectory.  Subtracting the $6$-bounce trajectory length from the sum of the two primitive lengths gives, $8.977479 + 8.601952 - 17.554815 = 0.024616$, a very small result.  The $6$-bounce value of ${\rm Det}\left({\bf M}-\mathbb{1}\right)=-3267.27$ also compares very closely to the product of the shorter trajectories' determinants, $68.35\times -48.05 = -3284.90$.}
\label{fig:cycle}
\end{figure}
Another stadium billiard example is given in Fig.~\ref{fig:cycle} using two of the shortest primitive periodic trajectories.  Abstractly labeling their symbolic codes $\overline{\gamma_1}$ (center panel) and $\overline{\gamma_2}$ (right panel), the symbolic code of the trajectory in the left panel is $\overline{\gamma_1\gamma_2}$.  The two primitive trajectories are associated with the two primary homoclinic trajectories of the homoclinic tangle of the shortest, most unstable periodic trajectory, i.e.~no.~$2$ of Fig.~\ref{fig:sospo8}.  There exist small deviations from either of the two triangular $3$-bounce trajectories, $\overline{\gamma_1}$ and $\overline{\gamma_2}$ , that lead to the $6$-bounce trajectory, $\overline{\gamma_1\gamma_2}$.  The curvature corrections have two main components, the classical action as in Eq.~\eqref{eq:hcf}, $W_{\gamma_1\gamma_2}$ is not quite equal to $W_{\gamma_1}+ W_{\gamma_2}$, and the stability ${\rm Det}({\bf M_{\gamma_1\gamma_2}}-\mathbb{1})$ is not quite equal to ${\rm Det}({\bf M_{\gamma_1}}-\mathbb{1})\times{\rm Det}({\bf M_{\gamma_2}}-\mathbb{1})$.  In this example, $W_{\gamma_1\gamma_2}$ is just a single part in $700$ shorter than the sum of the other two lengths, and the value of ${\rm Det}({\bf M}_{\gamma_1\gamma_2}-\mathbb{1})$ is about a half percent smaller than the product for the other two determinants.  Although, the differences are quite small, they may still be significant in a semiclassical expression, depending on the wave vector, i.e.~value of $\hbar$, especially the classical action correction.  

In~\cite{Li18} it is shown how to express the exact classical action correction, $\Delta W_{\gamma_1\gamma_2} = W_{\gamma_1\gamma_2} - W_{\gamma_1}- W_{\gamma_2}$, in terms of a phase space area bounded by invariant manifolds.  This leads to a dominant approximation for $\Delta W_{\gamma_1\gamma_2}$ in terms of an area, ${\cal A}^0$, that is somewhat like a slightly deformed parallelogram bounded by two unstable manifold segments, $U$, and two stable manifold segments, $S$, whose four corners and segments can be specified by particular points on four separate homoclinic trajectories.  The notation, $SUSU[ h^{(\gamma_1\gamma_2)}_0, h^{(\gamma_2\gamma_2)}_0, h^{(\gamma_2\gamma_1)}_0, h^{(\gamma_1\gamma_1)}_0 ]$, defines the deformed parallelogram starting at the homoclinic point $h^{(\gamma_1\gamma_2)}_0$, following the stable manifold to the point $h^{(\gamma_2\gamma_2)}_0$, switching to the unstable manifold, and so on until completing the loop using the four segments.  With the help of a symbolic dynamics, it turns out that the approximation is given by the area ${\cal A}^\circ$
\begin{equation}
\Delta W_{\gamma_1\gamma_2} \approx {\cal A}^\circ_{SUSU[ h^{(\gamma_1\gamma_2)}_0, h^{(\gamma_2\gamma_2)}_0, h^{(\gamma_2\gamma_1)}_0, h^{(\gamma_1\gamma_1)}_0 ]}\ ,
\end{equation}
which has exponentially small corrections that can be estimated.  Hence, the dominant part of the action curvature correction is given by a curvy parallelogram in phase space.  Of course, one could just calculate $\Delta W_{\gamma_1\gamma_2}$ directly, but due to the invariance of the unstable and stable manifolds, from the parallelogram area approximations it is possible to deduce all kinds of relationships between the various curvature corrections and combinations of simpler curvature corrections.  Without giving any details, note that in~\cite{Li19a} a method of calculating corrections to stabilities is given in terms of ratios of certain homoclinic phase point differences as well, again leading to various relationships between different stability curvature corrections.  The stability correction method also applies to the Sieber-Richter pairs discussed next.

\subsection{Sieber-Richter pairs}
\label{sec:srp}

\begin{figure}[t]
\centering
\includegraphics[width=0.99\textwidth]{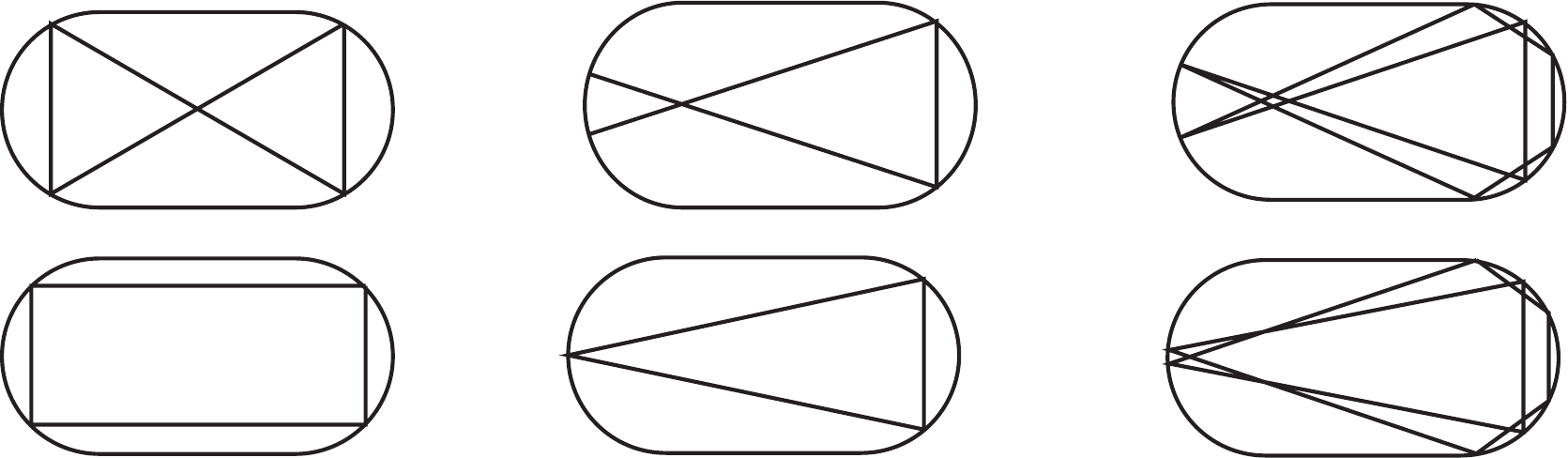}
\caption{Illustration of $4$-bounce, $6$-bounce, and $8$-bounce, left-to-right, respectively, Sieber-Richter pairs in the stadium billiard.  In each case, the crossing trajectory of the pair has the longer path length and is less unstable, and the Maslov indices of the pair are identical.  The action differences, $\Delta W$, and ratios, $R={\rm Tr}(M_1)/{\rm Tr}(M_2)$, for these three examples are: left) $\Delta W = 10.392305 - 9.656854 = 0.735451$, $R= 98.00/175.25=0.56$; center) $\Delta W = 18.081381 - 17.954958 = 0.126423$, $R= 4205.82/4947.41=0.85$; and right) $\Delta W = 18.506560 - 18.321952 = 0.184608$, $R= 11553.75/14854.78=0.78$.}
\label{fig:srp}
\end{figure}

Amongst the set of periodic trajectories in a chaotic system, there exist some that cross themselves in configuration space with small angle differences between their momenta (or almost in opposite directions); this is known as an encounter.  In time-reversal invariant systems they play an essential role in introducing correlations in classical actions~\cite{Sieber01}.  These correlations are linked to spectral fluctuation properties of quantum chaotic systems, which according to the Bohigas-Giannoni-Schmit conjecture imitate random matrix theory.  The correlations appear most directly in the squared modulus of the Fourier-transformed density of energy levels, i.e. the spectral form factor~\cite{Mehta04}.  They go beyond the leading contribution arising from the uniformity principle~\cite{Hannay84, Berry85, Mueller04, Heusler07, Haake18}.  In the simplest circumstances,  there exists an additional trajectory which does not cross itself and follows somewhat closely the crossing trajectory, but a portion of the trajectory is shadowed in the time-reversed direction.  This is illustrated by the left vertical pair of trajectories in Fig.~\ref{fig:srp}.  Starting in a clockwise motion on the left semicircle, the upper trajectory traverses the right semicircle in a counterclockwise direction, whereas the lower member of the pair follows a somewhat similar path in the clockwise direction on the right side; the diagram also represents the time-reversed pair as well,  i.e.~starting counterclockwise on the left.  In this example, the crossing angle is rather large, so the action correlation induced is not especially relevant for modifying the spectral form factor, but it serves to illustrate the basic idea.  In order to find the occurrence of a small angle, which leads to stronger correlations, much longer trajectories are required.  Another interesting circumstance is the self-retracing trajectory in the upper middle pair.  It winds both clockwise and counterclockwise around the triangular area on the right side of the stadium.  Below is its associated twice traversed non-crossing triangular periodic trajectory.  In this case, the self-retracing trajectory is associated with two non-crossing trajectories since there is both a clockwise and counterclockwise version of the lower trajectory.  Finally, a third example is shown with the right pair in which one of the trajectories follows the two excursions on the right in the same rotation direction whereas the other does so in opposite senses.  It can be difficult to imagine all the possibilities, and having a symbolic coding can be extremely helpful.

\begin{figure}[b]
\centering
\includegraphics[width=0.60\textwidth]{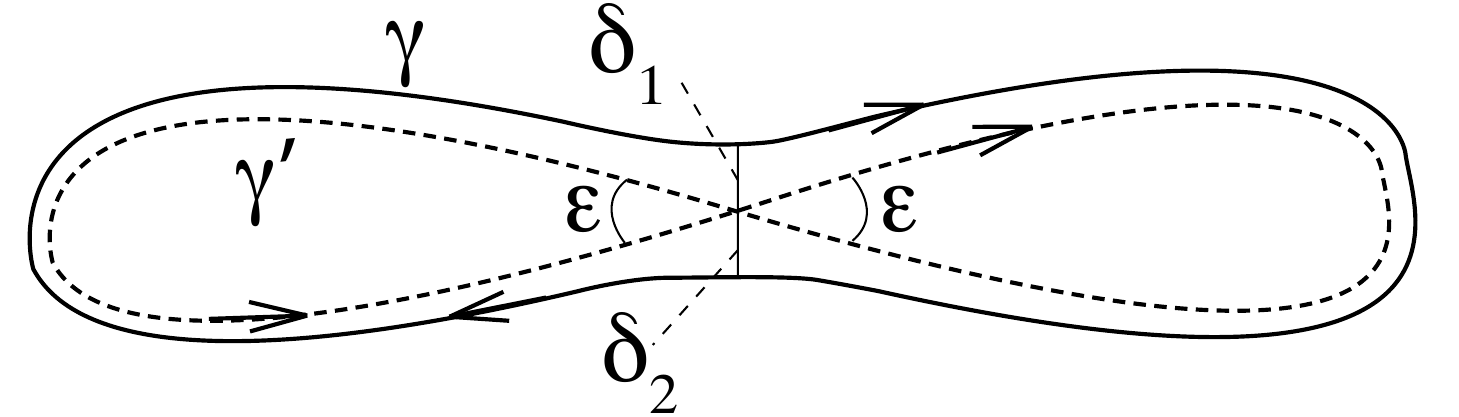}\qquad \includegraphics[width=0.36\textwidth]{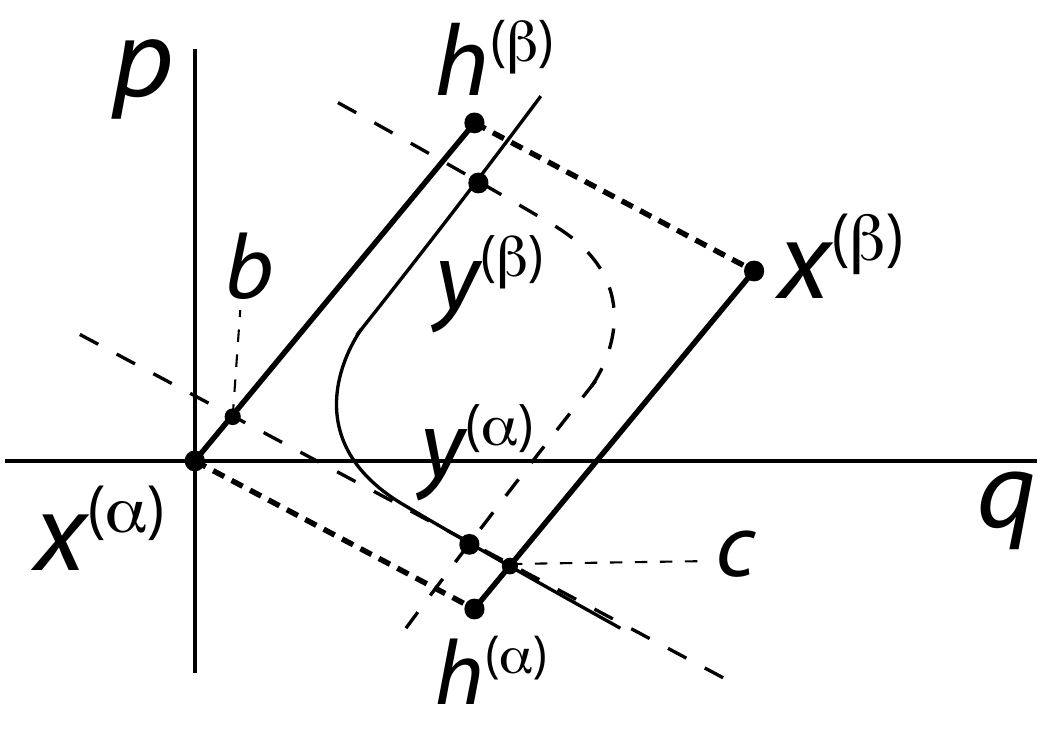}
\caption{Schematic illustration of useful local surface of section for a Sieber-Richter pair containing a single encounter.  The four phase points, $y^{(\beta)}, y^{(\alpha)},x^{(\beta)}, x^{(\alpha)}$, correspond to the crossing and non-crossing trajectories, plus their time-reversed partners.  The thicker solid and dashed lines correspond to unstable and stable manifold segments of $x^{(\beta)}, x^{(\alpha)}$, respectively.  The lighter solid and dashed lines correspond to Moser invariant curves.  Reproduced from Figs.~13 \& 15 of~\cite{Li17b}.}
\label{fig:srp2}
\end{figure}
Just as for the compound cycles in Sec.~\ref{sec:cycles}, the classical action difference between the trajectories of a Sieber-Richter pair can be given an exact relationship with an area in phase space bounded by segments of invariant manifolds~\cite{Li17b}.  The basic construction is illustrated in Fig~\ref{fig:srp2}.  A beneficial surface of section can be defined by a plane normal to the crossing trajectory's intersection point.   The crossing trajectory is labelled $y^{(\beta)}$ and it's time-reversed partner by  $y^{(\alpha)}$.  Likewise, the non-crossing trajectory and it's time-reversed partner correspond to the points $x^{(\beta)}$ and $x^{(\alpha)}$, respectively.  Segments of the unstable and stable manifolds of $x^{(\beta)}, x^{(\alpha)}$ are shown as well as the Moser invariant curves associated with $y^{(\beta)}, y^{(\alpha)}$.  Sieber and Richter gave an approximate, slight overestimate, classical action difference in terms of $p,\epsilon, \delta_1, \delta_2$, which happens to equal the area of the parallelogram formed by the unstable and stable manifold segments, i.e.
\begin{equation}
\Delta W \approx \frac{p\epsilon}{2}\left(\delta_1+\delta_2\right) =  {\cal A}^{\circ}_{USUS[x^{(\alpha)}h^{(\beta)}x^{(\beta)}h^{(\alpha)}]}
\end{equation}
For the left-most pair of trajectories in Fig.~\ref{fig:srp}, the Sieber-Richter estimate is given by $0.74048$ whereas the true value given in the caption is $0.73545$.  It is quite an excellent approximation.
The exponentially small correction to their estimate is given by the parallelogram area in Fig.~\ref{fig:srp2} involving the phase points $b,c,h^{(\alpha)},x^{(\alpha)}$.

\section{Perturbations}
\label{sec:strucstab}

There are many quantum chaos motivations for considering perturbations to chaotic dynamical systems, e.g.~the study of parametric fluctuations, such as universal conductance fluctuations~\cite{Lee85, Altshuler85, Goldberg91, Marcus92}, fundamental and dynamical weak symmetry breaking~\cite{Bohigas93, Bohigas95, Tomsovic00}, bifurcations~\cite{Arnold92}, and fidelity studies~\cite{Peres84, Jalabert01, Jacquod01b, Cerruti02, Cerruti03}, to name a few.  A key perturbation feature is the paradoxical structural stability of chaotic systems~\cite{Andronov37}.  Despite the exponential sensitivity of individual trajectories, it turns out that changes to structures such as unstable manifolds are very slight.  Curiously, from semiclassical methods it is clear that quantum mechanics does not depend on the exponential instability of individual trajectories, but rather on the structural stability of evolving Lagrangian manifolds.  Although individual trajectories appear in semiclassical summations, such as the van Vleck-Gutzwiller propagator~\cite{Vanvleck28, Gutzwiller71}, the nature of the two-point boundary value problem is such that the trajectory initial and final conditions under a small perturbation adjust in such a way that the action change can be calculated to first order along the unperturbed trajectory.  A small adjustment along the stable manifold near the initial boundary and a small adjustment along the unstable manifold near the final boundary occurs which effectively cancels out any exponential alteration of the classical actions.  The small action changes calculated along the unperturbed trajectories behave much like a random variable, and in semiclassical summations the statistical fluctuations of the classical actions becomes relevant.  

\subsection{Manifold stability}
\label{sec:ms}

\begin{figure}[t]
\centering
\includegraphics[width=1.0\textwidth]{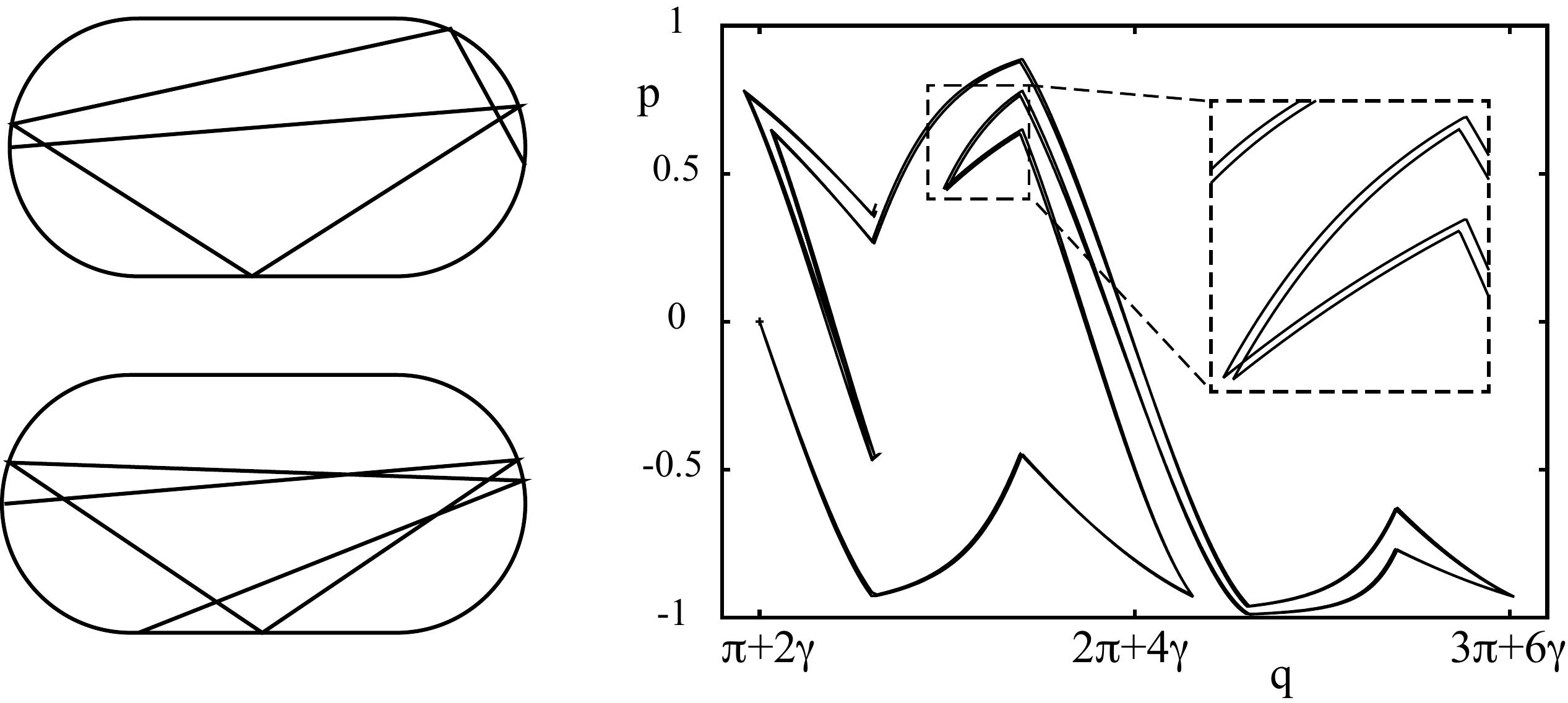}
\caption{Perturbation of trajectories and unstable manifold.  Five bounces of two trajectories starting from the same initial condition $(0, 0.075)$ are shown for $\gamma=1.0$ top left, and $\gamma=1.05$, bottom left.  Clearly, the two trajectories diverge exponentially rapidly from each other.  Roughly the same portion of the unstable manifold of trajectory no.~2 of Fig.~\ref{fig:sospo8} (horizontal bounce periodic) is shown for both values of $\gamma$.  The two manifolds are nearly indistinguishable in the right panel and that is a manifestation of the structural stability of the chaotic dynamics.  The inset magnifies the two manifolds where they differ the greatest in order to see the differences better.}
\label{fig:stabman}
\end{figure}

A good illustration of the structural stability can be made with a portion of the unstable manifold of the no.~$2$ (Fig.~\ref{fig:sospo8}) horizontal bounce periodic trajectory in the stadium billiard.  A convenient perturbation is to alter the length of the straight edges.  As shown in Fig.~\ref{fig:stabman} the beginnings of two trajectories using same initial condition (chosen close to the primary homoclinic trajectory a of Fig.~\ref{fig:soshet}) for two different values of $\gamma$ are shown on the left in Fig.~\ref{fig:stabman}.  The two diverge exponentially rapidly, are not close by the fourth bounce, and further propagation of the two leads to completely independent future exploration of the available phase space.  On the other hand, the segments of the unstable manifold associated with the horizontal bounce trajectory for the two values of $\gamma$ lie almost entirely on top of each other, and they are hard to distinguish in the figure.  At first it might seem paradoxical or hard to reconcile how all the trajectories are moving exponentially apart under perturbation, but the unstable manifold is barely affected.  However, consider the stadium trajectories in the neighborhood of the horizontal bounce periodic trajectory.  The Birkhoff normal coordinates discussed in Sec.~\ref{sec:bnc} and shown in Fig.~\ref{fig:sos} must differ slightly between the two different systems ($\gamma$ values).  Hence, the exact same initial condition appears in a slightly different location relative to the origin.  Or in other words, its projection along the the unstable and stable manifold has changed slightly (the values of $(Q,P)$ have shifted slightly).  Since two neighboring points on the unstable manifold diverge exponentially forward in time so must the identical initial condition diverge exponentially when propagated for two different values of $\gamma$, but that doesn't require more than a tiny alteration of the unstable manifold.  This situation is not only true near the horizontal bounce periodic trajectory, but everywhere in phase space.  Similarly, the same logic applies backward in time with respect to the stable manifold.  So the exponential divergence results forward or backward in time.

Thus, due to this structural stability the lowest order phase change of a trajectory's contribution in the van Vleck-Gutzwiller propagator is not exponentially sensitive to the strength of the perturbation, but rather linear. It turns out to be given by dividing the first order classical action correction by $\hbar$~\cite{Bohigas95}.  For the propagator the quantities being held fixed are the initial and final positions and time.  The new initial and final momenta of the stationary phase condition adjust depending on the perturbation, and the first order phase correction is calculated by integrating the new term in Hamilton's principal function along the original unperturbed trajectory labelled by $\alpha$, 
\begin{equation}
\label{eq:delaction}
\Delta S_\alpha(q, q^\prime; t) = \epsilon \int^t_0 {\partial {\cal L}_\alpha (q, q^\prime; t^\prime) \over \partial \epsilon}{\rm d}t^\prime \ .
\end{equation}
One application of this phase change is in the semiclassical theory of the quantum perturbative and Fermi Golden Rule regimes of the fidelity~\cite{Peres84, Cerruti02, Cerruti03}.  The basic idea is that since the trajectories all explore phase space in independent ways, for trajectories propagated longer than the logtime, small contributions from different trajectory segments are essentially randomized, and the total phase change is like a random walk in smaller phase changes.  This generates a classical action diffusion constant within an energy window centered at $E$ denoted $K(E)$~\cite{Bohigas95, Lakshminarayan99, Lakshminarayan00, Cerruti00}, 
\begin{equation}
<\Delta S^2_\alpha(q, q^\prime; t) >_\alpha - <\Delta S_\alpha(q, q^\prime; t) >_\alpha^2 = 2\epsilon^2 K(E) t
\end{equation}
where $K(E)$ is given by
\begin{equation}
\label{eq:ke}
K(E) = \int_0^\infty \left\langle {\partial {\cal L}_\alpha (q, q^\prime; 0) \over \partial \epsilon}{\partial {\cal L}_\alpha (q, q^\prime; t) \over \partial \epsilon}  \right\rangle dt
\end{equation}
with the averaging over all the contributing trajectories within the energy window.  This quantity is also linked to quantum level velocity variances under perturbation in the same works.

Another example arises in Gutzwiller's trace formula where the main contribution to the phase is based on Hamilton's characteristic function, $W_{\cal C}$, along the periodic trajectory.  The first order classical action correction is given by~\cite{Bohigas95},
\begin{equation} 
\label{e:pt}
\delta W_{\cal C}  =  \epsilon \oint {\partial {\cal L}_\alpha (q, q^\prime; t) \over \partial \epsilon} \, \rm{d} t\ ,
\end{equation}
where $\cal C$ is along the unperturbed periodic trajectory.  These action variations determine the precise changes in the quantum spectrum as a parameter is varied.  This expression leads to the same classical action diffusion constant and relationships with level velocities when treated statistically or summed over all periodic trajectories.

\subsection{Partition stability and bifurcations}
\label{sec:psb}

\begin{figure}[t]
\centering
\includegraphics[width=0.3\textwidth]{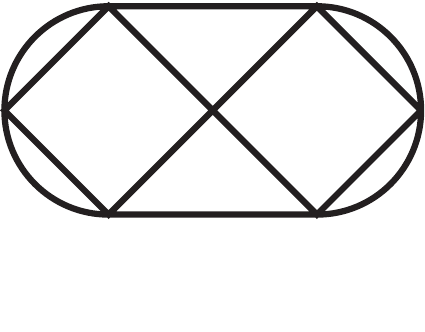}\qquad \includegraphics[width=0.65\textwidth]{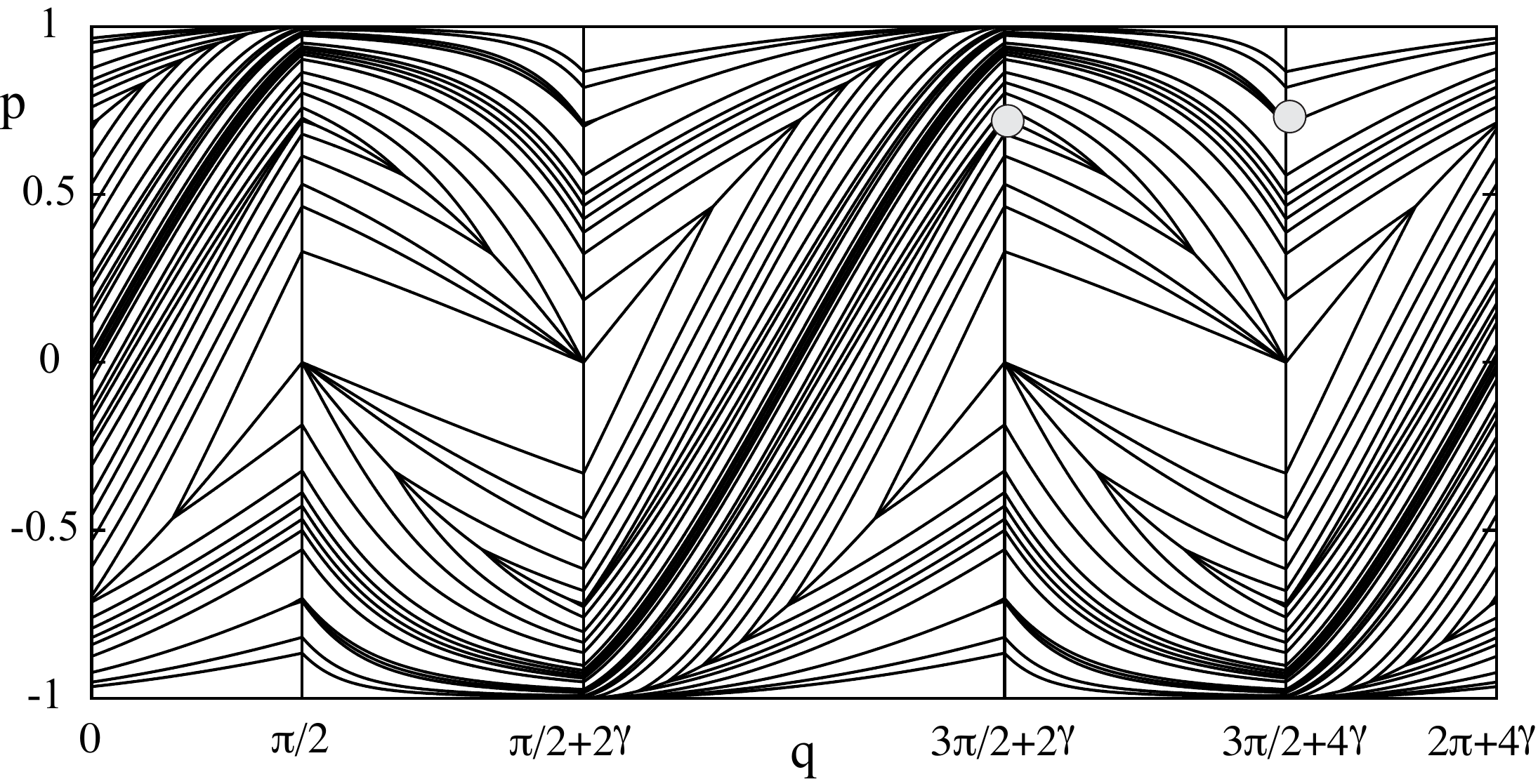}
\caption{Effect of perturbation on the partition.  Comparing to Fig.~\ref{fig:mrk} in which $\gamma=1.0$, here $\gamma=1.05$ and the differences are quite small.  The one and two bounce partitions have the exact same number, 16, 60, respectively.  The boundaries are slightly shifted, but the differences are imperceptible.  The three bounce partition has the first structural change.  Overall, it appears to the eye almost exactly like the right panel of Fig.~\ref{fig:mrk}  except for underneath the superposed two small grey circles, and their symmetry related counterparts.  A boundary extended across the semicircle-straight edge joint creating the need for new partition labels for 16 partitions.  All of them are related to a single trajectory bifurcation that occurs for $\gamma=1$ at which point the left pictured trajectory comes into being.  It has both reflection symmetries.  However, for $\gamma>1$, there exist 16 similar trajectories with various symmetries (one factor 2 is related to the direction of momentum, so geometrically 8).  The four bounces at the semicircle-straight edge joint, for $\gamma>1$ can deform slightly so as to be all four on the semicircle or all four on the straight edge, or two can be on the semicircle and two on the straight edge.}
\label{fig:pert}
\end{figure}
A further illustration of structural stability comes by looking at the partitions of the phase space depending on the number of map iterations as seen in Fig.~\ref{fig:pert}.  There are no discernible differences between the one and two bounce partitions for the two values of $\gamma$.  They both have the identical number of partitions, $16$ and $60$, respectively.  However, a tiny difference appears in the three bounce partition with the introduction of 16 new tiny partitions.  Their locations are underneath the light grey circles of the three bounce partition map in the right panel of Fig.~\ref{fig:pert} and those two locations' symmetry related images for a total of 16.  The new partitions are still very small for $\gamma=1.05$, difficult to notice, and grow with increasing $\gamma$.  They represent new symbolic codes for the dynamics, meaning that there are trajectories possessing new sequences in the order of sides hit.  

In this case all 16 tiny partitions are associated with the introduction of a single bifurcation.  It is also associated with a new periodic trajectory of six bounces that first appears at $\gamma =1$ pictured on the left side of Fig.~\ref{fig:pert}.  For $\gamma >1$, instead of a single closed curve (two periodic trajectories forward and backward in time), there exist $8$ closed curves, ($16$ periodic trajectories), some of which possess lower reflection symmetry.   This is a typical circumstance, i.e. that along with a perturbation there sometime comes a bifurcation.  It creates new dynamical possibilities and is reflected in the symbolic coding as an altered pruning or grammar rule.

The sudden appearance of new periodic trajectories under parameter variation has consequences for trace formulas.  It suggests a discontinuous alteration of a quantum spectrum under parameter variation, which is not possible.  In the stadium the joints of the semicircle and straight edges introduce a weak form of diffraction due to the discontinuity in the second derivative (curvature) along the boundary at that point.  The resolution of this problem there would be to introduce diffractive trajectories.  The diffractive periodic trajectories fade rapidly with increasing discontinuous angle changes.  Approaching the introduction of a new periodic trajectory would lead to an increasing diffractive contribution in such a way that the discontinuity would be smoothed out.  For smooth systems, the resolution depends on a complexification of Hamiltonian dynamics and led to the discovery of so-called ghost orbits~\cite{Kus93}.  They are discussed briefly in the next section.

\section{Complex trajectories}
\label{sec:complex}

It has long been known that quantum and wave mechanics encompasses phenomena, such as tunneling, diffraction, etc... that are classically forbidden~\cite{Merzbacher02, Nussenzweig92}.  The concept of analytically continuing classical mechanics to account for such phenomena goes back at least to 1972~\cite{Miller72}.  In ordinary (real) time-dependent WKB theory beginning with torus quantization~\cite{Maslov81}, the tori start out as real Lagrangian manifolds and the distinction between classically allowed and forbidden is determined by whether analytic continuation is required at some point in the semiclassical method.  A bit more recently, something as simple as propagating a wave packet or Glauber bosonic coherent state~\cite{Glauber63} in a time-dependent WKB theory has been discovered also to include necessarily complexified canonically conjugate position and momentum variables~\cite{Klauder78, Huber88, Baranger01}, but in this case does not, a priori, imply the presence of classically forbidden phenomena.  In essence, the required Lagrangian manifold underlying Gaussian states is necessarily complex~\cite{Huber88}.  Interestingly, the distinction of classically allowed and classically forbidden is not determined by real versus complex classical mechanics in this case, and the story is a bit more complicated.  In either case, there is a strong motivation to investigate complex classical dynamics.

\begin{figure}[t]
\centering
\includegraphics[width=0.59\textwidth]{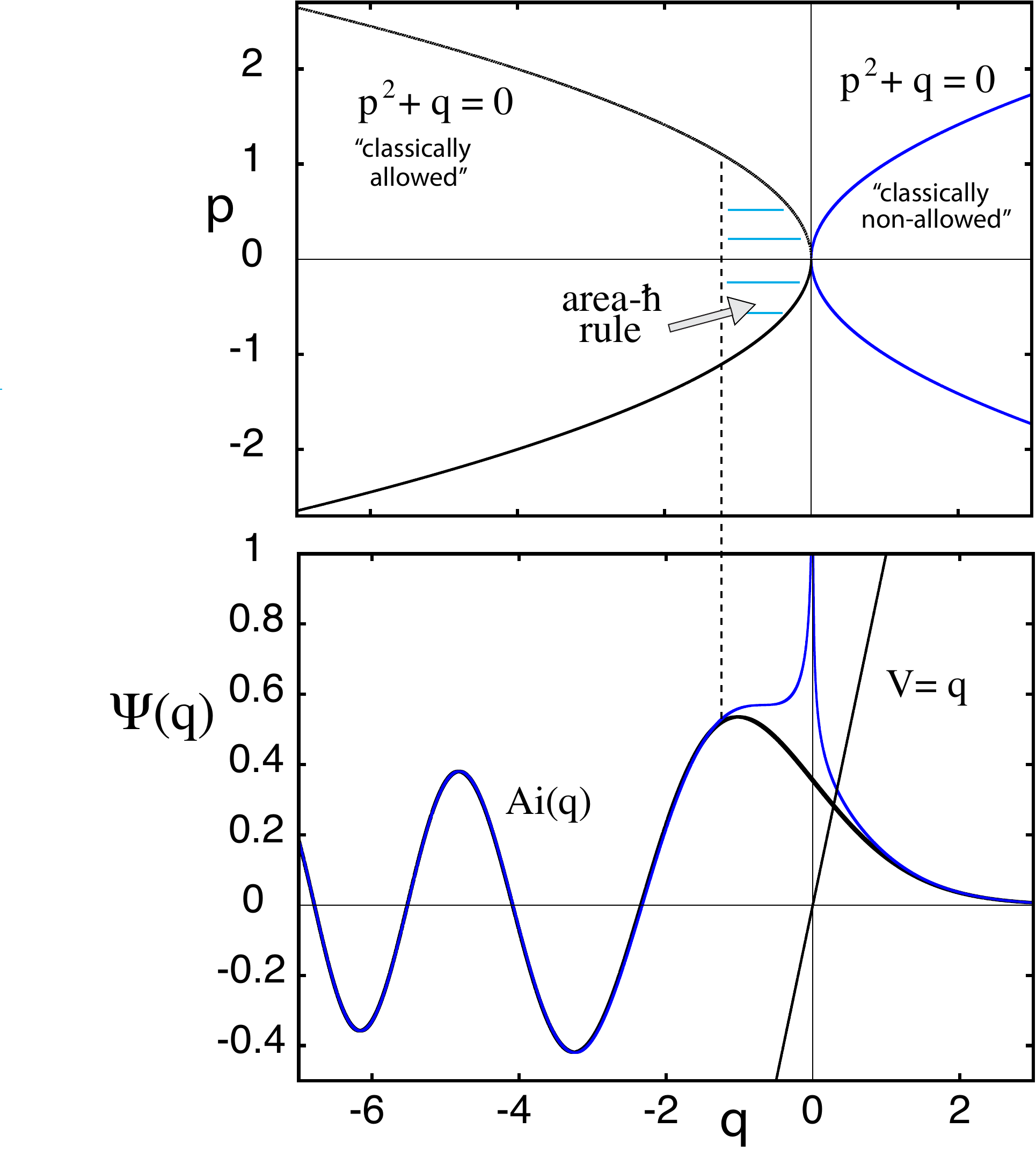}
\caption{Illustration of the utility of a complex trajectory for the Airy function.}
\label{fig:complexcontour}
\end{figure}

A rather straightforward example of the utility of the complexification of classical mechanics arises for the solutions of the Schr\"odinger equation with a linear ramp potential.  Let the Hamiltonian be
\begin{equation}
\hat H = \frac{\hat p^2}{2m} + \alpha \hat q \qquad{\rm  with} \ \{m=1/2,\ \alpha=1,\ \hbar=1\}
\end{equation}
For every continuous value of the energy, there is an Airy function eigensolution.  For the $E=0$ energy surface, the solution is $Ai(q)$.  The semiclassical description rests on the classical trajectories following the $E=0=p^2+q$ contours, which are drawn in Fig.~\ref{fig:complexcontour}.  For the $q\le0$ contour, the momentum is real.  The trajectory begins on the left moving to the right turning around at $q=0$ and gaining momentum moving back to the left.  The usual semiclassical construction for negative $q$ based on that classical trajectory is quite accurate as a function of $q$ up to the point where the area-$\hbar$ rule~\cite{Berry83} is violated (indicated by the dotted line).  On the other hand, there is no real trajectory for positive values of $q$.  Nevertheless, there is an equipotential contour with $p$ purely imaginary.  It is indicated in blue.  It turns out that this contour is also a solution of Hamilton's equations if time is integrated along a purely negative imaginary time path.  The trajectory starts at the far right for an initial condition of large negative imaginary $p=p(0)$ $[q(0)=-p(0)^2]$ and moves to the left until it arrives at $(p,q)=(0,0)$ for $t=p(0)$, at which time it begins moving back to the right with positive purely imaginary $p$.  In this case, the usual semiclassical construction generates an exponentially shrinking behavior that quite accurately describes the tail of $Ai(q)$ beyond the area-$\hbar$ rule (the boundary condition at infinity requires dropping the negative imaginary $p$ branch of that contour).    Hence, for this Hamiltonian, including an imaginary $p$ and $t$ trajectory results in an accurate approximation of the exponentially subdominant quantum behavior.  This subdominant behavior is often at the heart of tunneling and diffraction behaviors and complex classical mechanics contains the necessary trajectories to predict the phenomena.

For classically non-allowed processes, the dominant contributions come from the trajectories with the smallest imaginary classical actions due to the resulting exponential suppression.  Not too surprisingly then, in integrable dynamical systems, there tends to be a very small number of dominant trajectories, often just a single trajectory as typically happens with potential barrier tunneling.  This is also true for dynamical tunneling~\cite{Davis81} for which a potential barrier does not exist, but dynamical barriers do.  However, this is not true of chaotic dynamical systems.  Phenomena such as chaos-assisted tunneling~\cite{Tomsovic94, Frischat98}, resonance-assisted tunneling~\cite{Brodier01, Brodier02}, and chaotic tunneling~\cite{Shudo95, Shudo98, Creagh96, Creagh99, Creagh99b} may rely on many or even a vast number of trajectories with similar imaginary actions.  There are different mechanisms for generating an enormous increase in the number of trajectories with similar strength contributions due to the presence of chaos.  To mention one example, in the work of Creagh and Whelan on chaotic tunneling~\cite{Creagh96, Creagh99}, there are two wells in which the real time dynamics within each well is chaotic.  There is a dominant path underneath the barrier involving imaginary time propagation, which connects directly to a short real time periodic trajectory.  However, due to the chaos, the short periodic trajectory possesses a real time homoclinic tangle with the usual associated exponentially-proliferating-with-excursion-length number of homoclinic trajectories.  As there is a periodic trajectory which mimics every homoclinic trajectory excursion, there is an exponentially proliferating number of periodic trajectories with an exponentially close approach to the imaginary time path under the barrier connected to the short periodic trajectory.  A tiny deviation of each of the longer exponentially proliferating periodic trajectories can connect to a nearly imperceptibly different trajectory through the barrier, all possessing nearly the same imaginary part of their classical actions, and hence having nearly identical exponential suppressions.  On the other hand, the real parts of the actions vary considerably and generate complicated interference effects in the tunneling behaviors unlike the simple exponential decay seen for the Airy function and typical integrable system tunneling phenomena.

As mentioned above at the end of Sec.~\ref{sec:psb}, under perturbations chaotic systems sometimes generate bifurcations in their dynamics, which discontinuously adds (or subtracts) contributions to trace formulas, such as the Gutzwiller trace formula.  In a sense, real periodic trajectories ``disappear'' into complex phase space on the opposite sides of the bifurcations from which the real periodic trajectories exist.  In fact, their contributions show up in quantum spectra if not too far in parameter space away from the bifurcation.  Known as periodic ghost orbits~\cite{Kus93}, they are complex periodic trajectories whose properties determine extra contributions to trace formulas and they can also be used to provide uniform approximations.  In~\cite{Main97}, it is shown how to use the periodic real and ghost orbits to determine the parameters in three main types of bifurcation catastrophes, fold, cusp, and butterfly.

Along with analytic continuation to complex phase space variables and time comes a number of complications.  One of the most significant is that if Hamilton's equations are nonlinear, then there exist complex initial conditions for which the trajectory escapes to infinity in finite times~\cite{Huber88}.  Since time varies continuously, this leads to branch cuts in the semiclassical theory.  It also complicates numerically finding complex trajectories as there is a tendency for an algorithm, such as a variable time step Runge-Kutta, to fall into an infinite sequence of continually shrinking the time step from which it cannot escape if it is too close to the neighborhood of such a trajectory.  Even for a system as simple as a pure quartic oscillator, there is an infinity of branch cuts~\cite{Wang22a}.  These branch cuts are in addition to Stokes phenomena~\cite{Stokes50, Stokes57} as a complication regarding which saddle points actually contribute to quantum propagation.  Determining which complex trajectories properly contribute to classically allowed and non-allowed processes becomes much more difficult.  It also raises complications in determining the equivalent of a Maslov index~\cite{Wang22b}.  The folklore that it is possible just to follow continuously the phase of the relevant stability matrix determinants and cut it in half turns out to be incorrect under some circumstances.

For the semiclassical theory of coherent state and wave packet propagation, these complications arise with real time propagation.  We go into some detail here to illustrate the issues.  The multidimensional version of the basic bosonic coherent state definition,  
\begin{equation}
\label{eq:cs}
| z_1 \rangle = \exp \left(-\frac{\left| z_1 \right|^2}{2} + z_1 \hat a^\dagger \right)| 0\rangle \ ,
\end{equation}
in a quadrature representation generates a form equivalent to that of a Gaussian wave packet, 
\begin{equation}
\label{eq:wavepacket}
\phi_1(\vec x) = \left[\frac{{\rm Det}\left({\bf b}_1+{\bf b}_1^*\right)}{(2\pi\hbar)^D}\right]^{1/4} \exp\left[ - \left(\vec x - \vec q_1 \right) \cdot \frac{{\bf b}_1}{2\hbar} \cdot \left(\vec x - \vec q_1 \right) +\frac{i \vec p}{\hbar} \cdot \left(\vec x - \vec q_1 \right)\right]  \ ,
\end{equation}
except for generating a different global phase convention of no concern here.  The shape parameters are all contained in the matrix $\bf b$ (contains all the variance and covariance information), which is taken to be complex symmetric in order for the formulas ahead to be valid.  In order for the wave packet to be localized in space, the real parts of the eigenvalues of ${\bf b}_1$ must be positive, and in this case it is a square integrable function.  The $(\vec q_1,\vec p_1)$ values are the wave packet position and momentum centroids, respectively, with strictly real values.  

There are two extremely useful connections between these states and classical mechanics.  First consider the Wigner transform of $\phi_1(\vec x)$,
\begin{equation}
\label{eq:wtwp}
{\cal W}_1(\vec p, \vec q) = \frac{1}{(2\pi\hbar)^{D}} \int_{-\infty}^\infty {\rm d} \vec x \ {\rm e}^{i \vec p \cdot \vec x/\hbar} \phi_1 \left(q-\frac{\vec x}{2}\right)  \phi^*_1 \left(q+\frac{\vec x}{2}\right) 
= \left(\pi \hbar \right)^{-D} \exp \left[ - \left(\vec p - \vec p_1, \vec q - \vec q_1 \right) \cdot \frac{{\bf A}_1}{\hbar} \cdot \left(\vec p - \vec p_1, \vec q - \vec q_1 \right) \right] \ .
\end{equation}
where $D$ is the number of degrees of freedom, and ${\bf A}_1$ is a real symmetric, unit determinant, positive definite matrix,
\begin{equation}
\label{eq:mvg}
{\bf A}_1 = \left(\begin{array}{cc}
{\bf  c^{-1}} & {\bf  c}^{-1} \cdot {\bf  d}  \\
 {\bf  d} \cdot {\bf c}^{-1} & {\bf c} + {\bf  d} \cdot {\bf c}^{-1} \cdot {\bf  d} \end{array}  
\right)   
\end{equation}
constructed from the shape parameters using the notation 
\begin{equation}
\label{mvgwf}
{\bf b}_1 = {\bf c} + i {\bf d} \ .
\end{equation}
The Wigner transform, ${\cal W}_1(\vec p, \vec q)$, has a real, positive definite, localized, multivariate Gaussian form, which can be taken as a weighted density of real initial conditions, a Liouvillian density, for the usual classical propagation.   

The question is then how does that density evolve.  For extremely short times, much less than the logtime for a chaotic system, a local linearization is applicable and the result is given by replacing the $A_1$ matrix with
\begin{equation}
\label{eq:aevol}
{\bf A}_1 (t) = {{\bf M}_t^{-1}}^T \cdot {\bf A}_1 \cdot {\bf M}_t^{-1}
\end{equation}
in Eq.~\eqref{eq:wtwp} and substituting for $(\vec q_1, \vec p_1)$ the propagated coordinates $[\vec q_1(t), \vec p_1(t)]$~\cite{Tomsovic18b} to give a density ${\cal W}_1(\vec p, \vec q; t)$.  This is the exact classical equivalent of Heller's linearized wave packet dynamics~\cite{Heller75}.  At longer times for a chaotic system, assume the initial density ${\cal W}_1(\vec p, \vec q)$ is localized sufficiently to fit within the Birkhoff-Ozorio convergence zone of the $(\vec q_1, \vec p_1)$ trajectory.  All the neighboring trajectories have to stretch out along the unstable manifold of $(\vec q_1, \vec p_1)$ and converge onto it exponentially rapidly.  Geometrically, Berry et al.~\cite{Berry79b} described the structure as developing tendrils.  Furthermore, if there is interest in how the density ${\cal W}_1(\vec p, \vec q; t)$ overlaps with a second localized density ${\cal W}_2(\vec p, \vec q)$, it can be expressed as a sum over heteroclinic excursions using a similar kind of linearization that gave ${\bf A}_1 (t)$ in Eq.~\eqref{eq:aevol}, but individually for each heteroclinic excursion~\cite{Tomsovic91b, Oconnor92, Tomsovic93}; see the discussion in Sec.~\ref{sec:bnc} and Eq.~\eqref{eq:hetsumg}.  In the event that ${\cal W}_2(\vec p, \vec q)={\cal W}_1(\vec p, \vec q)$, then the sum is over homoclinic excursions.  This follows because just as the trajectories in ${\cal W}_1(\vec p, \vec q; t)$ collapse onto the unstable manifold forward in time, ${\cal W}_2(\vec p, \vec q; t)$ must collapse onto a stable manifold backward in time.  All the contributing trajectories must be in the neighborhood simultaneously of the unstable and stable manifolds, and each intersection gives a heteroclinic excursion whose local linearization captures the collective effects of the all the neighboring, i.e.~within the same partition, contributing trajectories.  The geometry is quite different in a generic integrable system where the evolution is found to develop whorls in~\cite{Berry79b}.  The overlap of ${\cal W}_1(\vec p, \vec q)$ with the various tori shears depending on the frequencies present in the relevant tori. 

The alternate corresponding classical picture, ``\`a la Maslov''~\cite{Maslov81}, comes from the identification of the appropriate Lagrangian manifolds for wave packets~\cite{Huber88},
\begin{equation}
\label{eq:constraints}
 {\bf b}_1 \cdot \left( \vec {\cal Q} - \vec q_1\right) + i \left( \vec {\cal P} - \vec p_1\right) = 0 \ ,
\end{equation}
which is a complex $D$-dimensional hyperplane in a complex $2D$-dimensional space.
The complex manifold associated with $\phi_1(\vec x)$ is propagated forward in time and used to construct the evolving wave function.  If a propagated coherent state/wave packet is overlapped with a final coherent state/wave packet, the evolved manifold is intersected with the manifold associated with  $\phi_2(\vec x)$.  This leads to the complex two-point boundary value equations, 
\begin{eqnarray}
\label{eq:sadcond}
{\bf b}_1 \cdot \left( \vec {\cal Q}_0 - \vec q_1\right) + i \left( \vec {\cal P}_0 - \vec p_1\right) &=& 0 \nonumber \\
{\bf b}^*_2 \cdot \left( \vec {\cal Q}_t - \vec q_2\right) - i \left( \vec {\cal P}_t - \vec p_2\right) &=& 0 \ ,
\end{eqnarray}
where in general the coordinates $(\vec {\cal Q}, \vec {\cal P})$ for all times must be complex.  The lower relation of Eq.~\eqref{eq:sadcond} has a slightly different form than the first to account for the difference between a ket vector and a bra vector.  

\begin{figure}[b]
\centering
\includegraphics[width=8.75cm]{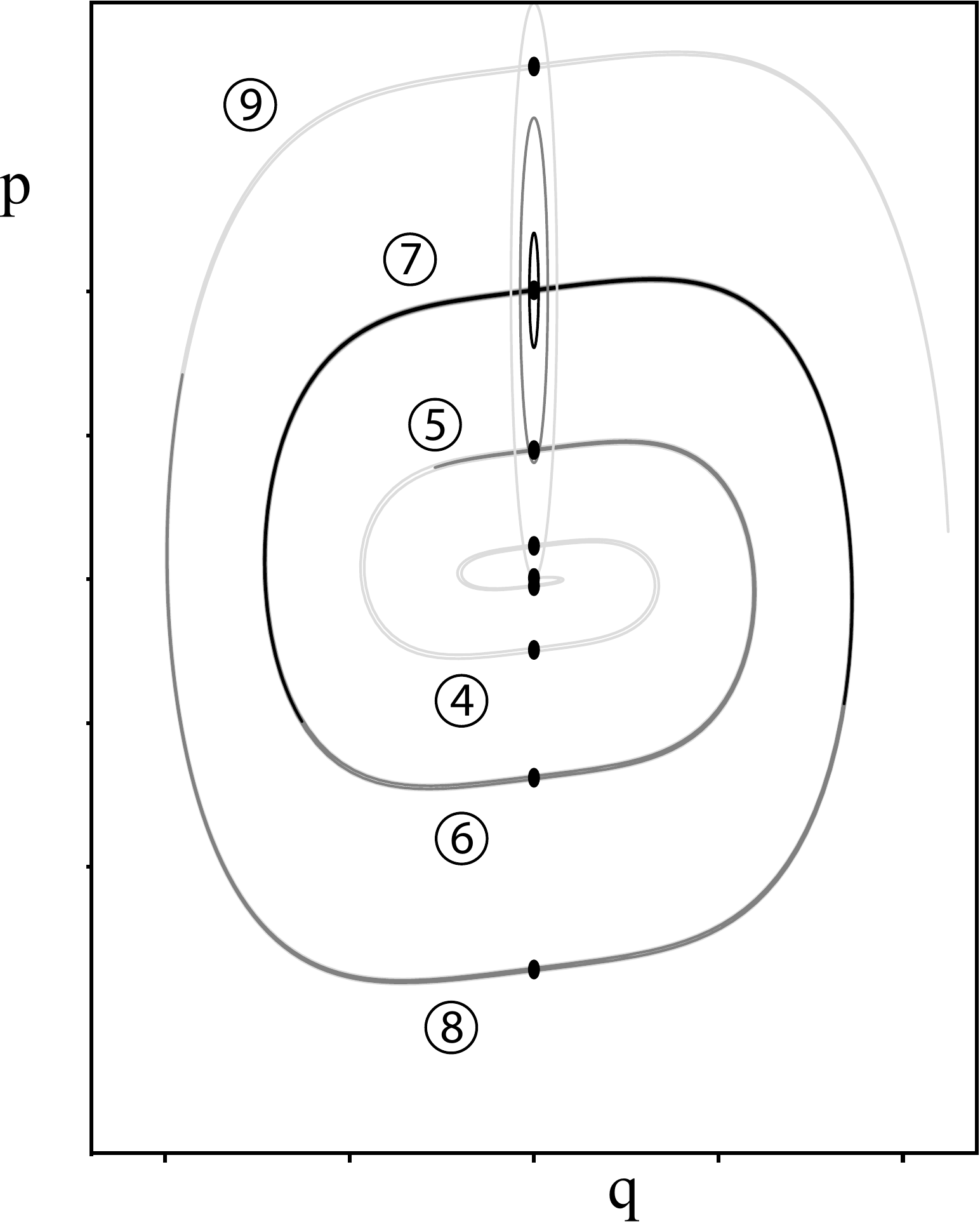}\qquad \includegraphics[width=7.25cm]{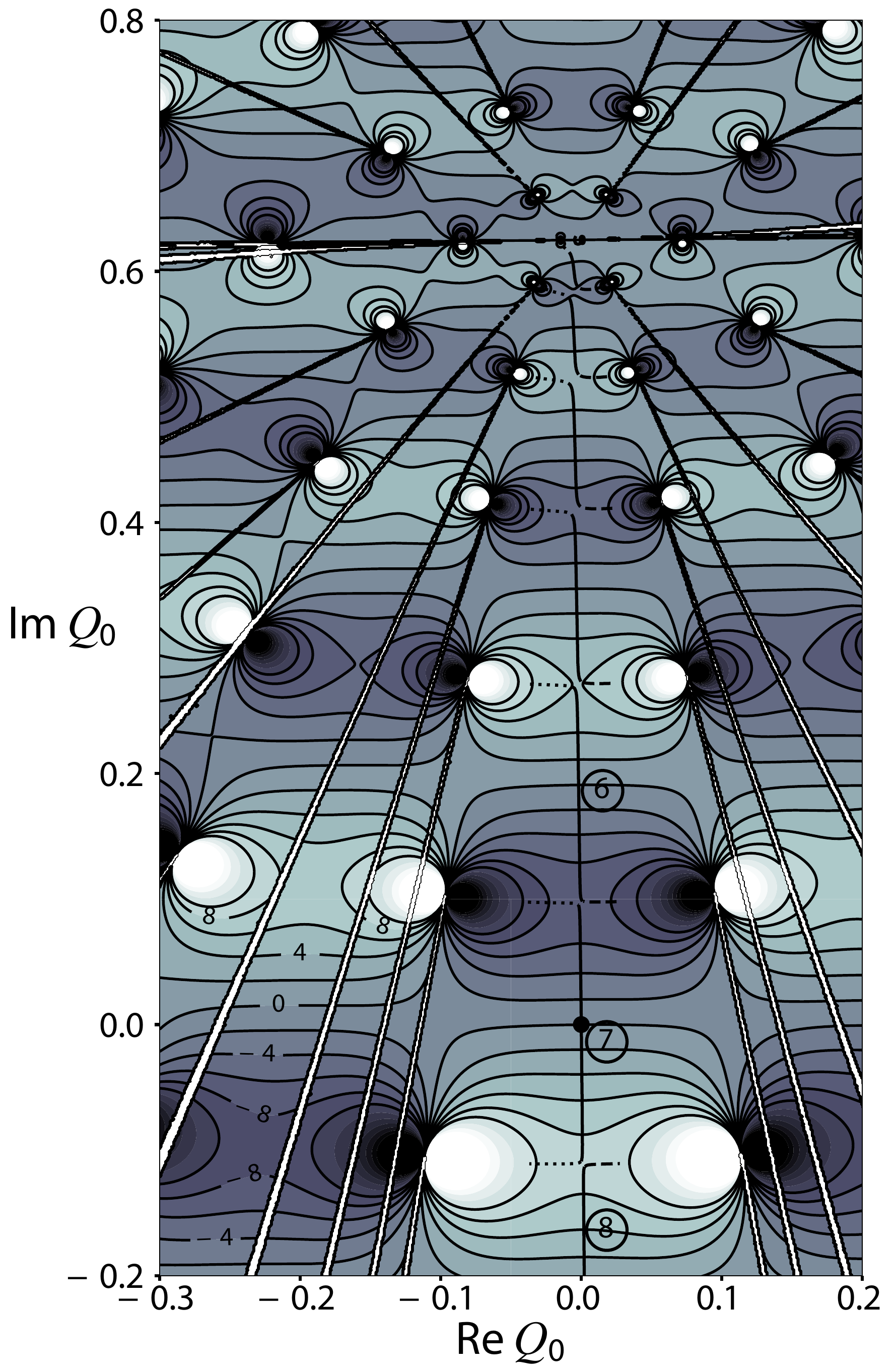}
\caption{Contours of the final position ${\cal Q}_t$'s real part  as a function of the trajectory initial conditions on the initial wave packet's Lagrangian manifold.  Real and imaginary parts of ${\cal P}_0$ follow using Eq.~\eqref{eq:constraints}; a similar figure (Fig.~3) was given in~\cite{Huber88}.  The black dot near \textcircled{\raisebox{-0.9pt}{7}} indicates the real initial condition of the Wigner density centroid. The initial conditions of exposed saddles associated with foliations \textcircled{\raisebox{-0.9pt}{6}}, \textcircled{\raisebox{-0.9pt}{7}}, \textcircled{\raisebox{-0.9pt}{8}} as a function of the propagated wave function's position $x$ appear as nearly vertical solid lines perpendicular to the contour lines of $Re\,x$; they are located on $Im\,x=0$ contours which are not shown.  The transition from exposed to hidden occurs at the {\it avoided crossing} of two saddles from different foliations, which occurs near the classical turning point.   The solid line turns to dashed or dotted where it transitions to hidden. The dashes are for saddles that crossed a Stokes line and must be excluded. Modified from Figs. 3 \& 4 of~\cite{Wang22a}.}
\label{fig:contourfig}
\end{figure}
The connection between these two pictures is illustrated in Fig.~\ref{fig:contourfig} for the one degree-of-freedom, purely quartic oscillator
\begin{equation}
H(q,p) = \frac{p^2}{2m} + \alpha q^4 \ ;
\end{equation}
see~\cite{Wang22a}.  In the left panel, an initial Gaussian ${\cal W}_1(\vec p, \vec q)$ is shown with its $1,3,5$ standard deviation contours.  It is centered on the black dot in the middle of branch \textcircled{\raisebox{-0.9pt}{7}}.   After three periods of the motion of its central trajectory, it has deformed into a whorl some of whose branches are labeled by the circled numbers.  It shows the Wigner phase space portrait of the classically evolved state.  On the right is a representation of the most relevant part of the propagated complex Lagrangian manifold, Eq.~\eqref{eq:constraints}, for the same state.  It is possible to associate each branch of the whorl with a contour appearing in the propagated Lagrangian manifold, three of which are labelled.  For chaotic systems, a similar association would also exist with the various branches of unstable and stable manifolds seen as tendrils and contours in the complex propagated manifold, but for the purpose of illustration it is not very helpful for illustration (hence the simpler integrable case given here).

Curiously, for nonlinear dynamical systems, there is an infinity of solutions for any fixed time, $t>0$, to the two-point boundary value problem defined for wave packets in Eq.~\eqref{eq:sadcond}, almost all of which must be rejected.  The solutions which must be kept are not necessarily easy to identify.  In a system with $D$ degrees of freedom, thus a phase space dimension of $2D$, each dimension is complex and requires $4D$ parameters to specify a point.  The constraints of Eq.~\eqref{eq:constraints} reduce the surface to an $D$ dimensional complex hyperplane.  After propagation, it becomes wildly more complicated and solutions appear at the intersections with the second $D$ dimensional complex hyperplane.  Figure~\ref{fig:contourfig} illustrates about the simplest case, an integrable one degree-of-freedom system.  The narrow blank lines that appear correspond to the aforementioned branch cuts.  Those are the initial conditions in which the system heads off to infinity on or before three periods of the central trajectory.  All of the saddle trajectories for classically allowed and non-allowed quantum processes are found within the lower central region bounded by the branch cuts.   The  branch cut and contour pattern visible keeps repeating off to infinity.  Nearly all of the Lagrangian manifold is irrelevant for semiclassical purposes.

There is a very interesting consequence of this circumstance in terms of the analytically continued complex trajectories.  Semiclassically, the overlap of a propagated initial localized wave packet with a final wave packet must be evaluated by the method of steepest descents.  All the saddle points correspond to complex trajectory solutions (with a single rare exception).  It turns out that for chaotic dynamical systems each heteroclinic contribution to an equivalent of the real trajectory classical summation, Eq.~\eqref{eq:hetsumg}, has a corresponding complex saddle point solution, which must be kept in the semiclassical summation~\cite{Pal16, Tomsovic18, Tomsovic18b}.   Each term in the saddle point summation must give nearly the identical contribution as each corresponding term given in the semiclassical heteroclinic trajectory summation given in~\cite{Tomsovic91b,Tomsovic93}.  In fact, starting with real heteroclinic trajectories and using a Newton-Raphson scheme, it is possible to identify the contributing complex saddle trajectories in a vastly simpler way than directly searching for those solutions, and without the wasted effort of identifying the infinity of saddle solutions that must be discarded.  Thus, there exists a close mathematical relationship between a complex saddle action and a real (stationary phase) action,  a complex stability matrix and real stability matrix, as well as Maslov indices for complex and real trajectories.  

These complex trajectories have nothing to do with classically forbidden processes, and they behave quite differently than the complex trajectories which are associated with forbidden processes.  For example, the classical action gives a complex result with the real part generating the phase and the imaginary part a decaying exponential behavior.  Under parameter variation, the real part of the complex action of a trajectory associated with a classically allowed process varies rapidly, hence rapidly varying the phase whereas the imaginary part is nearly stationary and the amplitude varies very slowly.  On the other hand, for a trajectory associated with a forbidden process, the imaginary part of the action varies quickly and the opposite occurs, the phase variation is slow whereas the amplitude changes rapidly.  

Only the saddles associated with classically allowed transport pathways can be found following real trajectories, the others cannot.   Additional contributions from saddles associated with forbidden processes, if any, require an alternative approach for their discovery~\cite{Wang22a}.  Since there are an infinity of saddles which must be thrown away, the forbidden associated saddles cannot just be directly found.  Instead, one workable search scheme starts with allowed associated saddles and follows them under parameter variation beyond where they have correspondence with a real trajectory, such as pushing them beyond a caustic barrier where there cannot be real trajectories.  In doing so, the dominant saddle solutions corresponding to non-allowed processes are identified, and also Stokes lines and anti-Stokes lines are encountered that must be taken into account in the usual way.

\begin{figure}[t]
\centering
\includegraphics[width=3 cm]{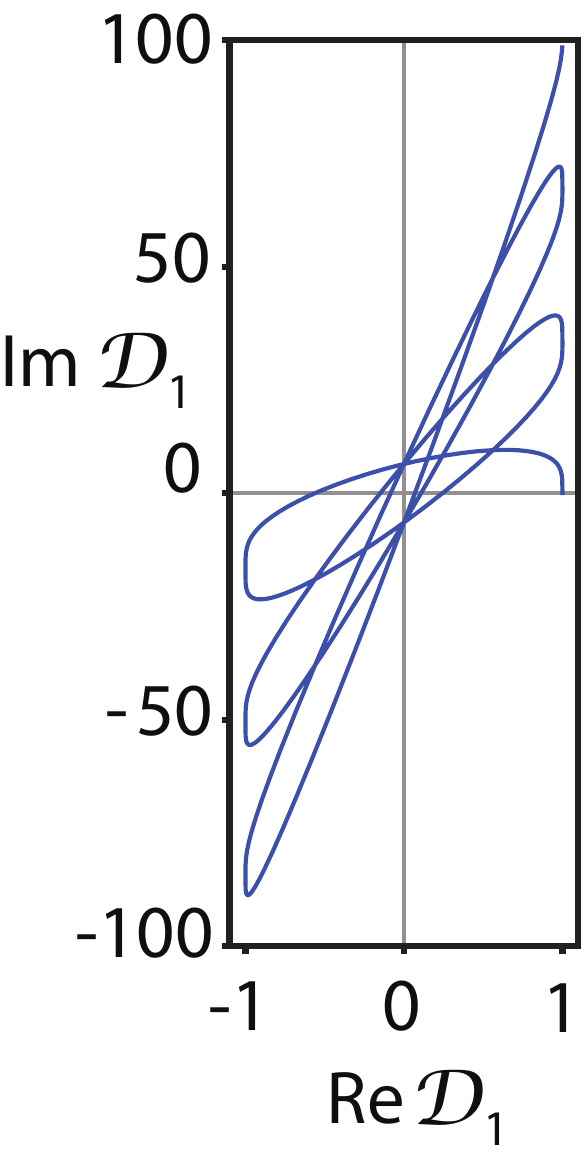} \includegraphics[width=3 cm]{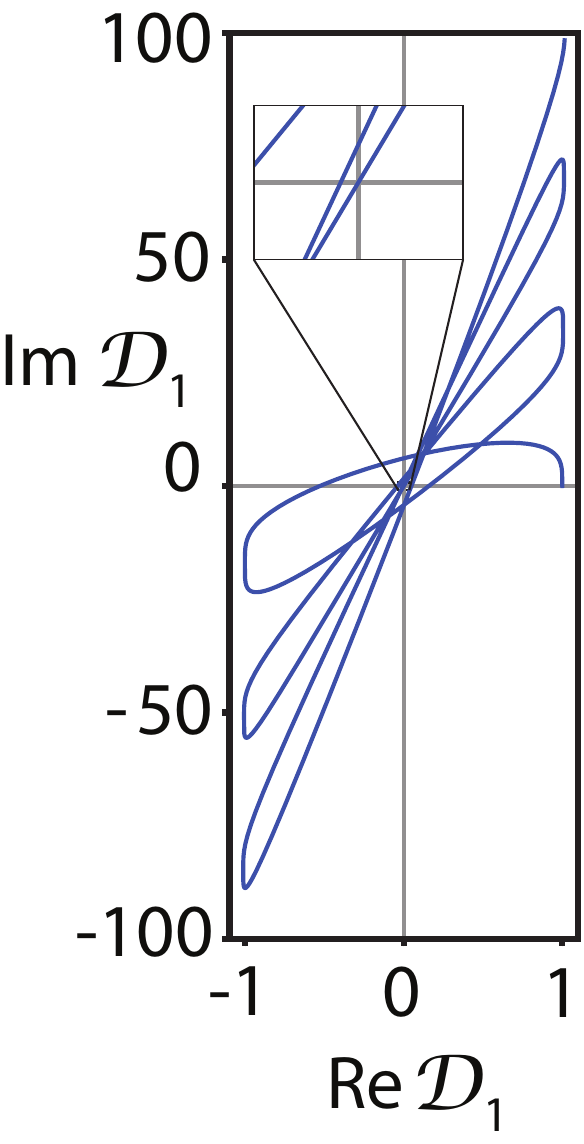}\ \ \ \includegraphics[width=10 cm]{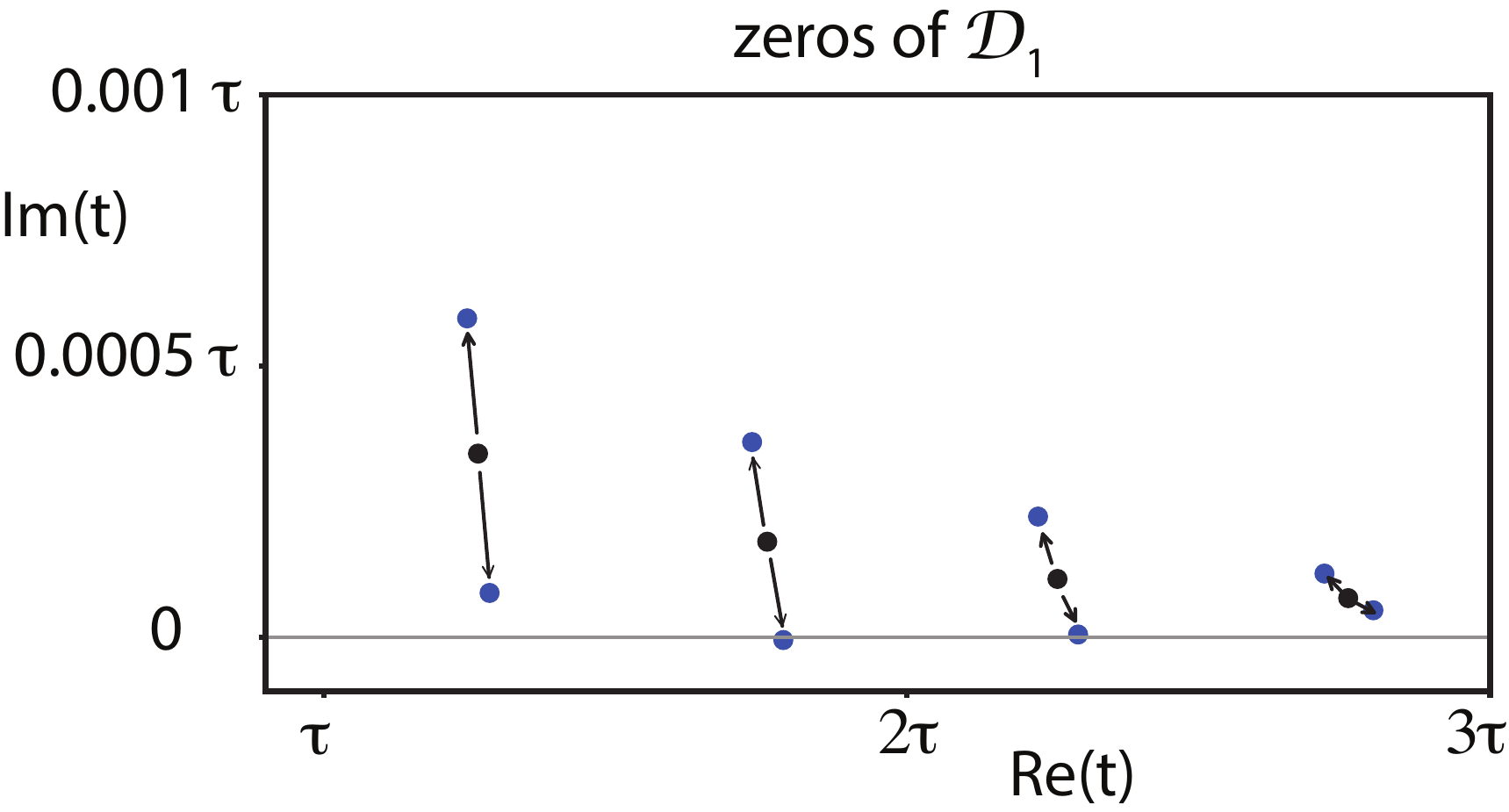} 
\caption{The evolution of the total accumulated phase for a time $t=3\tau$.  In the left panel, the real and imaginary parts of ${\cal D}_1$  are plotted for all times $0\le t \le 3\tau$.  Real trajectory initial conditions are used.  At $t=0$, the ${\cal D}_1$ curve starts at the point (1,0) and proceeds always in a counterclockwise direction as time increases up to the final time $3\tau$.  In the right panel, the phase for the critical saddle begins similarly, but the complexification of the stability matrix elements generates a point at which ${\cal D}_1=0$ for a real intermediate time that comes before the end of the propagation time.  In the right panel, the imaginary time zero's of ${\cal D}_1$ migrate under parameter variation.  Occasionally, one traverses the real time axis leading to a $\pi$ phase shift error and a breakdown of the common wisdom.  Reprinted from Figs. 3 \& 4 of~\cite{Wang22b}.
 \label{fig:fig3}} 
\end{figure}
Finally, it is necessary to understand how the phase of the resulting complex determinants behave as a function of time.  Recall that it is the inverse square root of the relevant determinant that enters a semiclassical expression.  The generally accepted common knowledge used to be that by following continuously the phase of this determinant's square root, there was no need to calculate a Maslov index.  However, this bit of common knowledge, while often giving the correct result, sometimes does not.  There are three main determinants that enter semiclassical expressions, ${\cal D}_0$ [Eq.~\eqref{eq:vanvleck}], ${\cal D}_1$, and ${\cal D}_2$, associated with the time Green function, coherent state/wave packet propagation, and transport coefficients, respectively, 
\begin{equation}
\label{eq:det}
{\cal D}_0 =  {\rm Det}\left({\bf M_{21}}\right) \quad \qquad {\cal D}_1 =  {\rm Det}\left({\bf M_{22}} + i {\bf M_{21}}\cdot {\bf b}_\alpha  \right) \quad \qquad {\cal D}_2 = {\rm Det}\left[{\bf M_{11}}\cdot {\bf b}_\alpha + {\bf b}^*_\beta \cdot {\bf M_{22}} + i \left({\bf b}^*_\beta\cdot  {\bf M_{21}}\cdot {\bf b}_\alpha - {\bf M_{12}} \right) \right]\ .
\end{equation}
Even though continuously following the relevant phase does not require a Maslov index, one can effectively  introduce an index by returning the final square root to a specific quadrant in the complex plane.  For each of the above determinants, the equivalent of a Maslov index can be calculated in this way.

For propagating the wave packet example of Fig.~\ref{fig:contourfig}, the determinant of interest is ${\cal D}_1$.  In this example, any real trajectory has a phase that evolves in time by rotating counterclockwise around zero.  See the illustration of this in the left panel of Fig.~\ref{fig:fig3}, which follows the 
phase accumulation for a real initial condition up to a time $t=3\tau$.  Often, the complex saddle trajectories behave exactly the same way, and the common wisdom works fine.  However, by altering a parameter in the calculation, such as final position, it occasionally happens that there is a sudden, discontinuous sign flip, i.e.~a $\pi$ phase shift, which is not possible.  It turns out this happens exactly when the complex deformation of the clockwise rotation of the determinant phase vanishes at a point in time; see the middle panel of Fig.~\ref{fig:fig3}.  On one side of this discontinuity, the total phase accumulated is different by $2\pi$ compared with the other side.  Hence, there is an abrupt $\pi$ phase shift for the square root determinant.  As discussed in \cite{Wang22b}, it turns out that for each rotation, there is a complex time value for which the determinant vanishes; see the right panel of Fig.~\ref{fig:fig3}.  As parameters are varied, these complex time zeros migrate.  If one of these zero's crosses the real time axis, it generates a $\pi$ phase shift error.  To correct the error properly, ideally the time propagation path would be deformed to a complex path that revolves around the zero in the correct orientation.  It is not absolutely necessary to go to the trouble of complex time paths as the alternative is to use parameter variation to identify possible candidates for phase errors and to just flip the sign each time a sudden sign flip is encountered (as long as one has a parameter value in which the determinant behaves well to begin with for each saddle contribution).

\section{Summary}
\label{sec:sum}

The semiclassical theory pillar of quantum chaos demands a rather extensive understanding of Hamiltonian chaos.  Relying on stationary phase and steepest descent asymptotic approximations, it expresses quantum mechanical quantities in terms of certain classical ingredients, such as classical actions, stability matrices, and Maslov indices.  The ingredients' behaviors become foremost in providing insight into a very broad range of phenomena, e.g.~level repulsion and rigidity, randomness of eigenstates, chaotic tunneling, diffraction, universal regimes of fidelity decay, parametric statistics, thermalization, etc...   A number of useful tools exist or have been developed in this general enterprise, the introduction of chaos paradigms, surfaces of section, dynamical maps, symbolic dynamics, stability analyses, action and uniformity principles, trace formulas, cycle expansions, catastrophe theory, etc..  The structural stability of chaotic systems also plays an important role.

In some cases, the questions that arise are not necessarily of direct obvious interest to a classical dynamics researcher.  An important extension for quantum phenomena is the study of classically non-allowed processes, which motivates analytically continuing classical mechanics to complex phase space variables and time.  In addition, even coherent state and wave packet dynamics lead us to complexified classical dynamics.  There are many complications that arise not present for real classical mechanics, such as trajectories that escape to infinity in finite times leading to branch cuts.  Nevertheless, complex mechanics adds immensely and frequently works beautifully in the description of non-allowed processes.  A great deal of progress has been made in the past $50$ or so years, which perhaps will continue.

\begin{ack}[Acknowledgments]
 
The author kindly acknowledges helpful discussions and a reading of the draft by Lukas Beringer, and Jizhou Li for help with a couple of figures.

\end{ack}

\bibliographystyle{JHEP}
\bibliography{classicalchaos, classicalchaos2, general_ref, manybody, nano, oceanacoustics, quantumchaos, rmtmodify}

\providecommand{\href}[2]{#2}\begingroup\raggedright\begin{thebibliography}{100}

\bibitem{Brody81}
T.A.~Brody, J.~Flores, J.B.~French, P.A.~Mello, A.~Pandey and S.S.M.~Wong,
  \emph{Random-matrix physics: spectrum and strength fluctuations},
  {\emph{Rev.~Mod.~Phys.} {\bfseries 53} (1981) 385}.

\bibitem{GutzwillerBook}
M.C.~Gutzwiller, \emph{Chaos in Classical and Quantum Mechanics},
  Springer-Verlag, New York (1990).

\bibitem{Chaosbook1}
P.~Cvitanovi\'c, R.~Artuso, P.~Dahlqvist, R.~Mainieri, G.~Tanner, G.~Vattay
  et~al., \emph{Chaos -- classical and quantum}, {\emph{chaosbook.org} 1}.

\bibitem{Stockmannbook}
H.-J.~St\"ockmann, \emph{Quantum Chaos: An Introduction}, Cambridge University
  Press, Cambridge (1999).

\bibitem{Dalessio16}
L.~D'Alessio, Y.~Kafri, A.~Polkovnikov and M.~Rigol, \emph{From quantum chaos
  and eigenstate thermalization to statistical mechanics and thermodynamics},
  {\emph{Adv.~Phys.} {\bfseries 65} (2016) 239}.

\bibitem{Haake18}
F.~Haake, S.~Gnutzmann and M.~Kus, \emph{Quantum signatures of chaos, fourth
  edition}, Springer, Heidelberg (2018).

\bibitem{Richter22}
K.~Richter, J.D.~Urbina and S.~Tomsovic, \emph{Semiclassical roots of
  universality in many-body quantum chaos}, {\emph{J.~Phys.~A: Math.~Theor.}
  {\bfseries 55} (2022) 453001}.

\bibitem{Altland23}
A.~Altland, B.~Post, J.~Sonner, J.~van~der Heijden and E.~Verlinde,
  \emph{Quantum chaos in 2d gravity}, {\emph{SciPost Phys.} {\bfseries 15}
  (2023) 064}.

\bibitem{Bohr13a}
N.~Bohr, \emph{I.~on the constitution of atoms and molecules},
  {\emph{Phil.~Mag.} {\bfseries 26} (1913) 1}.

\bibitem{Poincare92}
H.~Poincar\'e, \emph{Les m\'ethodes nouvelles de la m\'ecanique c\'eleste},
  vol.~1, Gauthier-Villars et fils, Paris (1892).

\bibitem{Poincare93}
H.~Poincar\'e, \emph{Les m\'ethodes nouvelles de la m\'ecanique c\'eleste},
  vol.~2, Gauthier-Villars et fils, Paris (1893).

\bibitem{Poincare99}
H.~Poincar\'e, \emph{Les m\'ethodes nouvelles de la m\'ecanique c\'eleste},
  vol.~3, Gauthier-Villars et fils, Paris (1899).

\bibitem{Lyapunov92}
A.M.~Lyapunov, \emph{The general problem of the stability of motion}, Ph.D.
  thesis, University of Kharkov, 1892.

\bibitem{Gutzwiller70}
M.C.~Gutzwiller, \emph{Energy spectrum according to classical mechanics},
  {\emph{Journal of Mathematical Physics} {\bfseries 11} (1970) 1791}.

\bibitem{Gutzwiller71}
M.C.~Gutzwiller, \emph{Periodic orbits and classical quantization conditions},
  {\emph{J.~Math.~Phys.} {\bfseries 12} (1971) 343}.

\bibitem{Balian71}
R.~Balian and C.~Bloch, \emph{Asymptotic evaluation of the green's functions
  for large quantum numbers}, {\emph{Ann.~Phys. (N.Y.)} {\bfseries 63} (1971)
  592}.

\bibitem{Maslov81}
V.P.~Maslov and M.V.~Fedoriuk, \emph{Semiclassical approximation in quantum
  mechanics}, Reidel Publishing Company, Dordrecht (1981).

\bibitem{Goldstein80}
H.~Goldstein, \emph{Classical mechanics}, Addison-Wesley, Reading (1980).

\bibitem{Heller91}
E.J.~Heller, \emph{Wavepacket dynamics and quantum chaology},  in \emph{Chaos
  and Quantum Physics}, M.J.~Giannoni, A.~Voros and J.~Zinn-Justin, eds.,
  (Amsterdam), pp.~547--663, North-Holland (1991).

\bibitem{Tomsovic18b}
S.~Tomsovic, \emph{Complex saddle trajectories for multidimensional quantum
  wave packet/coherent state propagation: application to a many-body system},
  {\emph{Phys.~Rev.~E} {\bfseries 98} (2018) 023301}.

\bibitem{Keller58}
J.B.~Keller, \emph{Corrected {Bohr}-{Sommerfeld} quantum conditions for
  nonseparable systems}, {\emph{Ann.~Phys. (N.Y.)} {\bfseries 4} (1958) 180}.

\bibitem{Birkhoff27}
G.D.~Birkhoff, \emph{On the periodic motions of dynamical systems}, {\emph{Acta
  Math.} {\bfseries 50} (1927) 359}.

\bibitem{Lichtenberg92}
A.J.~Lichtenberg and M.A.~Lieberman, \emph{Regular and Chaotic Dynamics},
  Springer, New York (1992).

\bibitem{Kolmogorov54}
A.N.~Kolmogorov, \emph{On the conservation of conditionally periodic motions
  under small perturbation of the hamiltonian}, {\emph{Dokl.~Akad.~Nauk SSSR}
  {\bfseries 98} (1954) 527}.

\bibitem{Arnold63}
V.I.~Arnol'd, \emph{Proof of a theorem of a.~n.~kolmogorov on the preservation
  of conditionally periodic motions under a small perturbation of the
  hamiltonian}, {\emph{Russ.~Math.~Surv.} {\bfseries 18} (1963) 9}.

\bibitem{Moser62}
J.~Moser, \emph{On invariant curves of area-preserving mappings of an annulus},
  {\emph{Nachr.~Akad.~Wiss.~G{\"o}ttingen,} {\bfseries II} (1962) 673}.

\bibitem{Thom89}
R.~Thom, \emph{Structural stability and morphogenesis}, CRC Press, Boca Raton
  (1989).

\bibitem{Arnold92}
V.I.~Arnol'd, \emph{Catastrophe Theory}, Springer-Verlag, Berlin (1992).

\bibitem{Ozoriobook}
A.M.~Ozorio~de Almeida, \emph{Hamiltonian systems: Chaos and quantization},
  Cambridge University Press, Cambridge (1988).

\bibitem{Ullmo16}
D.~Ullmo, \emph{Bohigas-giannoni-schmit conjecture},
  \href{https://doi.org/10.4249/scholarpedia.31721}{\emph{Scholarpedia}
  {\bfseries 11} (2016) 31721}.

\bibitem{Boltzmann71}
L.~Boltzmann, \emph{Einige allgemeine s\"atze \"uber w\"armegleichgewicht},
  {\emph{Wiener Berichte} {\bfseries 63} (1871) 679}.

\bibitem{Koopman32}
B.O.~Koopman and J.~{von Neumann}, \emph{Dynamical systems of continuous
  spectra}, {\emph{Proc.~Natl.~Acad.~Sci.} {\bfseries 18} (1932) 255}.

\bibitem{Kolmogorov58}
A.N.~Kolmogorov, \emph{New metric invariant of transitive dynamical systems and
  endomorphisms of {L}ebesgue spaces}, {\emph{Doklady of Russian Academy of
  Sciences} {\bfseries 119} (1958) 861}.

\bibitem{Sinai59}
Y.G.~Sinai, \emph{On the notion of entropy of a dynamical system},
  {\emph{Doklady of Russian Academy of Sciences} {\bfseries 124} (1959) 768}.

\bibitem{Anosov67}
D.V.~Anosov and Y.G.~Sinai, \emph{Some smooth ergodic systems},
  {\emph{Russ.~Math.~Surv.} {\bfseries 22} (1967) 103}.

\bibitem{Bohigas84}
O.~Bohigas, M.-J.~Giannoni and C.~Schmit, \emph{Characterization of chaotic
  quantum spectra and universality of level fluctuation laws},
  {\emph{Phys.~Rev.~Lett.} {\bfseries 52} (1984) 1}.

\bibitem{Andronov37}
A.A.~Andronov and L.~Pontryagin, \emph{Syst{\`e}mes grossiers},
  {\emph{Dokl.~Akad.~Nauk.~SSSR} {\bfseries 14} (1937) 247}.

\bibitem{Devaney22}
R.L.~Devaney, \emph{An Introduction to Chaotic Dynamical Systems}, CRC Press,
  Boca Raton, 3rd~ed. (2022).

\bibitem{Messiah61}
A.~Messiah, \emph{Quantum Mechanics}, Dover, New York (2014).

\bibitem{Merzbacher02}
E.~Merzbacher, \emph{The early history of quantum tunneling}, {\emph{Physics
  Today} {\bfseries 55} (2002) 44}.

\bibitem{Miller72}
W.H.~Miller and T.F.~George, \emph{Analytic continuation of classical mechanics
  for classically forbidden collision processes}, {\emph{J.~Chem.~Phys.}
  {\bfseries 56} (1972) 5668}.

\bibitem{Creagh98}
S.C.~Creagh, \emph{Tunneling in two dimensions},  in \emph{Tunneling in complex
  systems, Proceedings from the Institute for Nuclear Theory: Volume 5},
  S.~Tomsovic, ed., (Singapore), pp.~35--100, World Scientific (1998).

\bibitem{Klauder78}
J.R.~Klauder, \emph{Continuous representations and path integrals, revisited},
  in \emph{Proceedings of the NATO Advanced Study Institute on Path Integrals
  and their Applications in Quantum, Statistical, and Solid State Physics},
  G.J.~Papadopoulos and J.T.~Devresse, eds., (New York), pp.~5--38, Plenum
  (1978).

\bibitem{Baranger01}
M.~Baranger, M.A.M.~de~Aguiar, F.~Keck, H.J.~Korsch and B.~Schellhaass,
  \emph{Semiclassical approximations in phase space with coherent states},
  {\emph{J.~Phys.~A:~Math.~Gen.} {\bfseries 34} (2001) 7227}.

\bibitem{Huber87}
D.~Huber and E.J.~Heller, \emph{Generalized gaussian wave packet dynamics},
  {\emph{J.~Chem.~Phys.} {\bfseries 87} (1987) 5302}.

\bibitem{Huber88}
D.~Huber, E.J.~Heller and R.G.~Littlejohn, \emph{Generalized {G}aussian wave
  packet dynamics, {S}chr{\"o}dinger equation, and stationary phase
  approximation}, {\emph{J.~Chem.~Phys.} {\bfseries 89} (1988) 2003}.

\bibitem{Pal16}
H.~Pal, M.~Vyas and S.~Tomsovic, \emph{Generalized gaussian wave packet
  dynamics: Integrable and chaotic systems}, {\emph{Phys.~Rev.~E} {\bfseries
  93} (2016) 012213}.

\bibitem{Tomsovic18}
S.~Tomsovic, P.~Schlagheck, D.~Ullmo, J.-D.~Urbina and K.~Richter,
  \emph{Post-{Ehrenfest} many-body quantum interferences in ultracold atoms
  far-out-of-equilibrium}, {\emph{Phys.~Rev.~A} {\bfseries 97} (2018)
  061606(R)}.

\bibitem{Nonnenmacher26}
S.~Nonnenmacher, \emph{Quantum ergodicity and semiclassical measures:
  mathematical results}, {\emph{arXiv:2604.12098 [quant-ph] (to be published in
  this volume)} (2026) }.

\bibitem{Urbina26}
J.-D.~Urbina and K.~Richter, \emph{Quantum chaos in many-body systems of
  indistinguishable particles}, {\emph{arXiv:2604.12745 [quant-ph] (to be
  published in this volume)} (2026) }.

\bibitem{Mueller26}
S.~M{\"u}ller and M.~Sieber, \emph{Semiclassical periodic-orbit theory for
  quantum spectra}, {\emph{arXiv:2605.19019 [quant-ph] (to be published in this
  volume)} (2026) }.

\bibitem{Harayama26}
T.~Harayama, \emph{Chaotic billiard lasers}, {\emph{arXiv:2604.23614 [quant-ph]
  (to be published in this volume)} (2026) }.

\bibitem{Lakshminarayan26}
A.~Lakshminarayan, \emph{Quantum chaos and quantum information: Interactions
  and implications}, {\emph{arXiv:2604.12267 [quant-ph] (to be published in
  this volume)} (2026) }.

\bibitem{Hentschel26}
M.~Hentschel, \emph{Quantum chaos in phase space}, {\emph{arXiv:2604.12741
  [quant-ph] (to be published in this volume)} (2026) }.

\bibitem{Novaes26}
M.~Novaes, \emph{Semiclassical theory of transport}, {\emph{arXiv:2604.13193
  [quant-ph] (to be published in this volume)} (2026) }.

\bibitem{Garcia26}
I.~Garc{\'i}a-Mata and D.~Wisniacki, \emph{Quantum analogues of exponential
  sensitivity: from loschmidt echo to krylov complexity},
  {\emph{arXiv:2604.12707 [quant-ph] (to be published in this volume)} (2026)
  }.

\bibitem{Shudo26}
A.~Shudo, \emph{Chaos and quantum tunneling}, {\emph{arXiv:2604.12926
  [quant-ph] (to be published in this volume)} (2026) }.

\bibitem{Dietz26}
B.~Dietz, \emph{Relativistic quantum billiards}, {\emph{arXiv:2604.13003
  [quant-ph] (to be published in this volume)} (2026) }.

\bibitem{Benenti26}
G.~Benenti, G.~Casati, J.~Gong and Z.~Zou, \emph{The quantum kicked rotor: A
  paradigm of quantum chaos. foundational aspects and new perspectives},
  {\emph{arXiv:2604.12316 [quant-ph] (to be published in this volume)} (2026)
  }.

\bibitem{Sauer97}
T.~Sauer, C.~Grebogi and J.A.~Yorke, \emph{How long do numerical chaotic
  solutions remain valid?}, {\emph{Phys.~Rev.~Lett.} {\bfseries 97} (1997) 59}.

\bibitem{Richter14}
M.~Richter, S.~Lange, A.~B\"acker and R.~Ketzmerick, \emph{Visualization and
  comparison of classical structures and quantum states of four-dimensional
  maps}, {\emph{Phys.~Rev.~E} {\bfseries 89} (2014) 022902}.

\bibitem{Firmbach18}
M.~Firmbach, S.~Lange, A.~B\"acker and R.~Ketzmerick, \emph{Three-dimensional
  billiards: Visualization of regular structures and their hierarchy},
  {\emph{Phys.~Rev.~E} {\bfseries 98} (2018) 022214}.

\bibitem{Chirikov79}
B.V.~Chirikov, \emph{A universal instability of many-dimensional oscillator
  systems}, {\emph{Phys.~Rep.} {\bfseries 52} (1979) 263}.

\bibitem{Arnold67}
V.I.~Arnol'd and A.~Avez, \emph{Probl\`emes ergodiques de la m\'ecanique
  classique}, Gauthier-Villars, Paris (1967).

\bibitem{Arnold78}
V.I.~Arnol'd, \emph{Mathematical Methods of Classical Mechanics}, Springer,
  Berlin (1978).

\bibitem{Greene79}
J.M.~Greene, \emph{A method for determining a stochastic transition},
  {\emph{J.~Math.~Phys.} {\bfseries 20} (1979) 1183}.

\bibitem{Greene81}
J.M.~Greene, R.S.~MacKay, F.~Vivaldi and M.J.~Feigenbaum, \emph{Universal
  behaviour in families of area-preserving maps}, {\emph{Physica~D} {\bfseries
  3} (1981) 468}.

\bibitem{MacKay84a}
R.S.~MacKay, J.D.~Meiss and I.C.~Percival, \emph{Transport in hamiltonian
  systems}, {\emph{Physica~D} {\bfseries 13} (1984) 55}.

\bibitem{MacKay84b}
R.S.~MacKay, J.D.~Meiss and I.C.~Percival, \emph{Stochasticity and transport in
  hamiltonian systems},
  \href{https://doi.org/10.1103/PhysRevLett.52.697}{\emph{Phys.~Rev.~Lett.}
  {\bfseries 52} (1984) 697}.

\bibitem{Channon80}
S.R.~Channon and J.L.~Lebowitz, \emph{Numerical experiments in stochasticity
  and homoclinic oscillation}, {\emph{Ann.~NY~Acad.~Sci.} {\bfseries 357}
  (1980) 108}.

\bibitem{Bensimon84}
D.~Bensimon and L.P.~Kadanoff, \emph{Extended chaos and disappearance of kam
  trajectories}, {\emph{Physica~D} {\bfseries 13} (1984) 82}.

\bibitem{MacKay87}
R.S.~MacKay, J.D.~Meiss and I.C.~Percival, \emph{Resonances in area-preserving
  maps}, {\emph{Physica} {\bfseries 27D} (1987) 1}.

\bibitem{Meiss15}
J.D.~Meiss, \emph{Thirty years of turnstiles and transport}, {\emph{Chaos}
  {\bfseries 25} (2015) 097602}.

\bibitem{Geisel86}
T.~Geisel, G.~Radons and J.~Rubner, \emph{Kolmogorov-arnol'd-moser barriers in
  the quantum dynamics of chaotic systems}, {\emph{Phys.~Rev.~Lett.} {\bfseries
  57} (1986) 2883}.

\bibitem{Brown86}
R.C.~Brown and R.E.~Wyatt, \emph{Quantum mechanical manifestation of cantori:
  wave-packet localization in stochastic regions}, {\emph{Phys.~Rev.~Lett.}
  {\bfseries 57} (1986) 1}.

\bibitem{Radons88}
G.~Radons and R.E.~Prange, \emph{Wave functions at the critical
  kolmogorov-arnol'd-moser surface}, {\emph{Phys.~Rev.~Lett.} {\bfseries 61}
  (1988) 1691 }.

\bibitem{Bohigas93}
O.~Bohigas, S.~Tomsovic and D.~Ullmo, \emph{Manifestations of classical phase
  space structures in quantum mechanics}, {\emph{Phys.~Rep.} {\bfseries 223}
  (1993) 43}.

\bibitem{Karney77}
C.F.F.~Karney and A.~Bers, \emph{Stochastic ion heating by a perpendicularly
  propagating electrostatic wave}, {\emph{Phys.~Rev.~Lett.} {\bfseries 39}
  (1977) 550}.

\bibitem{Oberthaler99}
M.K.~Oberthaler, R.M.~Godun, M.B.~d'Arcy, G.S.~Summy and K.~Burnett,
  \emph{Observation of quantum accelerator modes}, {\emph{Phys.~Rev.~Lett.}
  {\bfseries 83} (1999) 4447}.

\bibitem{Fishman03}
S.~Fishman, I.~Guarneri and L.~Rebuzzini, \emph{A theory for quantum
  accelerator modes in atom optics}, {\emph{J.~Stat.~Phys.} {\bfseries 110}
  (2003) 911}.

\bibitem{Fishman82}
S.~Fishman, D.R.~Grempel and R.E.~Prange, \emph{Chaos, quantum recurrences, and
  {A}nderson localization}, {\emph{Phys.~Rev.~Lett.} {\bfseries 49} (1982)
  509}.

\bibitem{Lloyd69}
P.~Lloyd, \emph{Exactly solvable model of electronic states in a
  three-dimensional disordered hamiltonian: non-existence of localized states},
  {\emph{J.~Phys.~C: Solid State Phys.} {\bfseries 2} (1969) 1717}.

\bibitem{Moore95}
F.L.~Moore, J.C.~Robinson, C.F.~Bharucha, B.~Sundaram and M.G.~Raizen,
  \emph{Atom optics realization of the quantum $\delta$-kicked rotor},
  {\emph{Phys.~Rev.~Lett.} {\bfseries 75} (1995) 4598}.

\bibitem{Shepelyansky83}
D.L.~Shepelyansky, \emph{Some statistical properties of simple classically
  stochastic quantum systems}, {\emph{Physica D} {\bfseries 8} (1983) 208}.

\bibitem{Dahlqvist90}
P.~Dahlqvist and G.~Russberg, \emph{Existence of stable orbits in the x**2 y**2
  potential}, {\emph{Phys.~Rev.~Lett.} {\bfseries 65} (1990) 2837}.

\bibitem{Tomsovic07}
S.~Tomsovic and A.~Lakshminarayan, \emph{Fluctuations of finite-time stability
  exponents in the standard map and the detection of small islands},
  {\emph{Phys.~Rev.~E} {\bfseries 76} (2007) 036207}.

\bibitem{Manchein09a}
C.~Manchein, M.W.~Beims and J.-M.~Rost, \emph{Footprints of sticky motion in
  the phase space of higher dimensional nonintegrable conservative systems},
  {\emph{arXiv:} {\bfseries 0907.4181 [nlin.CD]} (2009) }.

\bibitem{Lakshminarayan11}
A.~Lakshminarayan and S.~Tomsovic, \emph{Kolmogorov-{Sinai} entropy of
  many-body {Hamiltonian} systems}, {\emph{Phys.~Rev.~E} {\bfseries 84} (2011)
  016218}.

\bibitem{Kolmogorov59}
A.N.~Kolmogorov, \emph{Entropy per unit time as a metric invariant of
  automorphism}, {\emph{Doklady of Russian Academy of Sciences} {\bfseries 124}
  (1959) 754}.

\bibitem{Pesin77}
Y.B.~Pesin, \emph{Characteristic {L}yapunov exponents and smooth ergodic
  theory}, {\emph{Russ.~Math.~Surv.} {\bfseries 32} (1977) 55}.

\bibitem{Kolovsky04}
A.R.~Kolovsky and A.~Buchleitner, \emph{Quantum chaos in the {Bose}-{Hubbard}
  model}, {\emph{Europhys. ~Lett.} {\bfseries 68} (2004) 632}.

\bibitem{Wisdom87}
J.~Wisdom, \emph{Chaotic behavior in the solar system}, {\emph{Nucl.~Phys.~B,
  Proc.~Suppl.} {\bfseries 2} (1987) 391}.

\bibitem{Sussman92}
G.J.~Sussman and J.~Wisdom, \emph{Chaotic evolution of the solar system},
  {\emph{Science} {\bfseries 257} (1992) 56}.

\bibitem{Laskar94}
J.~Laskar, \emph{Large-scale chaos in the solar system}, {\emph{Astron.~\&
  Astrophys.} {\bfseries 287} (1994) L9}.

\bibitem{Lecar01}
M.~Lecar, F.A.~Franklin, M.J.~Holman and N.J.~Murray, \emph{Chaos in the solar
  system}, {\emph{Annu.~Rev.~Astron.~Astrophys.} {\bfseries 39} (2001) 581}.

\bibitem{Shinbrot92}
T.~Shinbrot, C.~Grebogi, J.~Wisdom and J.A.~Yorke, \emph{Chaos in a double
  pendulum}, {\emph{Amer.~J.~Phys.} {\bfseries 60} (1992) 491}.

\bibitem{Gutzwiller73}
M.C.~Gutzwiller, \emph{The anisotropic {K}epler problem in two dimensions},
  {\emph{J.~Math.~Phys.} {\bfseries 14} (1973) 139}.

\bibitem{Devaney78}
R.L.~Devaney, \emph{Nonregularizability of the anisotropic {K}epler problem},
  {\emph{J.~Differ.~Equ.} {\bfseries 29} (1978) 253}.

\bibitem{Delande86}
D.~Delande and J.C.~Gay, \emph{Quantum chaos and statistical properties of
  energy levels: Numerical study of the hydrogen atom in a magnetic field},
  \href{https://doi.org/10.1103/PhysRevLett.57.2006}{\emph{Phys. Rev. Lett.}
  {\bfseries 57} (1986) 2006}.

\bibitem{Holle88}
A.~Holle, J.~Main, G.~Wiebusch, H.~Rottke and K.H.~Welge, \emph{Quasi-landau
  spectrum of the chaotic diamagnetic hydrogen atom}, {\emph{Phys.~Rev.~Lett.}
  {\bfseries 61} (1988) 161}.

\bibitem{Friedrich89}
H.~Friedrich and D.~Wintgen, \emph{The hydrogen atom in a uniform magnetic
  field--an example of chaos}, {\emph{Phys.~Rep.} {\bfseries 183} (1989) 37}.

\bibitem{Iu91}
C.-h.~Iu, G.R.~Welch, M.M.~Kash, D.~Kleppner, D.~Delande and J.C.~Gay,
  \emph{Diamagnetic rydberg atom: Confrontation of calculated and observed
  spectra}, \href{https://doi.org/10.1103/PhysRevLett.66.145}{\emph{Phys. Rev.
  Lett.} {\bfseries 66} (1991) 145}.

\bibitem{Bayfield74}
J.E.~Bayfield and P.M.~Koch, \emph{Multiphoton ionization of highly excited
  hydrogen atoms}, {\emph{Phys.~Rev.~Lett.} {\bfseries 33} (1974) 258}.

\bibitem{Galvez88}
E.J.~Galvez, B.E.~Sauer, L.~Moorman, P.M.~Koch and D.~Richards, \emph{Microwave
  ionization of h-atoms - breakdown of classical dynamics for
  high-frequencies}, {\emph{Phys.~Rev.~Lett.} {\bfseries 61} (1988) 2011}.

\bibitem{Koch95}
P.M.~Koch and K.A.H.~van Leeuwen, \emph{The importance of resonances in
  microwave ionization of excited hydrogen atoms}, {\emph{Phys.~Rep.}
  {\bfseries 255} (1995) 289}.

\bibitem{Sinai70}
Y.G.~Sinai, \emph{Dynamical systems with elastic reflections. ergodic
  properties of dispersing billiards.}, {\emph{Russ.~Math.~Surv.} {\bfseries
  25} (1970) 137}.

\bibitem{Lorentz05a}
H.A.~Lorentz, \emph{The motion of electrons in metallic bodies i},
  {\emph{Proc.~K.~Ned.~Akad.~Wet.} {\bfseries 7} (1905) 438}.

\bibitem{Lorentz05b}
H.A.~Lorentz, \emph{The motion of electrons in metallic bodies ii},
  {\emph{Proc.~K.~Ned.~Akad.~Wet.} {\bfseries 7} (1905) 585}.

\bibitem{Lorentz05c}
H.A.~Lorentz, \emph{The motion of electrons in metallic bodies iii},
  {\emph{Proc.~K.~Ned.~Akad.~Wet.} {\bfseries 7} (1905) 684}.

\bibitem{Bunimovich74}
L.A.~Bunimovich, \emph{On ergodic properties of certain billiards},
  {\emph{Func. ~Anal.~Appl.} {\bfseries 8} (1974) 254}.

\bibitem{Hopf37}
E.~Hopf, \emph{Ergodentheorie}, Springer-Verlag, Berlin (1937).

\bibitem{Vanvleck28}
J.H.V.~Vleck, \emph{The correspondence principle in the statistical
  interpretation of quantum mechanics}, {\emph{Proc.~Natl.~Acad.~Sci.~U.S.A.}
  {\bfseries 14} (1928) 178}.

\bibitem{Einstein17}
A.~Einstein, \emph{Zum quantensatz von sommerfeld und epstein},
  {\emph{Verh.~Deutsch.~Phys.~Ges.~Berlin} {\bfseries 19} (1917) 82}.

\bibitem{Brillouin26}
L.~Brillouin, \emph{La m{\'e}canique ondulatoire de sch{r\"o}dinger: une
  m{\'e}thode g{\'e}n{\'e}rale de resolution par approximations successives},
  {\emph{Comptes Rendus de l'Acad.~des Sci.} {\bfseries 183} (1926) 24}.

\bibitem{Li18}
J.~Li and S.~Tomsovic, \emph{Exact relations between homoclinic and periodic
  orbit actions in chaotic systems}, {\emph{Phys.~Rev.~E} {\bfseries 97} (2018)
  022216}.

\bibitem{Ozorio89}
A.M.~Ozorio~de Almeida, \emph{On the quantisation of homoclinic motion},
  {\emph{Nonlinearity} {\bfseries 2} (1989) 519}.

\bibitem{Li17}
J.~Li and S.~Tomsovic, \emph{Accurate determination of heteroclinic orbits in
  chaotic dynamical systems}, {\emph{J.~Phys.~A: Math.~Theor.} {\bfseries 50}
  (2017) 135101}.

\bibitem{Cvitanovic88}
P.~Cvitanovi\'{c}, \emph{Invariant measurement of strange sets in terms of
  cycles}, {\emph{Phys.~Rev.~Lett.} {\bfseries 61} (1988) 2729}.

\bibitem{Cvitanovic89}
P.~Cvitanovi\'{c} and B.~Eckhardt, \emph{Periodic-orbit quantization of chaotic
  systems}, {\emph{Phys.~Rev.~Lett.} {\bfseries 63} (1989) 823}.

\bibitem{Sieber01}
M.~Sieber and K.~Richter, \emph{Correlations between periodic orbits and their
  role in spectral statistics}, {\emph{Physica Scripta} {\bfseries T90} (2001)
  128}.

\bibitem{Li17b}
J.~Li and S.~Tomsovic, \emph{Geometric determination of classical actions of
  heteroclinic and unstable periodic orbits}, {\emph{arXiv:} {\bfseries
  1703.07045v2 [nlin.CD]} (2022) }.

\bibitem{Ott97}
E.~Ott, \emph{Chaos in Dynamical Systems}, Cambridge University Press, New York
  (1997).

\bibitem{Wolfson01}
M.A.~Wolfson and S.~Tomsovic, \emph{On the stability of long-range sound
  propagation through a structured ocean}, {\emph{J.~Acoust.~Soc.~Am.}
  {\bfseries 109} (2001) 2693}.

\bibitem{Grassberger88}
P.~Grassberger, R.~Badii and A.~Politi, \emph{Scaling laws for invariant
  measures on hyperbolic and nonhyperbolic attractors}, {\emph{J.~Stat.~Phys.}
  {\bfseries 51} (1988) 135}.

\bibitem{Sepulveda89}
M.A.~Sep\'ulveda, R.~Badii and E.~Pollak, \emph{Spectral analysis of
  conservative dynamical systems}, {\emph{Phys.~Rev.~Lett.} {\bfseries 63}
  (1989) 1226}.

\bibitem{Amitrano92}
C.~Amitrano and R.S.~Berry, \emph{Probability distributions of local lyapunov
  exponents for small clusters}, {\emph{Phys.~Rev.~Lett.} {\bfseries 68} (1992)
  729}.

\bibitem{Amitrano93}
C.~Amitrano and R.S.~Berry, \emph{Probability distributions of local lyapunov
  exponents for hamiltonian systems}, {\emph{Phys.~Rev.~E} {\bfseries 47}
  (1993) 3158}.

\bibitem{Crisanti93}
A.~Crisanti, G.~Paladin and A.~Vulpiani, \emph{Products of Random Matrices in
  Statistical Physics}, Springer-Verlag, Berlin (1993).

\bibitem{Schomerus02}
H.~Schomerus and M.~Titov, \emph{Statistics of finite-time lyapunov exponents
  in a random time-dependent potential}, {\emph{Phys.~Rev.~E} {\bfseries 66}
  (2002) 066207}.

\bibitem{Beims07}
M.W.~Beims, C.~Manchein and J.-M.~Rost, \emph{Origin of chaos in soft
  interactions and signatures of non-ergodicity}, {\emph{Phys.~Rev.~E}
  {\bfseries 76} (2007) 056203}.

\bibitem{Selberg56}
A.~Selberg, \emph{Harmonic analysis and discontinuous groups in weakly
  symmetric riemannian spaces with applications to dirichlet series},
  {\emph{J.~Indian Math.~Soc., New Series} {\bfseries 20} (1956) 47}.

\bibitem{Tabor89a}
M.~Tabor, \emph{Chaos and Integrability in Nonlinear Dynamics: An
  Introduction}, Wiley, New York (1989).

\bibitem{Oseledec68}
V.I.~Oseledec, \emph{A multiplicative ergodic theorem. lyapunov characteristic
  numbers for dynamical systems}, {\emph{Trudy~Moskov.~Mat.~Ob\v{s}\v{c}.}
  {\bfseries 19} (1968) 197}.

\bibitem{Tall24}
J.~Tall and S.~Tomsovic, \emph{Reduced dimensional {M}onte {C}arlo method:
  Preliminary integrations}, {\emph{Phys.~Rev.~E} {\bfseries 109} (2024)
  045308}.

\bibitem{Cvitanovic91}
P.~Cvitanovi\'{c}, \emph{Periodic orbits as the skeleton of classical and
  quantum chaos}, {\emph{Physica~D} {\bfseries 51} (1991) 138}.

\bibitem{Tanner97}
G.~Tanner, \emph{How chaotic is the stadium billiard? a semiclassical
  analysis}, {\emph{J.~Phys.~A: Math.~Gen.} {\bfseries 30} (1997) 2863}.

\bibitem{McDonald79}
S.W.~McDonald and A.N.~Kaufman, \emph{Spectrum and eigenfunctions for a
  hamiltonian with stochastic trajectories},
  \href{https://doi.org/10.1103/PhysRevLett.42.1189}{\emph{Phys. Rev. Lett.}
  {\bfseries 42} (1979) 1189}.

\bibitem{McDonaldthesis}
S.W.~McDonald, \emph{Wave dynamics of regular and chaotic rays}, Ph.D. thesis,
  University of California, Lawrence Berkeley Laboratory, 1983.

\bibitem{Heller84}
E.J.~Heller, \emph{Bound-state eigenfunctions of classically chaotic
  {H}amiltonian systems: Scars of periodic orbits},
  \href{https://doi.org/10.1103/PhysRevLett.53.1515}{\emph{Phys. Rev. Lett.}
  {\bfseries 53} (1984) 1515}.

\bibitem{Hummel23}
Q.~Hummel, K.~Richter and P.~Schlagheck, \emph{Genuine many-body quantum scars
  along unstable modes in bose-hubbard systems}, {\emph{Phys.~Rev.~Lett.}
  {\bfseries 130} (2023) 250402}.

\bibitem{Hannay84}
J.H.~Hannay and A.M.O.~de~Almeida, \emph{Periodic orbits and a correlation
  function for the semiclassical density of states}, {\emph{J.~Phys.~A}
  {\bfseries 17} (1984) 3429}.

\bibitem{Sieber99}
M.~Sieber, \emph{Geometrical theory of diffraction and spectral statistics},
  {\emph{J.~Phys.~A: Math.~Gen.} {\bfseries 32} (1999) 7679}.

\bibitem{Argaman96}
N.~Argaman, \emph{Semiclassical analysis of the quantum interference
  corrections to the conductance of mesoscopic systems},
  \href{https://doi.org/10.1103/PhysRevB.53.7035}{\emph{Phys. Rev. B}
  {\bfseries 53} (1996) 7035}.

\bibitem{Richter02}
K.~Richter and M.~Sieber, \emph{Semiclassical theory of chaotic quantum
  transport},
  \href{https://doi.org/10.1103/PhysRevLett.89.206801}{\emph{Phys.~Rev.~Lett.}
  {\bfseries 89} (2002) 206801}.

\bibitem{Berry85}
M.V.~Berry, \emph{Semiclassical theory of spectral rigidity},
  {\emph{Proc.~R.~Soc.~A} {\bfseries 400} (1985) 229}.

\bibitem{Elton10}
J.R.~Elton, A.~Lakshminarayan and S.~Tomsovic, \emph{Fluctuations in classical
  sum rules}, {\emph{Phys.~Rev.~E} {\bfseries 82} (2010) 046223}.

\bibitem{Pollicott11}
M.~Pollicott and R.~Sharp, \emph{On the {H}annay-{O}zorio de {A}lmeida sum
  formula},  in \emph{Dynamics, Games and Science II. Springer Proceedings in
  Mathematics, vol 2}, M.~Peixoto, M.~Pinto and A.~Rand, eds., (Berlin),
  pp.~575--590, Springer (2011).

\bibitem{Lakshminarayan93}
A.~Lakshminarayan and N.L.~Balazs, \emph{The classical and quantum mechanics of
  lazy baker maps}, {\emph{Ann.~Phys. (N.Y.)} {\bfseries 226} (1993) 350}.

\bibitem{Pollicott85}
M.~Pollicott, \emph{On the rate of mixing of axiom a flows},
  {\emph{Invent.~Math.} {\bfseries 81} (1985) 413}.

\bibitem{Pollicott86}
M.~Pollicott, \emph{Meromorphic extensions of generalised zeta functions},
  {\emph{Invent.~Math.} {\bfseries 85} (1986) 147}.

\bibitem{Ruelle86}
D.~Ruelle, \emph{Resonances of chaotic dynamical systems},
  {\emph{Phys.~Rev.~Lett.} {\bfseries 56} (1986) 405}.

\bibitem{Ruelle87}
D.~Ruelle, \emph{Resonances for axiom a flow}, {\emph{J.~Differ.~Geom.}
  {\bfseries 25} (1987) 99}.

\bibitem{Wiggins88}
S.~Wiggins, \emph{Global Bifurcations and Chaos}, Springer-Verlag, New York,
  Berlin, Heidelberg (1988).

\bibitem{Ozorio88}
A.M.~Ozorio~de Almeida, \emph{Hamiltonian systems: chaos and quantization},
  Cambridge University Press, Cambridge (1988).

\bibitem{Li17a}
J.~Li and S.~Tomsovic, \emph{Geometric determination of classical actions of
  heteroclinic and unstable periodic orbits}, {\emph{Phys.~Rev.~E} {\bfseries
  95} (2017) 062224}.

\bibitem{Moser56}
J.~Moser, \emph{The analytic invariants of an area-preserving mapping near a
  hyperbolic fixed point}, {\emph{Commun.~Pure Appl.~Math.} {\bfseries II}
  (1956) 673}.

\bibitem{Silva87}
G.L.~da~Silva~Ritter, A.M.~Ozorio~de Almeida and R.~Douady, \emph{Analytical
  determination of unstable periodic orbits in area preserving maps},
  {\emph{Physica~D} {\bfseries 29} (1987) 181}.

\bibitem{Tomsovic93}
S.~Tomsovic and E.J.~Heller, \emph{The long-time semiclassical dynamics of
  chaos: the stadium billiard}, {\emph{Phys.~Rev.~E} {\bfseries 47} (1993)
  282}.

\bibitem{Brodier01}
O.~Brodier, P.~Schlagheck and D.~Ullmo, \emph{Resonance-assisted tunneling in
  near-integrable systems}, {\emph{Phys.~Rev.~Lett.} {\bfseries 87} (2001)
  064101}.

\bibitem{Brodier02}
O.~Brodier, P.~Schlagheck and D.~Ullmo, \emph{Resonance-assisted tunneling},
  {\emph{Ann.~Phys. (N.Y.)} {\bfseries 300} (2002) 88}.

\bibitem{Percival80}
I.C.~Percival, \emph{Variational principles for invariant tori and cantori},
  {\emph{AIP Conf.~Proc.} {\bfseries 57} (1980) 302}.

\bibitem{Michler12}
M.~Michler, A.~B\"acker, R.~Ketzmerick, H.-J.~St\"ockmann and S.~Tomsovic,
  \emph{Universal quantum localizing transition of a partial barrier in a
  chaotic sea}, {\emph{Phys.~Rev.~Lett.} {\bfseries 109} (2012) 234101}.

\bibitem{Smale65}
S.~Smale, \emph{Diffeomorphisms with many periodic points},  in
  \emph{Differential and Combinatorial Topology: A Symposium in Honor of
  Marston Morse}, S.S.~Cairns, ed., (Princeton), Princeton University Press
  (1965).

\bibitem{Smale67}
S.~Smale, \emph{Differentiable dynamical systems},
  {\emph{Bull.~Amer.~Math.~Soc.} {\bfseries 73} (1967) 747}.

\bibitem{Robinson99}
C.~Robinson, \emph{Dynamical Systems: Stability, Symbolic Dynamics, and Chaos},
  CRC Press, Boca Raton, FL, 2~ed. (1999).

\bibitem{Morse38}
M.~Morse and G.A.~Hedlund, \emph{Symbolic dynamics}, {\emph{Amer.~J.~Math.}
  {\bfseries 60} (1938) 815}.

\bibitem{Christiansen96}
F.~Christiansen and A.~Politi, \emph{Symbolic encoding in symplectic maps},
  {\emph{Nonlinearity} {\bfseries 9} (1996) 1623}.

\bibitem{Biham92}
O.~Biham and M.~Kvale, \emph{The numerical computation of connecting orbits in
  dynamical systems}, {\emph{Phys.~Rev.~A} {\bfseries 46} (1992) 6334}.

\bibitem{Hansen95}
K.T.~Hansen and P.~Cvitanovic, \emph{Symbolic dynamics and markov partitions
  for the stadium billiard}, {\emph{arXiv:} {\bfseries 9502005 [chao-dyn]}
  (1995) }.

\bibitem{Hadamard1898}
J.~Hadamard, \emph{Les surfaces \`{a} courbures oppos\'{e}es et leurs lignes
  g\'{e}odesiques}, {\emph{J.~Math.~Pures~Appl.~series 5} {\bfseries 4} (1898)
  27}.

\bibitem{Coven08}
E.~Coven and Z.~Nitecki, \emph{On the genesis of symbolic dynamics as we know
  it}, \href{https://doi.org/10.4064/cm110-2-1}{\emph{Colloquium Mathematicae}
  {\bfseries 110} (2008) 227}.

\bibitem{Hedlund44}
G.~Hedlund, \emph{Sturmian minimal sets}, {\emph{Amer.~J.~Math.} {\bfseries 66}
  (1944) 605}.

\bibitem{Morse21}
H.M.~Morse, \emph{Recurrent geodesics on a surface of negative curvature},
  {\emph{Trans.~Am.~Math.~Soc.} {\bfseries 22} (1921) 84}.

\bibitem{Hirata23}
Y.~Hirata and J.M.~Amigo, \emph{A review of symbolic dynamics and symbolic
  reconstruction of dynamical systems}, {\emph{Chaos} {\bfseries 33} (2023)
  052101}.

\bibitem{Li20}
J.~Li and S.~Tomsovic, \emph{Homoclinic orbit expansion of arbitrary
  trajectories in chaotic systems: classical action function and its memory},
  {\emph{arXiv:} {\bfseries 2009.12224 [nlin.CD]} (2020) }.

\bibitem{Li19}
J.~Li and S.~Tomsovic, \emph{Exact decomposition of homoclinic orbit actions in
  chaotic systems: information reduction}, {\emph{Phys.~Rev.~E} {\bfseries 99}
  (2019) 032212}.

\bibitem{Li19a}
J.~Li and S.~Tomsovic, \emph{Asymptotic relationship between homoclinic and
  periodic orbit stability exponents}, {\emph{Phys.~Rev.~E} {\bfseries 100}
  (2019) 052202}.

\bibitem{Mitchell03a}
K.A.~Mitchell, J.P.~Handley, B.~Tighe, J.B.~Delos and S.K.~Knudson,
  \emph{Geometry and topology of escape. i. epistrophes}, {\emph{Chaos}
  {\bfseries 13} (2003) 880}.

\bibitem{Mitchell03b}
K.A.~Mitchell, J.P.~Handley, J.B.~Delos and S.K.~Knudson, \emph{Geometry and
  topology of escape. ii. homotopic lobe dynamics}, {\emph{Chaos} {\bfseries
  13} (2003) 892}.

\bibitem{Mitchell06}
K.A.~Mitchell and J.B.~Delos, \emph{A new topological technique for
  characterizing homoclinic tangles}, {\emph{Physica~D} {\bfseries 221} (2006)
  170}.

\bibitem{Mehta04}
M.L.~Mehta, \emph{Random Matrices (Third Edition)}, Elsevier, Amsterdam (2004).

\bibitem{Mueller04}
S.~M\"uller, S.~Heusler, P.~Braun, F.~Haake and A.~Altland, \emph{Semiclassical
  foundation of universality in quantum chaos},
  \href{https://doi.org/10.1103/PhysRevLett.93.014103}{\emph{Phys. Rev. Lett.}
  {\bfseries 93} (2004) 014103}.

\bibitem{Heusler07}
S.~Heusler, S.~M\"uller, A.~Altland, P.~Braun and F.~Haake,
  \emph{Periodic-orbit theory of level correlations}, {\emph{Phys.~Rev.~Lett.}
  {\bfseries 98} (2007) 044103}.

\bibitem{Lee85}
P.A.~Lee and A.D.~Stone, \emph{Universal conductance fluctuations in metals},
  {\emph{Phys.~Rev.~Lett.} {\bfseries 55} (1985) 1622}.

\bibitem{Altshuler85}
B.L.~Al'tshuler, \emph{Fluctuations in the extrinsic conductivity of disordered
  conductors}, {\emph{JETP Lett.} {\bfseries 41} (1985) 648}.

\bibitem{Goldberg91}
J.~Goldberg, U.~Smilansky, M.V.~Berry, W.~Schweizer, G.~Wunner and G.~Zeller,
  \emph{The parametric number variance}, {\emph{Nonlinearity} {\bfseries 4}
  (1991) 1}.

\bibitem{Marcus92}
C.M.~Marcus, A.J.~Rimberg, R.M.~Westervelt, P.F.~Hopkins and A.C.~Gossard,
  \emph{Conductance fluctuations and chaotic scattering in ballistic
  microstructures}, {\emph{Phys.~Rev.~Lett.} {\bfseries 69} (1992) 506}.

\bibitem{Bohigas95}
O.~Bohigas, M.-J.~Giannoni, A.M.~Ozorio~de Almeida and C.~Schmit, \emph{Chaotic
  dynamics and the {GOE-GUE} transition}, {\emph{Nonlinearity} {\bfseries 8}
  (1995) 203}.

\bibitem{Tomsovic00}
S.~Tomsovic, M.B.~Johnson, A.~Hayes and J.D.~Bowman, \emph{Statistical theory
  of parity nonconservation in compound nuclei}, {\emph{Phys.~Rev.~C}
  {\bfseries 62} (2000) 054607}.

\bibitem{Peres84}
A.~Peres, \emph{Ergodicity and mixing in quantum theory},
  \href{https://doi.org/10.1103/PhysRevA.30.504}{\emph{Phys.~Rev.~A} {\bfseries
  30} (1984) 504}.

\bibitem{Jalabert01}
R.A.~Jalabert and H.M.~Pastawski, \emph{Environment-independent decoherence
  rate in classically chaotic systems}, {\emph{Phys.~Rev.~Lett.} {\bfseries 86}
  (2001) 2490}.

\bibitem{Jacquod01b}
P.~Jacquod, P.G.~Silvestrov and C.W.J.~Beenakker, \emph{Golden rule decay
  versus lyapunov decay of the quantum loschmidt echo}, {\emph{Phys.~Rev.~E}
  {\bfseries 64} (2001) 055203}.

\bibitem{Cerruti02}
N.R.~Cerruti and S.~Tomsovic, \emph{Sensitivity of wave field evolution and
  manifold stability in chaotic systems}, {\emph{Phys.~Rev.~Lett.} {\bfseries
  88} (2002) 054103}.

\bibitem{Cerruti03}
N.R.~Cerruti and S.~Tomsovic, \emph{A uniform approximation for the fidelity in
  chaotic systems}, {\emph{J.~Phys.~A: Math.~Gen.} {\bfseries 36} (2003) 3451}.

\bibitem{Lakshminarayan99}
A.~Lakshminarayan, N.R.~Cerruti and S.~Tomsovic, \emph{Classical diffusion and
  quantum level velocities: Systematic deviations from random matrix theory},
  {\emph{Phys.~Rev.~E} {\bfseries 60} (1999) 3992}.

\bibitem{Lakshminarayan00}
A.~Lakshminarayan, N.R.~Cerruti and S.~Tomsovic, \emph{Phase space localization
  of chaotic eigenstates: violating ergodicity}, {\emph{Phys.~Rev.~E}
  {\bfseries 63} (2000) 016209}.

\bibitem{Cerruti00}
N.R.~Cerruti, A.~Lakshminarayan, J.H.~Lefebvre and S.~Tomsovic, \emph{Exploring
  phase space localization of chaotic eigenstates via parametric variation},
  {\emph{Phys.~Rev.~E} {\bfseries 63} (2000) 016208}.

\bibitem{Kus93}
M.~Ku{\'s}, F.~Haake and D.~Delande, \emph{Prebifurcation periodic ghost orbits
  in semiclassical quantization}, {\emph{Phys.~Rev.~Lett.} {\bfseries 71}
  (1993) 2167}.

\bibitem{Nussenzweig92}
H.M.~Nussenzweig, \emph{Diffraction Effects in Semiclassical Scattering},
  Cambridge University Press, Cambridge (1992).

\bibitem{Glauber63}
R.J.~Glauber, \emph{Coherent and incoherent states of the radiation field},
  {\emph{Phys.~Rev.} {\bfseries 131} (1963) 2766}.

\bibitem{Berry83}
M.V.~Berry, J.H.~Hannay and A.M.~Ozorio~de Almeida, \emph{Intensity moments of
  semiclassical wavefunctions}, {\emph{Physica} {\bfseries 17} (1983) 229}.

\bibitem{Davis81}
M.J.~Davis and E.J.~Heller, \emph{Quantum dynamical tunneling in bound states},
  {\emph{J.~Chem.~Phys.} {\bfseries 75} (1981) 246}.

\bibitem{Tomsovic94}
S.~Tomsovic and D.~Ullmo, \emph{Chaos-assisted tunneling}, {\emph{Phys.~Rev.~E}
  {\bfseries 50} (1994) 145}.

\bibitem{Frischat98}
S.D.~Frischat and E.~Doron, \emph{Dynamical tunneling in mixed systems},
  {\emph{Phys.~Rev.~E} {\bfseries 57} (1998) 1421}.

\bibitem{Shudo95}
A.~Shudo and K.S.~Ikeda, \emph{Complex classical trajectories and chaotic
  tunneling}, {\emph{Phys.~Rev.~Lett.} {\bfseries 74} (1995) 682}.

\bibitem{Shudo98}
A.~Shudo and K.S.~Ikeda, \emph{Chaotic tunneling: A remarkable manifestation of
  complex classical dynamics in non-integrable quantum phenomena},
  {\emph{Physica D} {\bfseries 115} (1998) 234}.

\bibitem{Creagh96}
S.C.~Creagh and N.D.~Whelan, \emph{Complex periodic orbits and tunneling in
  chaotic potentials}, {\emph{Phys.~Rev.~Lett.} {\bfseries 77} (1996) 4975}.

\bibitem{Creagh99}
S.C.~Creagh and N.D.~Whelan, \emph{A matrix element for chaotic tunnelling
  rates and scarring intensities}, {\emph{Ann.~Phys. (N.Y.)} {\bfseries 272}
  (1999) 196}.

\bibitem{Creagh99b}
S.C.~Creagh and N.D.~Whelan, \emph{Homoclinic structure controls chaotic
  tunneling}, {\emph{Phys.~Rev.~Lett.} {\bfseries 82} (1999) 5237}.

\bibitem{Main97}
J.~Main and G.~Wunner, \emph{Hydrogen atom in a magnetic field: Ghost orbits,
  catastrophes, and uniform semiclassical approximations}, {\emph{Phys.~Rev.~A}
  {\bfseries 55} (1997) 1743}.

\bibitem{Wang22a}
H.~Wang and S.~Tomsovic, \emph{Semiclassical propagation of coherent states and
  wave packets: hidden saddles}, {\emph{Phys.~Rev.~E} {\bfseries 105} (2022)
  054206}.

\bibitem{Stokes50}
G.G.~Stokes, \emph{On the numerical calculation of a class of definite
  integrals and infinite series}, {\emph{Trans.~Cambridge Philos.~Soc.}
  {\bfseries 9} (1850) 166}.

\bibitem{Stokes57}
G.G.~Stokes, \emph{On the discontinuity of arbitrary constants which appear in
  divergent developments}, {\emph{Trans.~Cambridge Philos.~Soc.} {\bfseries 10}
  (1857) 105}.

\bibitem{Wang22b}
H.~Wang and S.~Tomsovic, \emph{A corrected maslov index for complex saddle
  trajectories}, {\emph{Phys.~Rev.~E} {\bfseries 105} (2022) 054207}.

\bibitem{Heller75}
E.J.~Heller, \emph{Time-dependent approach to semiclassical dynamics},
  {\emph{J.~Chem.~Phys.} {\bfseries 62} (1975) 1544}.

\bibitem{Berry79b}
M.V.~Berry and N.L.~Balazs, \emph{Evolution of semiclassical quantum states in
  phase space}, {\emph{J.~Phys.~A} {\bfseries 12} (1979) 625}.

\bibitem{Tomsovic91b}
S.~Tomsovic and E.J.~Heller, \emph{Semiclassical dynamics of chaotic motion:
  Unexpected long time accuracy}, {\emph{Phys.~Rev.~Lett.} {\bfseries 67}
  (1991) 664}.

\bibitem{Oconnor92}
P.W.~O'Connor, S.~Tomsovic and E.J.~Heller, \emph{Semiclassical dynamics in the
  strongly chaotic regime - breaking the log-time barrier}, {\emph{Physica D}
  {\bfseries 55} (1992) 340}.

\end{thebibliography}\endgroup

\end{document}